\documentclass[twocolumn,preprintnumbers,superscriptaddress,amsmath,amssymb,nofootinbib]{revtex4}

\usepackage{graphicx}
\usepackage{dcolumn}
\usepackage{bm}
\usepackage{amsmath, amssymb}

\usepackage[normalem]{ulem}  

\usepackage{siunitx}

\renewcommand\sout{\bgroup \color{red} \ULdepth=-.5ex \ULset}

\newcommand{\Slash}[1]{\ooalign{\hfil/\hfil\crcr$#1$}}
\newcommand{\Psfig}[2]{\includegraphics[width=#1]{#2}}
\newcommand{\PsfigII}[2]{\includegraphics[scale=#1]{#2}}

\allowdisplaybreaks

\def\Kaellen{K\"{a}llen }
\def\Schr{Schr\"{o}dinger }

\def\prt{\partial}

\def\Rho{\text{P}}

\begin{document}

\preprint{}

\title{\boldmath Two-body wave functions and compositeness from
  scattering amplitudes:\\ II.~Application to the physical $N ^{\ast}$
  and $\Delta ^{\ast}$ resonances}

\author{Takayasu Sekihara} 
\email{sekihara@post.j-parc.jp}
\affiliation{Graduate School of Life and Environmental Science, 
  Kyoto Prefectural University, 
  Sakyo-ku, Kyoto 606-8522, Japan}
\affiliation{Research Center for Nuclear Physics
  (RCNP), Osaka University, Ibaraki, Osaka, 567-0047, Japan}
\affiliation{RIKEN Cluster for Pioneering Research, RIKEN, Wako
  351-0198, Japan}

\date{\today}

\begin{abstract}

  The meson--baryon molecular components for the $N^{\ast}$ and
  $\Delta ^{\ast}$ resonances are investigated in terms of the
  compositeness, which is defined as the norm of the two-body wave
  function from the meson--baryon scattering amplitudes.  The
  scattering amplitudes are constructed in a $\pi N$-$\eta N$-$\sigma
  N$-$\rho N$-$\pi \Delta$ coupled-channels problem in a meson
  exchange model together with several bare $N^{\ast}$ and $\Delta
  ^{\ast}$ states, and parameters are fitted so as to reproduce the
  on-shell $\pi N$ partial wave amplitudes up to the center-of-mass
  energy $\SI{1.9}{GeV}$ with the orbital angular momentum $L \le 2$.
  As a result, the Roper resonance $N (1440)$ is found to be dominated
  by the $\pi N$ and $\sigma N$ molecular components while the
  bare-state contribution is small.  The squared wave functions in
  coordinate space imply that both in the $\pi N$ and $\sigma N$
  channels the separation between the meson and baryon is about more
  than $\SI{1}{fm}$ for the $N (1440)$ resonance.  On the other hand,
  dominant meson--baryon molecular components are not observed in any
  other $N^{\ast}$ and $\Delta ^{\ast}$ resonances in the present
  model, although they have some fractions of the meson--baryon
  clouds.

\end{abstract}

\maketitle

\section{Introduction}

The spectroscopy of the nucleon resonances $N^{\ast}$ and $\Delta
^{\ast}$ is a key clue to understand nonperturbative aspects of
quantum chromodynamics (QCD), which is the fundamental theory of
hadrons and strong interactions~\cite{Zyla:2020zbs}.  Historically,
the $\Delta (1232)$ resonance opens the door to the color degrees of
freedom~\cite{Han:1965pf}, which is an essential idea of QCD.  The
masses, widths, and other properties such as transition strength of
the nucleon resonances have been good tests to examine the behavior of
constituent quarks inside them.  Today rich spectra of the nucleon
resonances have been revealed in the $\pi N$ coupled-channels
scattering amplitudes by dynamical coupled-channels
models~\cite{Krehl:1999km, Chen:2007cy, Ronchen:2012eg,
  Kamano:2013iva} as well as the $K$-matrix
approaches~\cite{Anisovich:2011fc, Workman:2012hx}.  In addition,
results from the lattice QCD simulations have been used to discuss the
$N^{\ast}$ and $\Delta ^{\ast}$ physics~\cite{Lang:2016hnn,
  Liu:2015ktc, Liu:2016uzk, Wu:2016ixr, Wu:2017qve}.  Recent studies
on the $N^{\ast}$ and $\Delta ^{\ast}$ spectroscopy can be found in,
e.g., Refs.~\cite{Segovia:2015hra, Ronchen:2014cna, Mart:2015jof,
  Ronchen:2015vfa, Beck:2016hcy, Kamano:2016bgm, Chen:2017pse,
  Ronchen:2018ury, Chen:2019fzn}.  Motivated by these rich spectra of
the nucleon resonances, we are now in a phase of clarifying their
internal structure.

In this study I utilize the $\pi N$ scattering amplitudes for the
calculations of the meson--baryon molecular components for the
$N^{\ast}$ and $\Delta ^{\ast}$ resonances.  This can be done in a
strategy developed in my first paper of a
series~\cite{Sekihara:2016xnq}, where I have shown that the two-body
wave function of the bound state, both in the stable and decaying
cases, can be extracted from the residue of the off-shell scattering
amplitude at the bound-state pole.  Furthermore, the normalization of
the two-body wave function from the residue is automatically achieved.
In this sense, once the $\pi N$ coupled-channels interactions are
fixed so as to reproduce the empirical $\pi N$ scattering amplitudes,
one can discuss the internal structure of the $N^{\ast}$ and $\Delta
^{\ast}$ resonances with the meson--baryon wave functions from the
off-shell parts of the scattering amplitudes and its norm, which is
called compositeness~\cite{Hyodo:2011qc, Aceti:2012dd, Hyodo:2013nka,
  Sekihara:2014kya}.  (For calculations of the compositeness, see,
e.g., Refs.~\cite{Aceti:2014ala, Nagahiro:2014mba, Kamiya:2015aea,
  Sekihara:2015gvw, Guo:2016wpy, Guo:2015daa, Lu:2016gev,
  Kang:2016ezb, Albaladejo:2016hae, Kamiya:2016oao, Kamiya:2017pcq,
  Tsuchida:2017gpb, Oller:2017alp, Guo:2019kdc}.  See also review
articles on hadronic molecules~\cite{Guo:2017jvc} and hadron--hadron
scattering~\cite{Oller:2019opk}.) In particular, from the $\pi N$
scattering amplitudes I can unveil amounts of the meson--baryon
molecular components of $N^{\ast}$ and $\Delta ^{\ast}$ resonances
which are claimed to be ``dynamically generated'' without bare
$N^{\ast}$ and $\Delta ^{\ast}$ states.  Furthermore, even for the
resonances which originate from bare states, I can evaluate how much
the bare states are dressed in the meson--baryon clouds in terms of
the compositeness.  These are the aim of the present paper, the second
paper of a series for the two-body wave function and compositeness in
general quantum systems following the first
paper~\cite{Sekihara:2016xnq}.

This paper is organized as follows.  First, I briefly show the
formulae of the two-body wave function and compositeness from the
scattering amplitude of constituents in Sec.~\ref{sec:2}.  Next, in
Sec.~\ref{sec:3} I construct an effective model to describe the $\pi
N$ scattering amplitudes and $N^{\ast}$ and $\Delta ^{\ast}$
resonances.  In Sec.~\ref{sec:4} I utilize the $\pi N$ scattering
amplitudes in the effective model for the calculations of the two-body
wave functions and compositeness for the $N^{\ast}$ and $\Delta
^{\ast}$ resonances, and I discuss their meson--baryon molecular
components.  Section~\ref{sec:5} is devoted to the conclusion of this
study.

\section{Formulae of compositeness}
\label{sec:2}

First of all, I briefly summarize the formulae of the two-body wave
functions and compositeness from the scattering amplitudes, which were
expressed in detail in Refs.~\cite{Sekihara:2015gvw,
  Sekihara:2016xnq}.  Below I focus on unstable resonance states, but
formulations and equations are applicable to stable bound states as
well.  After showing general formulae in Sec.~\ref{sec:2A}, I rewrite
them in terms of the complex scaling method for numerical calculations
of resonance wave functions in Sec.~\ref{sec:2B}.

\subsection{Two-body wave functions from scattering amplitudes}
\label{sec:2A}

In this study I consider a two-body to two-body scattering, $k (
\bm{q} ) \to j ( \bm{q}^{\prime} )$, where $k ( \bm{q} )$ and $j (
\bm{q}^{\prime} )$ denote the channels (relative three-momenta) of the
two-body initial and final states, respectively.  The scattering is
governed by the two-body interaction $V_{\alpha , j k} ( E ;
q^{\prime} , q)$ in momentum space, where $E$ is the center-of-mass
energy of the system, $q^{( \prime )} \equiv | \bm{q}^{( \prime )} |$,
and partial-wave projection to a certain quantum number $\alpha$ of
the system was performed.  I allow that the interaction may
intrinsically depend on the energy $E$.  Then, the scattering
amplitude $T_{\alpha , j k} ( E ; q^{\prime} , q)$ in a
coupled-channels problem is a solution of the Lippmann--Schwinger
equation:
\begin{align}
  & T_{\alpha , j k} ( E ; q^{\prime} , q ) =
  V_{\alpha , j k} ( E ; q^{\prime} , q )
  \notag \\
  & + \sum _{l = 1}^{N_{\rm chan}} \int _{0}^{\infty} d k \, k^{2}
  \frac{V_{\alpha , j l} ( E ; q^{\prime} , k )
    T_{\alpha , l k} ( E ; k , q )}{E - \mathcal{E}_{l} ( k ) + i 0} .
  \label{eq:LSeq}
\end{align}
Here $N_{\rm chan}$ is the number of channels and $\mathcal{E}_{j} ( q
)$ is the on-shell energy of the system in the $j$th channel as a
function of the relative three-momentum $q$, for which I take the
semirelativistic option:
\begin{equation}
  \mathcal{E}_{j} ( q ) = \sqrt{m_{j} + q^{2}} + \sqrt{M_{j} + q^{2}} ,
\end{equation}
for stable particle channels, and
\begin{equation}
  \mathcal{E}_{j} ( E; q ) = \sqrt{m_{j} + q^{2}} + \sqrt{M_{j} + q^{2}}
  + \Sigma _{j} ( E ; q ) ,
  \label{eq:Ej_SE}
\end{equation}
for unstable particle channels with the self-energy $\Sigma _{j}$
whose practical form is given in Sec.~\ref{sec:3C}.  The masses of two
particles in the $j$th channel are $m_{j}$ and $M_{j}$.

In my model space I treat only two-body to two-body scattering.  In
the study of the $N^{\ast}$ and $\Delta ^{\ast}$ resonances below, I
will implement one-body states (bare $N^{\ast}$ and $\Delta ^{\ast}$)
into the two-body interaction $V_{\alpha , j k}$ and three-body state
($\pi \pi N$) into the self-energy $\Sigma _{j}$.

In the physical scattering, the initial and final states are on their
mass shell, $E = \mathcal{E}_{j} ( q^{\prime} ) = \mathcal{E}_{k} ( q
)$, and hence the scattering amplitude $T_{\alpha , j k} ( E ;
q^{\prime} , q )$ is a function only of the energy $E$, which is
called the on-shell amplitude.  The on-shell amplitude is a solution
of the Lippmann--Schwinger equation~\eqref{eq:LSeq} and can be
evaluated by the Heitler equation
\begin{align}
  & T_{\alpha , j k}^{\text{on-shell}} ( E ) = K_{\alpha , j k}^{\text{on-shell}} ( E )
  \notag \\
  & \quad + \sum _{l = \text{stable}}
  K_{\alpha , j l}^{\text{on-shell}} ( E )
  \left [ - i \frac{\rho _{l} ( E )}{2} \right ]
  T_{\alpha , l k}^{\text{on-shell}} ( E ) ,
\end{align}
with the on-shell $K$-matrix $K_{\alpha , j k}^{\text{on-shell}}$ and
phase space $\rho _{l}$ where $l$ represents channels of stable
particles.  The $K$-matrix is a solution of the following equation
\begin{align}
  & K_{\alpha , j k} ( E ; q^{\prime} , q ) =
  V_{\alpha , j k} ( E ; q^{\prime} , q )
  \notag \\
  & + \sum _{l = 1}^{N_{\rm chan}} \mathcal{P} \int _{0}^{\infty} d k \, k^{2}
  \frac{V_{\alpha , j l} ( E ; q^{\prime} , k )
    K_{\alpha , l k} ( E ; k , q )}{E - \mathcal{E}_{l} ( k )} ,
  \label{eq:Kmateq}
\end{align}
and its on-shell part, $K_{\alpha , j k}^{\text{on-shell}}$, is
obtained by taking the on-shell momenta $q^{\prime}$ and $q$ for the
parameters of $K_{\alpha , j k}$ as $\mathcal{E}_{j} ( q^{\prime} ) =
\mathcal{E}_{k} ( q ) = E$.  By means of $\mathcal{P}$ I take the
Cauchy principal value for the integral over the momentum variable in
stable open channels, but it returns to the usual integral in closed
channels or unstable channels.  The phase space is defined for
stable channels as
\begin{equation}
  \rho _{j} ( E ) = \frac{E^{4} - ( m_{j}^{2} - M_{j}^{2} )^{2}}
       {4 \pi E^{3}} k_{j} ( E ) \theta ( E - m_{j} - M_{j} ) ,
       \label{eq:PSrho}
\end{equation}
with the Heaviside step function $\theta ( x )$ and the on-shell
relative three-momentum
\begin{equation}
  k_{j} ( E ) 
  =
  \frac{\lambda ^{1/2}( E^{2} , m_{j}^{2} , M_{j}^{2} )}{2 E} .
  \label{eq:k_mom_on}
\end{equation}
Here $\lambda ( x, y, z ) \equiv x^{2} + y^{2} + z^{2} - 2 x y - 2 y z
- 2 z x$ is the \Kaellen function.

Besides the on-shell amplitude, mathematically one may treat
$T_{\alpha , j k} ( E ; q^{\prime} , q )$ as a function of the three
independent variables $E$, $q^{\prime}$, and $q$ as an off-shell
amplitude.  In particular, calculation of the off-shell amplitude with
the complex energy $E$ is essential to extract the two-body wave
function from the scattering amplitude, as seen below.

The scattering amplitude may have resonance poles in the complex
energy plane.  Each pole position $E_{\rm pole}$ coincides with an
eigenvalue of the \Schr equation for a resonance state as an
eigenstate.  The pole for a certain resonance state exists at the same
position in the on-shell and off-shell amplitudes.  In particular, the
resonance pole in the off-shell amplitude can be described as
\begin{equation}
  T_{\alpha , j k} ( E ; q^{\prime} , q ) =
  \frac{\gamma _{j} ( q^{\prime} ) \gamma _{k} ( q )}{E - E_{\rm pole}}
  + ( \text{regular at } E = E_{\rm pole} ) ,
\end{equation}
with the residue $\gamma _{j} ( q^{\prime} ) \gamma _{k} ( q )$.

As pointed out first by Weinberg~\cite{Weinberg:1965zz} and discussed
in Refs.~\cite{Sekihara:2015gvw, Sekihara:2016xnq}, the residue of the
off-shell amplitude contains information on the two-body wave function
of the corresponding resonance state as
\begin{equation}
  \gamma _{j} ( q ) = \frac{1}{( 2 \pi )^{3/2}}
         [ E_{\rm pole} - \mathcal{E}_{j} ( q ) ] 
         \tilde{R}_{j} ( q ) ,
         \label{eq:gamma_q}
\end{equation}
where $\tilde{R}_{j} ( q )$ is the radial part of the resonance wave
function in the $j$th channel in momentum space.  An important point
is that one cannot introduce any scaling factor for this wave function
$\tilde{R}_{j} ( q )$ because the Lippmann--Schwinger
equation~\eqref{eq:LSeq} is an inhomogeneous integral equation.
Therefore, the wave function in Eq.~\eqref{eq:gamma_q} should be
automatically scaled.  Indeed, the normalization of the wave function
in Eq.~\eqref{eq:gamma_q} is guaranteed by the fact that the residue
of the resonance propagator $1 / ( E - \hat{H} )$, where $\hat{H}$
is the full Hamiltonian, is chosen to be exactly unity in the present
formulation~\cite{Sekihara:2016xnq}.

From the residue, one can calculate the norm of the two-body wave
function in the $j$th channel.  In the present notation, the
expression of the norm is
\begin{equation}
  X_{j} = \int _{0}^{\infty} d q \, \tilde{\Rho} _{j} ( q ) ,
  \label{eq:Xj}
\end{equation}
with the density distribution $\tilde{\Rho} _{j} ( q )$ defined as
\begin{equation}
  \tilde{\Rho} _{j} ( q ) \equiv \frac{q^{2}}{( 2 \pi )^{3}}
  \tilde{R}_{j} ( q )^{2}
  = q^{2} \left [ \frac{\gamma _{j} ( q )}
    {E_{\rm pole} - \mathcal{E}_{j} ( q )} \right ] ^{2} .
  \label{eq:Rhoj}
\end{equation}
I call $X_{j}$ the compositeness of the channel $j$ for the resonance
state.  Here I note that in Eq.~\eqref{eq:Rhoj} one should calculate
the complex number squared rather than the absolute value squared so
as to achieve the normalizability for resonance wave
functions~\cite{Hernandez:1984zzb}.

The wave function in coordinate space can be calculated by the Fourier
transformation, which brings the radial part of the wave function
in coordinate space $R_{j} ( r )$:
\begin{equation}
  R_{j} ( r ) 
  = \frac{i^{L}}{2 \pi ^{2}} 
  \int _{0}^{\infty} d q \, q^{2} \tilde{R}_{j} ( q ) j_{L} ( q r ) ,
\end{equation}
where $j_{L}( x ) $ is the spherical Bessel function with the orbital
angular momentum $L$ for the channel $j$.  Therefore, after omitting
the irrelevant factor $i^{L}$ from the angular part, the density
distribution in coordinate space becomes
\begin{align}
  \Rho _{j} ( r )
  = & r^{2} \left [ \frac{1}{2 \pi ^{2}}
    \int _{0}^{\infty} d q \, q^{2} \tilde{R}_{j} ( q )
    j_{L} ( q r ) \right ] ^{2}
  \notag \\
  = & \frac{2 r^{2}}{\pi} \left [ 
    \int _{0}^{\infty} d q \, q^{2} \frac{\gamma _{j} ( q )}
    {E_{\rm pole} - \mathcal{E}_{j} ( q )}
    j_{L} ( q r ) \right ] ^{2}
  ,
\end{align}
which is related to the compositeness as
\begin{equation}
  X_{j} = \int _{0}^{\infty} d r \, \Rho _{j} ( r ) .
  \label{eq:Xj_coord}
\end{equation}

Because the wave function from the scattering amplitude is
automatically scaled, the compositeness~\eqref{eq:Xj} and
\eqref{eq:Xj_coord} should be normalized correctly.  Indeed, for
energy-independent interactions, the compositeness was proved to be
exactly unity in Ref.~\cite{Hernandez:1984zzb} in the nonrelativistic
single-channel resonances, and the validity was extended to
coupled-channels problems in semirelativistic cases in
Ref.~\cite{Sekihara:2016xnq}.

However, when one treats energy dependent interaction and/or unstable
constituent with its self-energy, the sum of the compositeness of all
channels deviates from unity.  This can be interpreted as a
missing-channel contribution, which is not explicit degrees of freedom
in the model space of the two-body channels but is implemented into
the interaction and/or into the self-energy.  I represent the
missing-channel contribution as $Z$:
\begin{equation}
  Z \equiv 1 - \sum _{j = 1}^{N_{\rm chan}} X_{j} .
  \label{eq:missing}
\end{equation}
Here I note that both the compositeness $X_{j}$ and the
missing-channel contribution $Z$ are model dependent quantities
because they are not physical observables (see discussion in
Ref.~\cite{Sekihara:2016xnq}).

For resonance states, the compositeness $X_{j}$ and the
missing-channel contribution $Z$ become complex in general.
Therefore, in contrast to the stable bound states, one cannot make a
probabilistic interpretation for them.  Actually, for stable bound
states, $X_{j}$ and $Z$ are real and bound in the range $[0, 1]$ and
hence a sum rule
\begin{equation}
  \sum _{j = 1}^{N_{\rm chan}} | X_{j} | + | Z | = 1
  \label{eq:absX}
\end{equation}
is satisfied.  On the other hand, this relation is not satisfied for
resonance states because both $X_{j}$ and $Z$ are complex.  To
interpret such complex values, I introduce a quantity $U$ for
resonance states to measure the deviation from the sum
rule~\eqref{eq:absX}, according to Refs.~\cite{Kamiya:2015aea,
  Kamiya:2016oao}
\begin{equation}
  U \equiv \sum _{j = 1}^{N_{\rm chan}} | X_{j} | + | Z | - 1 .
  \label{eq:XtildeU}
\end{equation}
Because of the definition of $Z$ in Eq.~\eqref{eq:missing}, $U$
satisfies $U \ge 0$.  Furthermore, $U$ becomes small if $\text{Im} \,
X_{j}$ and $\text{Im} \, Z$ are negligible and $\text{Re} \, X_{j}$
and $\text{Re} \, Z$ are not negatively large.  In such a case, the
wave function of the resonance state considered is similar to that of
a certain stable bound state.  In particular, $U$ goes to zero in a
stable bound-state limit, so $U$ is understood as an uncertainty of
the interpretation of the complex-valued $X_{j}$ and $Z$, as discussed
in Refs~\cite{Kamiya:2015aea, Kamiya:2016oao}.

In this line, I employ quantities introduced in
Refs.~\cite{Sekihara:2015gvw, Sekihara:2016xnq}:
\begin{equation}
  \tilde{X}_{j} \equiv \frac{ | X_{j} | }{1 + U} ,
  \quad
  \tilde{Z} \equiv \frac{ | Z | }{1 + U} .
  \label{eq:XtildeZtilde}
\end{equation}
The quantities $\tilde{X}_{j}$ and $\tilde{Z}$ are real, bound in the
range $[0, 1]$, and automatically satisfy the sum rule:
\begin{equation}
  \sum _{j = 1}^{N_{\rm chan}} \tilde{X}_{j} + \tilde{Z} = 1 .
  \label{eq:Xtilde_sum}
\end{equation}
Then, to estimate uncertainties of the probabilistic interpretation of
$\tilde{X}_{j}$ and $\tilde{Z}$, I utilize the
relation~\eqref{eq:XtildeU}, which means that $U$ measures the
deviation from sum rule for a bound state~\eqref{eq:absX}.  Therefore,
a contribution from each $X_{j}$ or $Z$ to the deviation $U$ can be
estimated by $U$ divided by the number of the degrees of freedom
$N_{\rm chan} + 1$, to which I refer as the reduced uncertainty
$U_{\rm r}$:
\begin{equation}
  U_{\rm r} \equiv \frac{U}{N_{\rm chan} + 1} .
  \label{eq:Ur}
\end{equation}
In this sense, if and only if $U_{\rm r} \ll 1$, one can interpret
$\tilde{X}_{j}$ ($\tilde{Z}$) as the probability of finding the
composite (missing) part, and $U_{\rm r}$ can be considered as the
uncertainty of the probabilities $\tilde{X}_{j}$ and $\tilde{Z}$.

\begin{figure*}[!p]
  \centering
  \PsfigII{0.175}{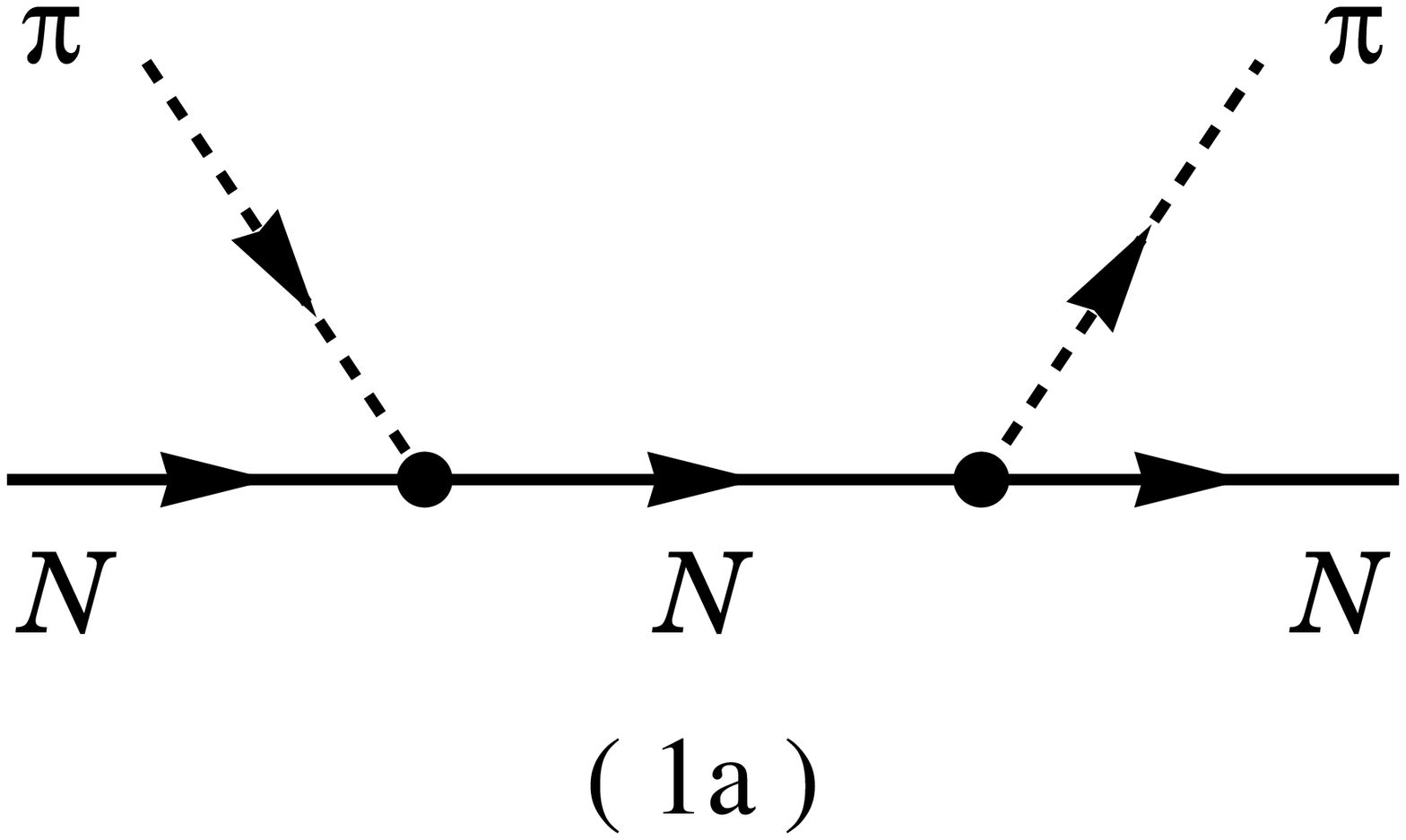} ~
  \PsfigII{0.175}{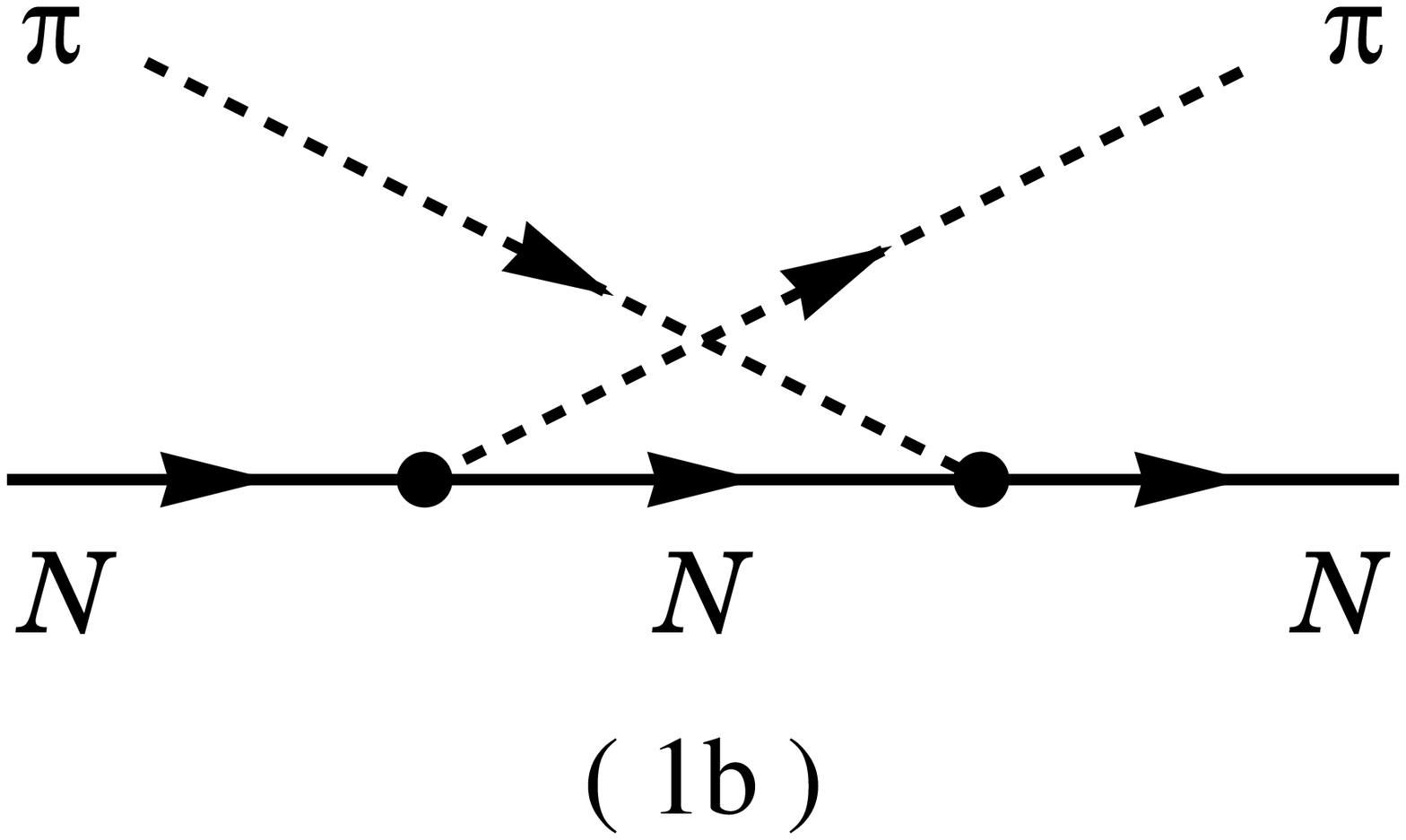} ~
  \PsfigII{0.175}{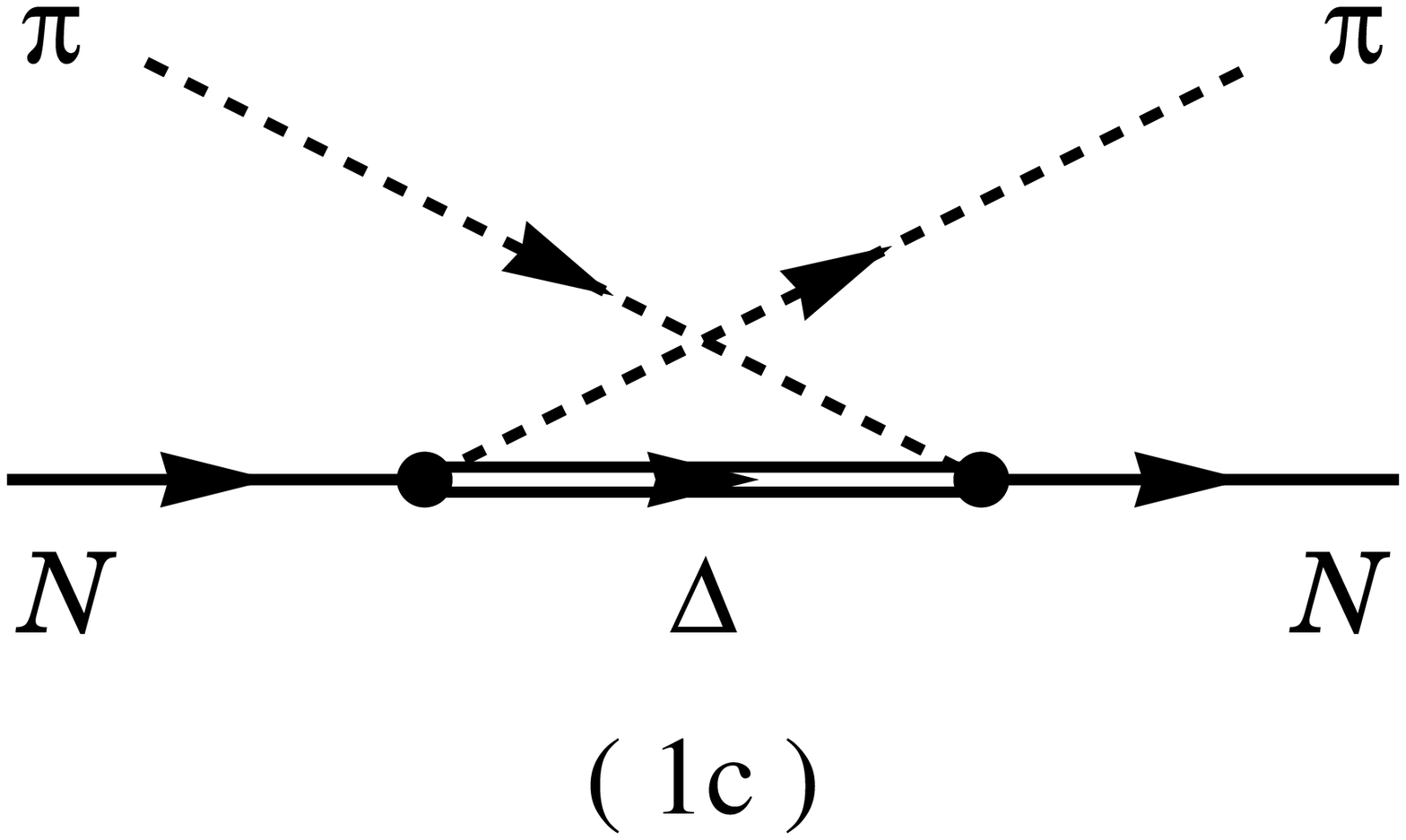} ~
  \PsfigII{0.175}{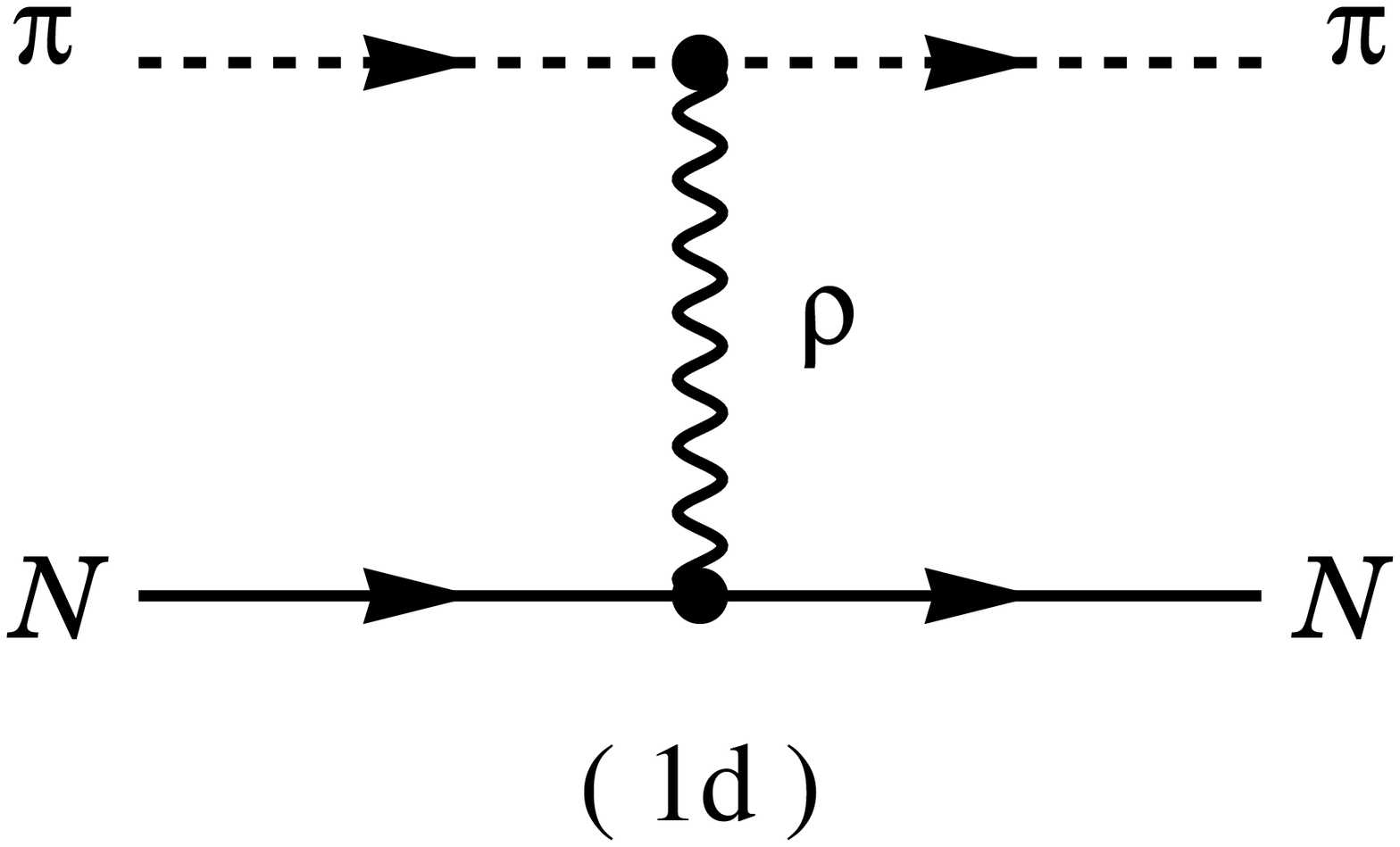} ~
  \PsfigII{0.175}{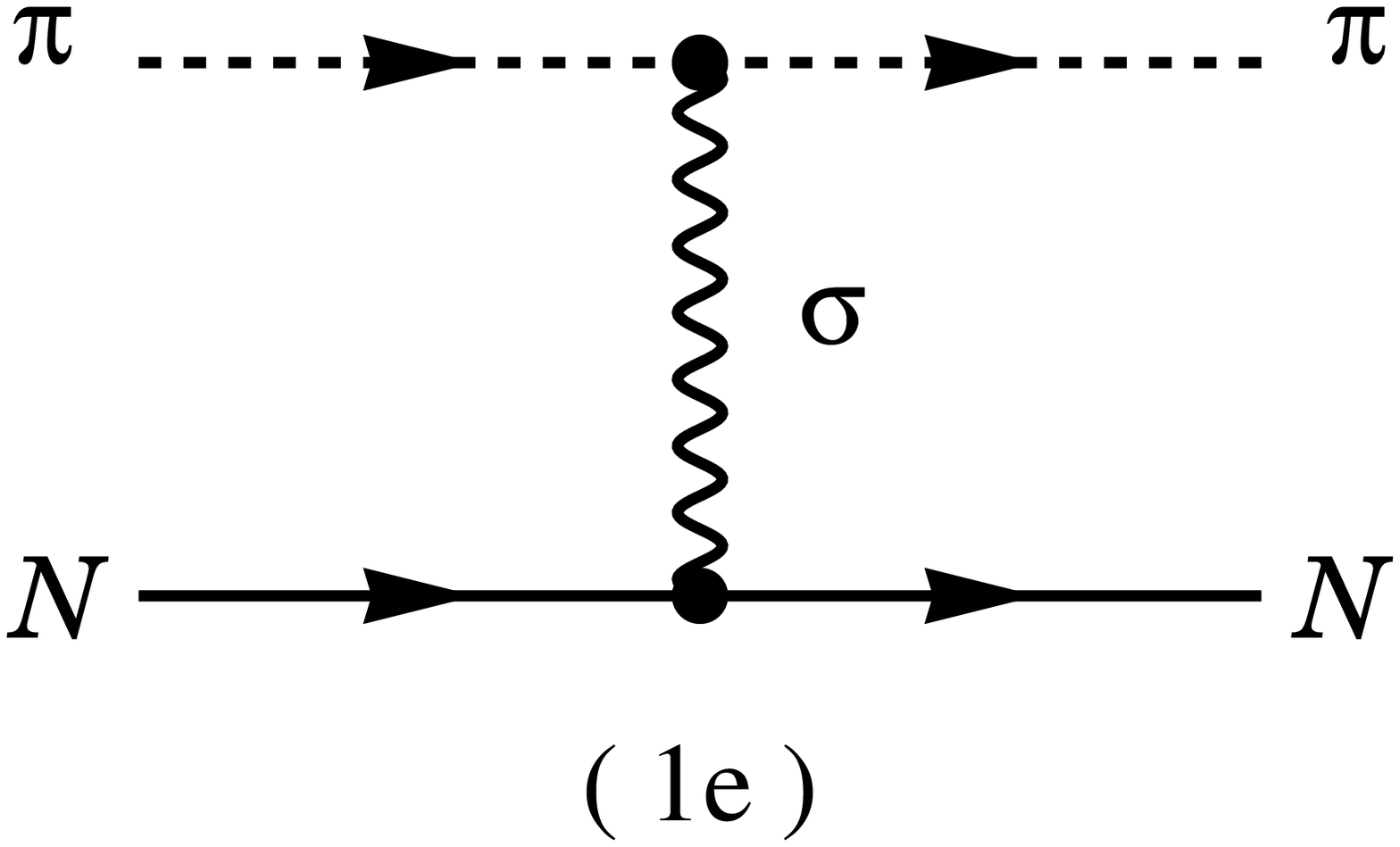} ~
  \PsfigII{0.175}{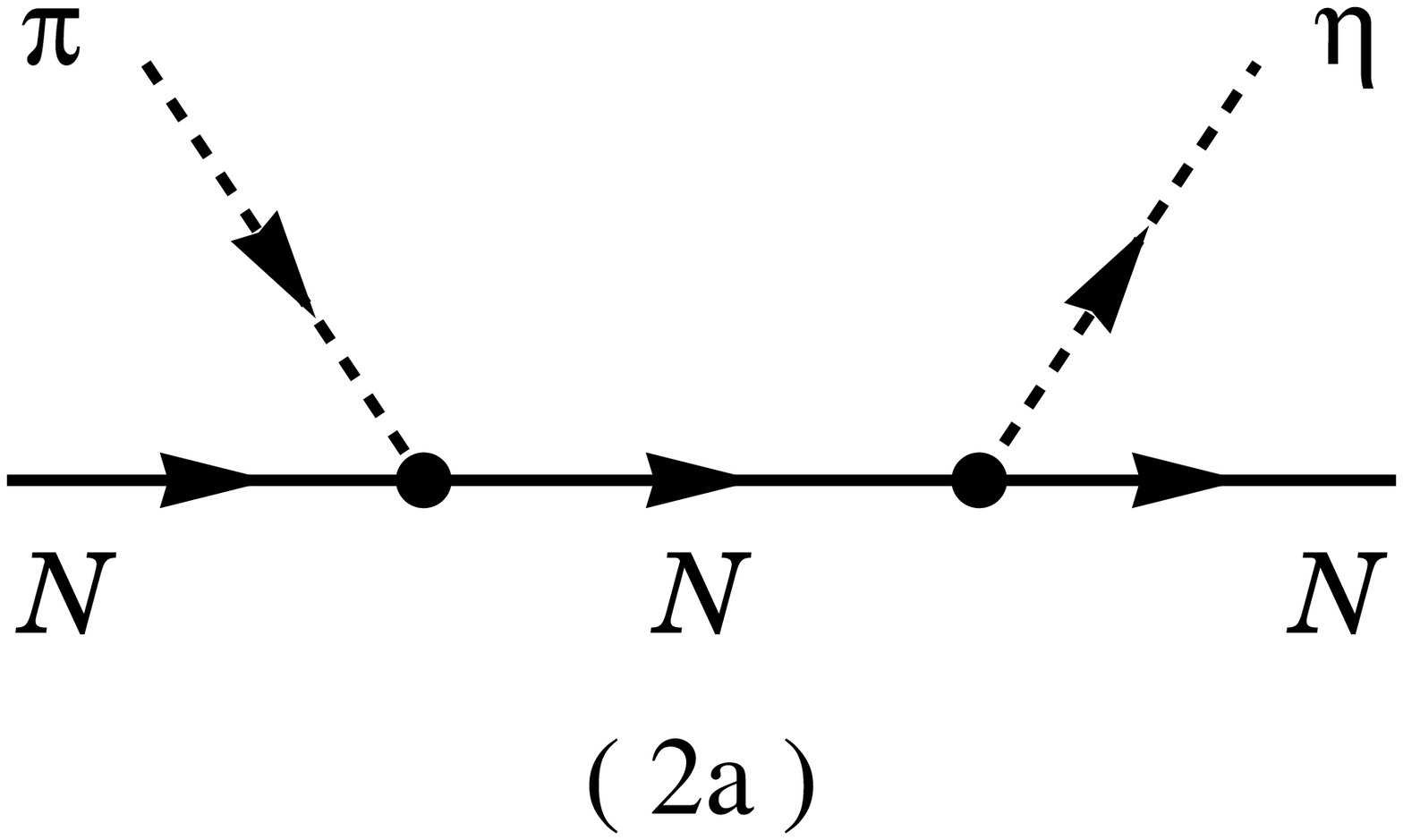} ~
  \PsfigII{0.175}{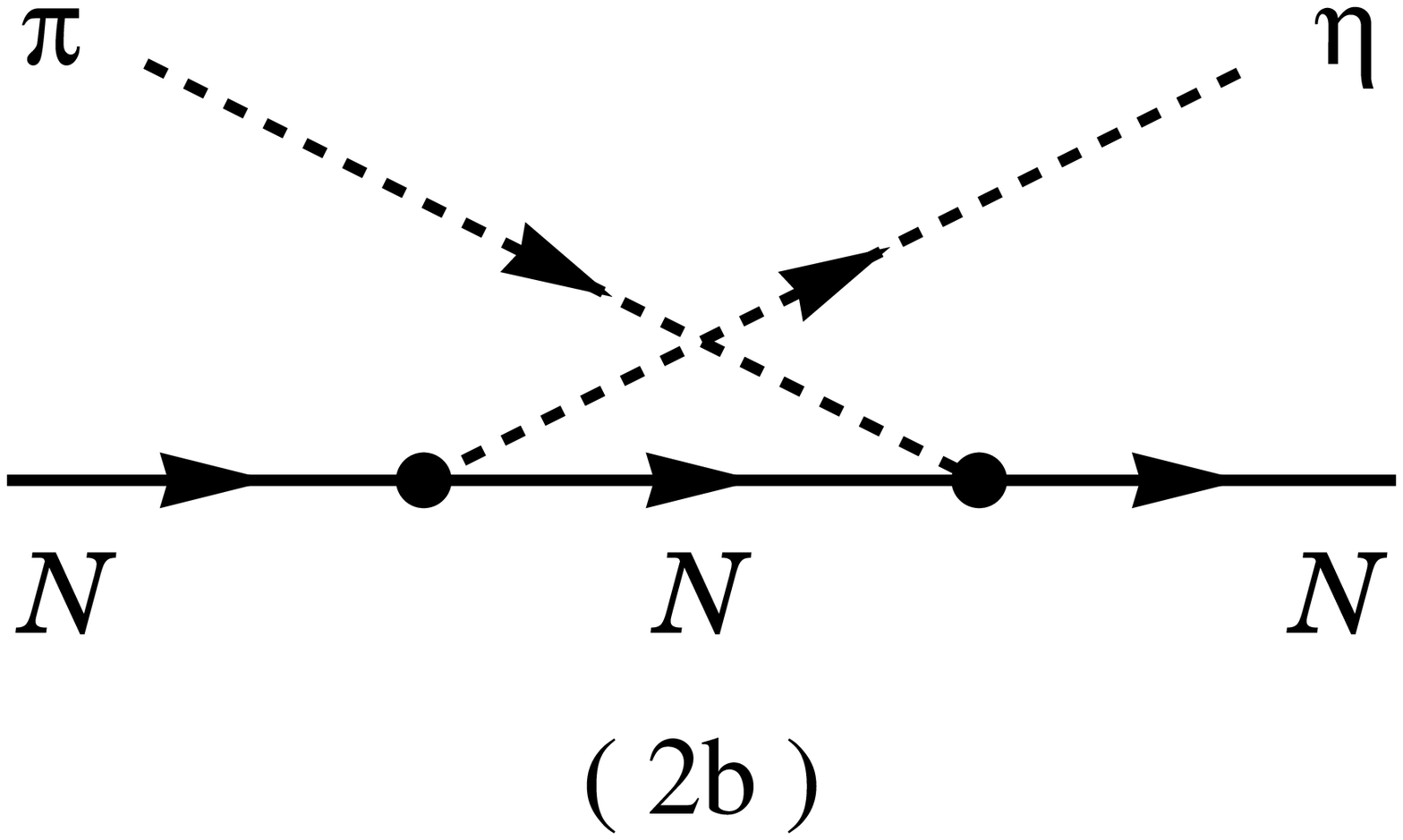} ~
  \PsfigII{0.175}{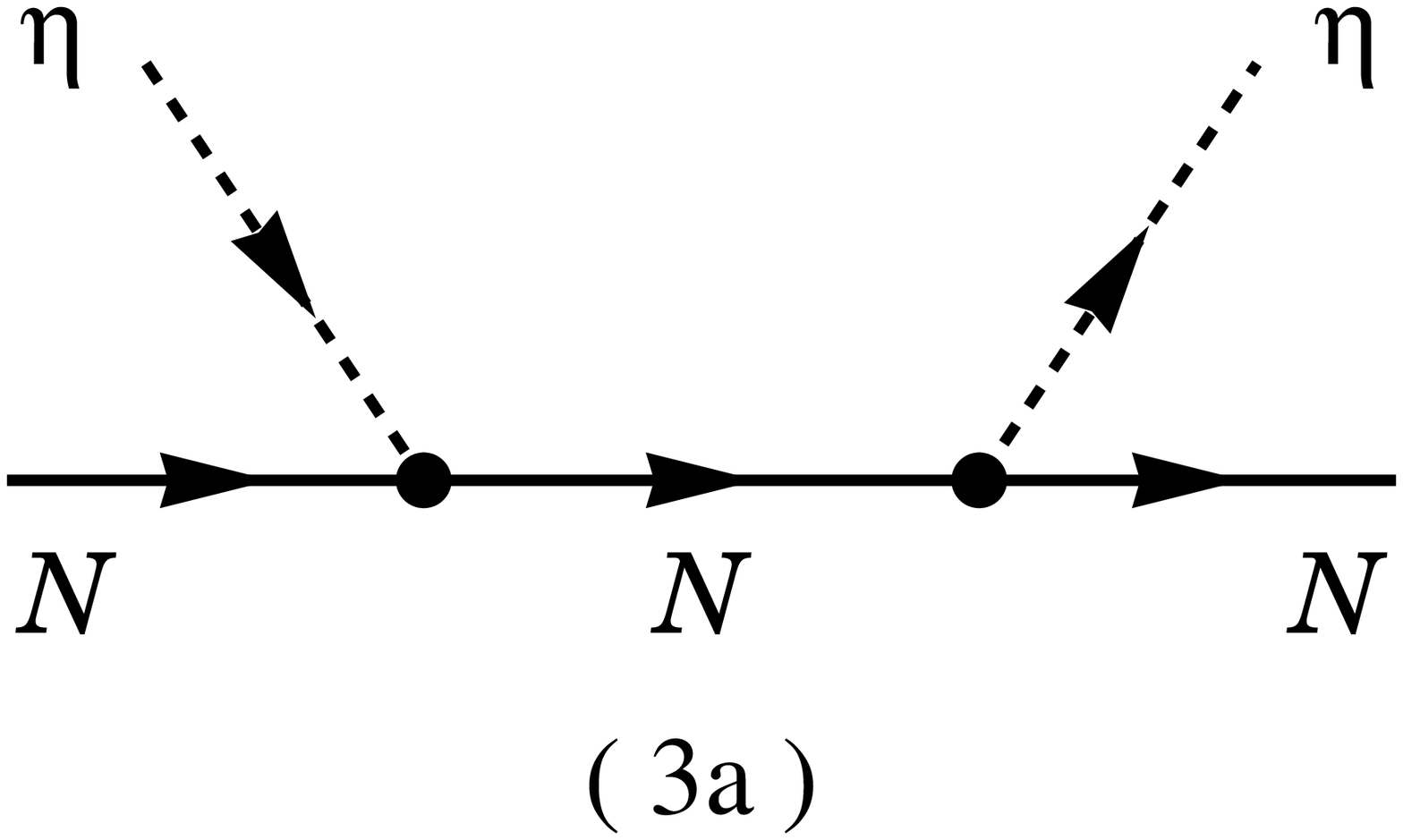} ~
  \PsfigII{0.175}{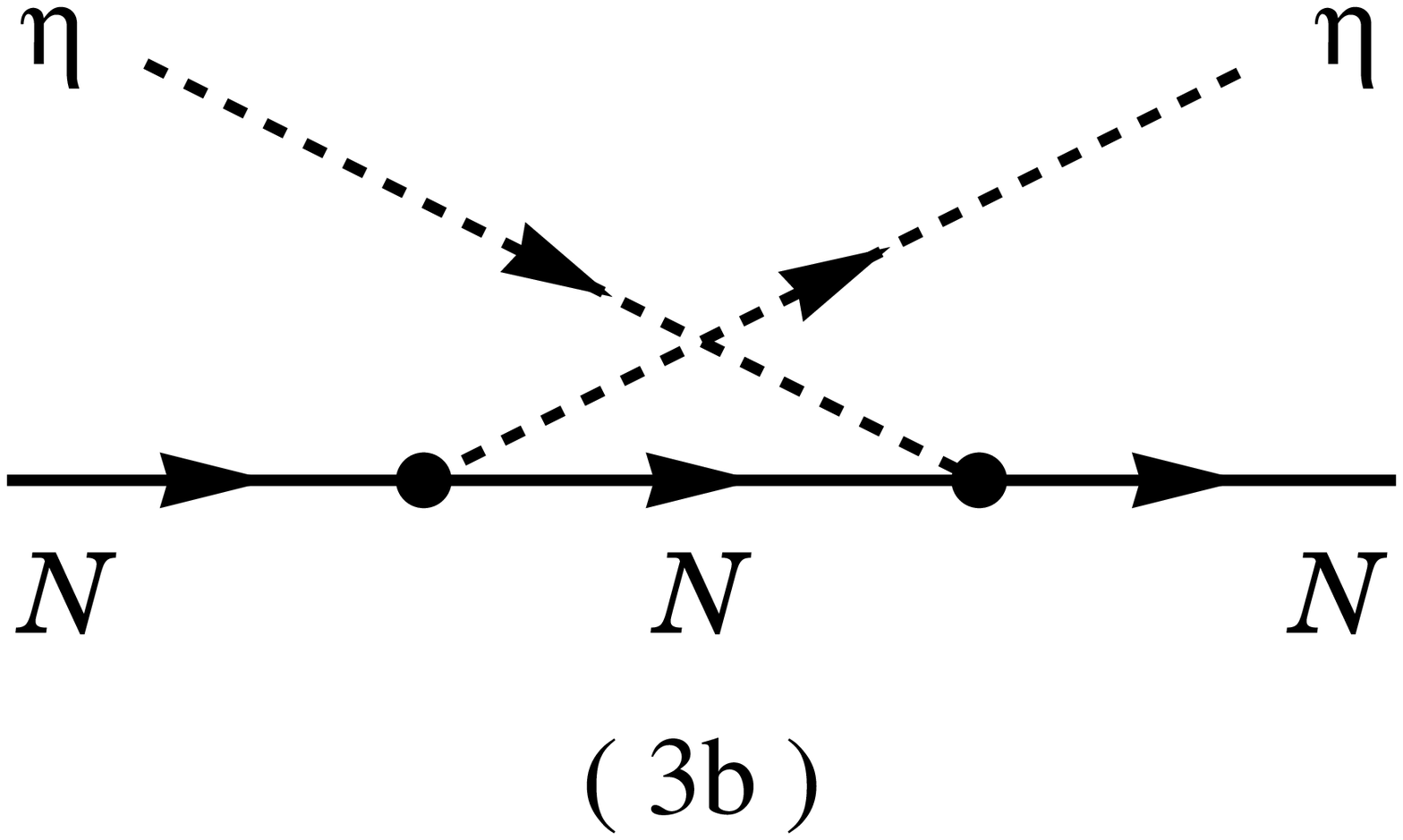} ~
  \PsfigII{0.175}{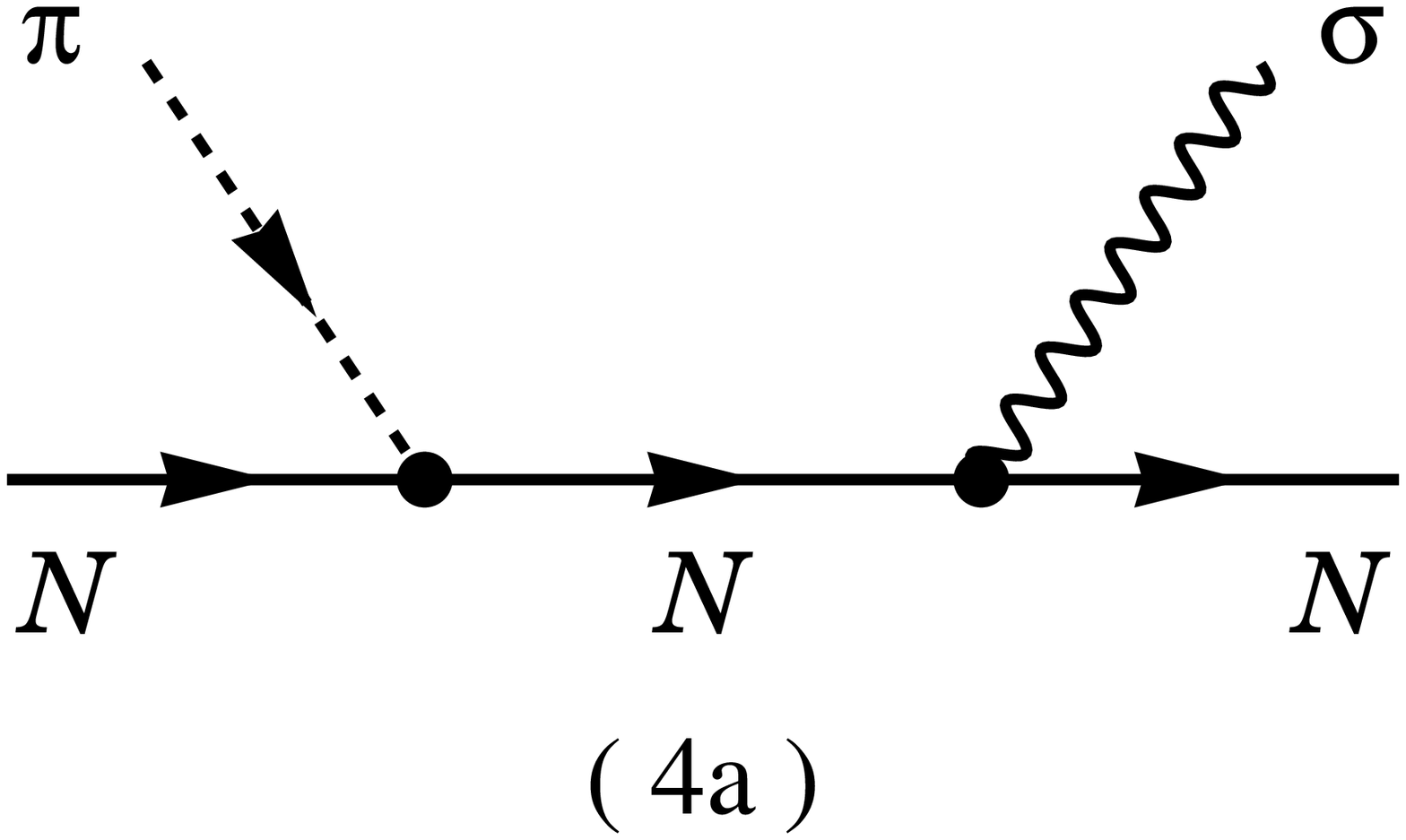} ~
  \PsfigII{0.175}{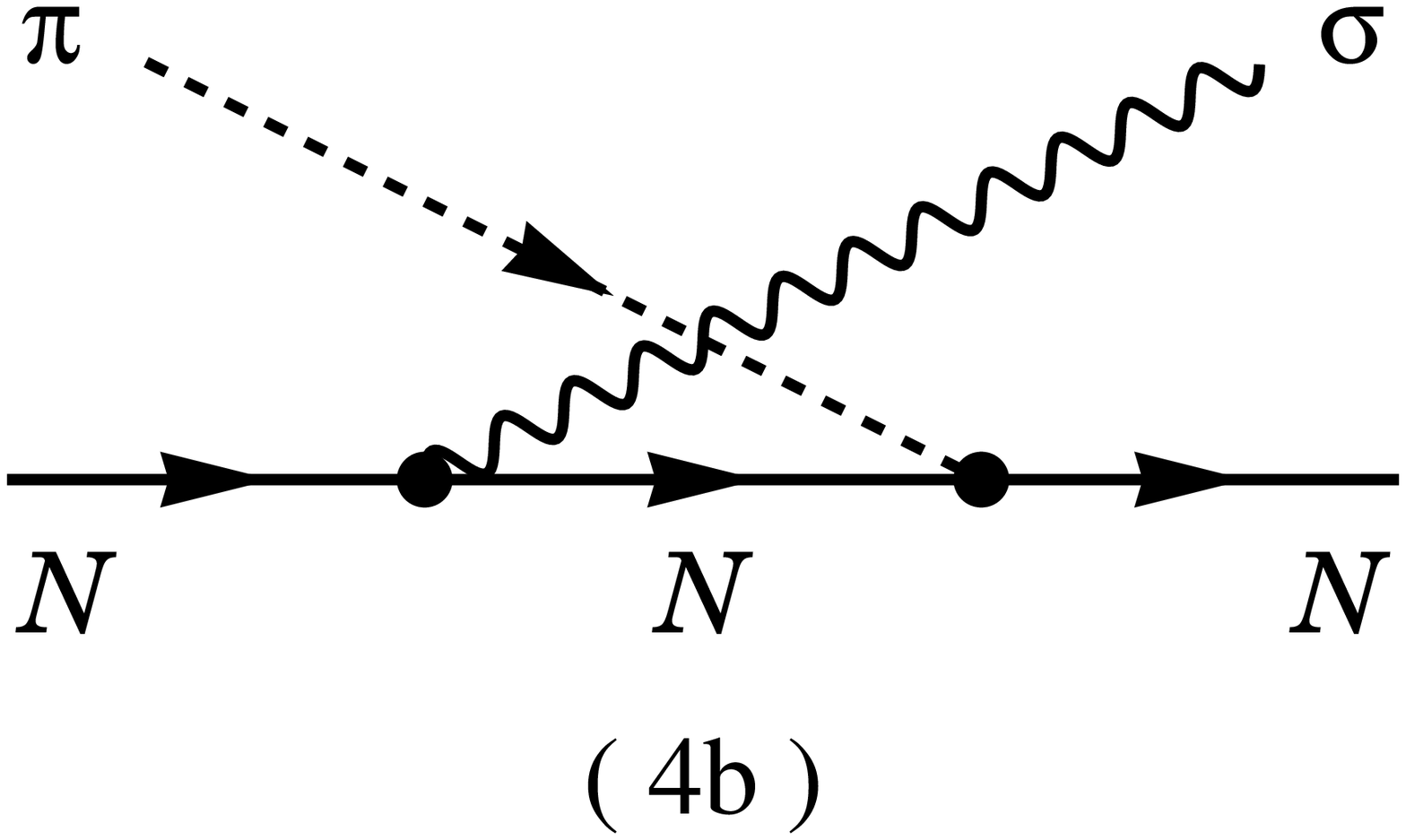} ~
  \PsfigII{0.175}{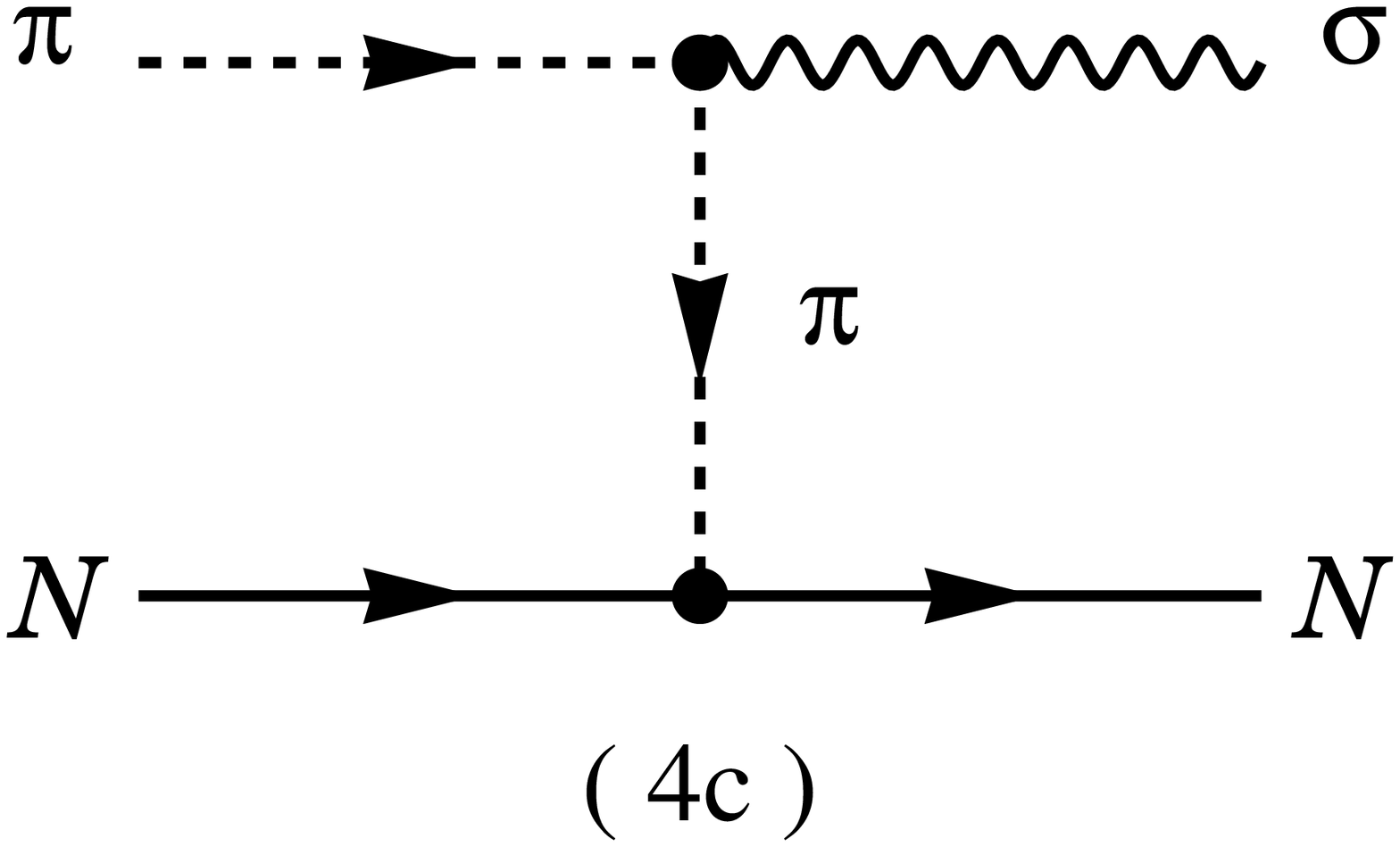} ~
  \PsfigII{0.175}{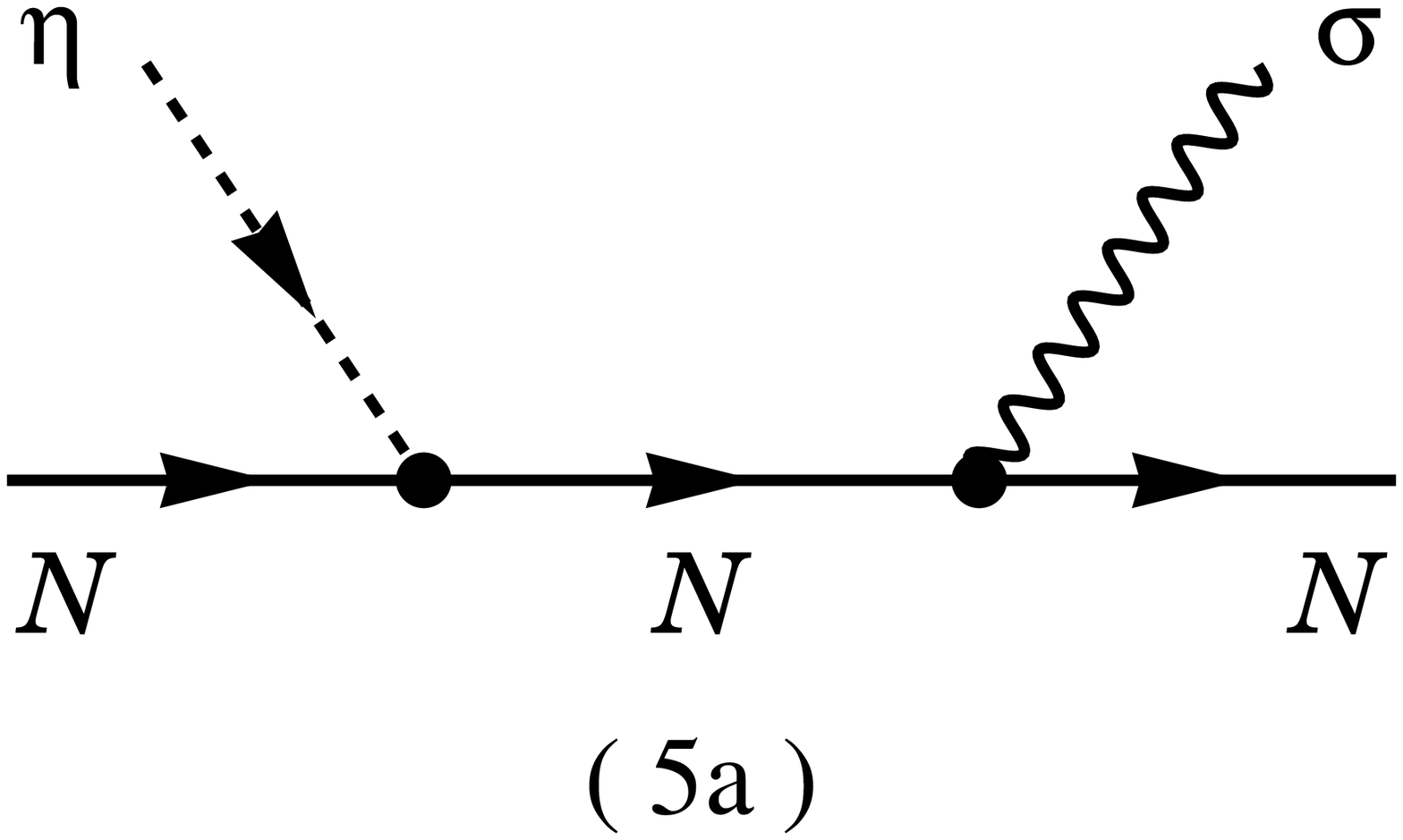} ~
  \PsfigII{0.175}{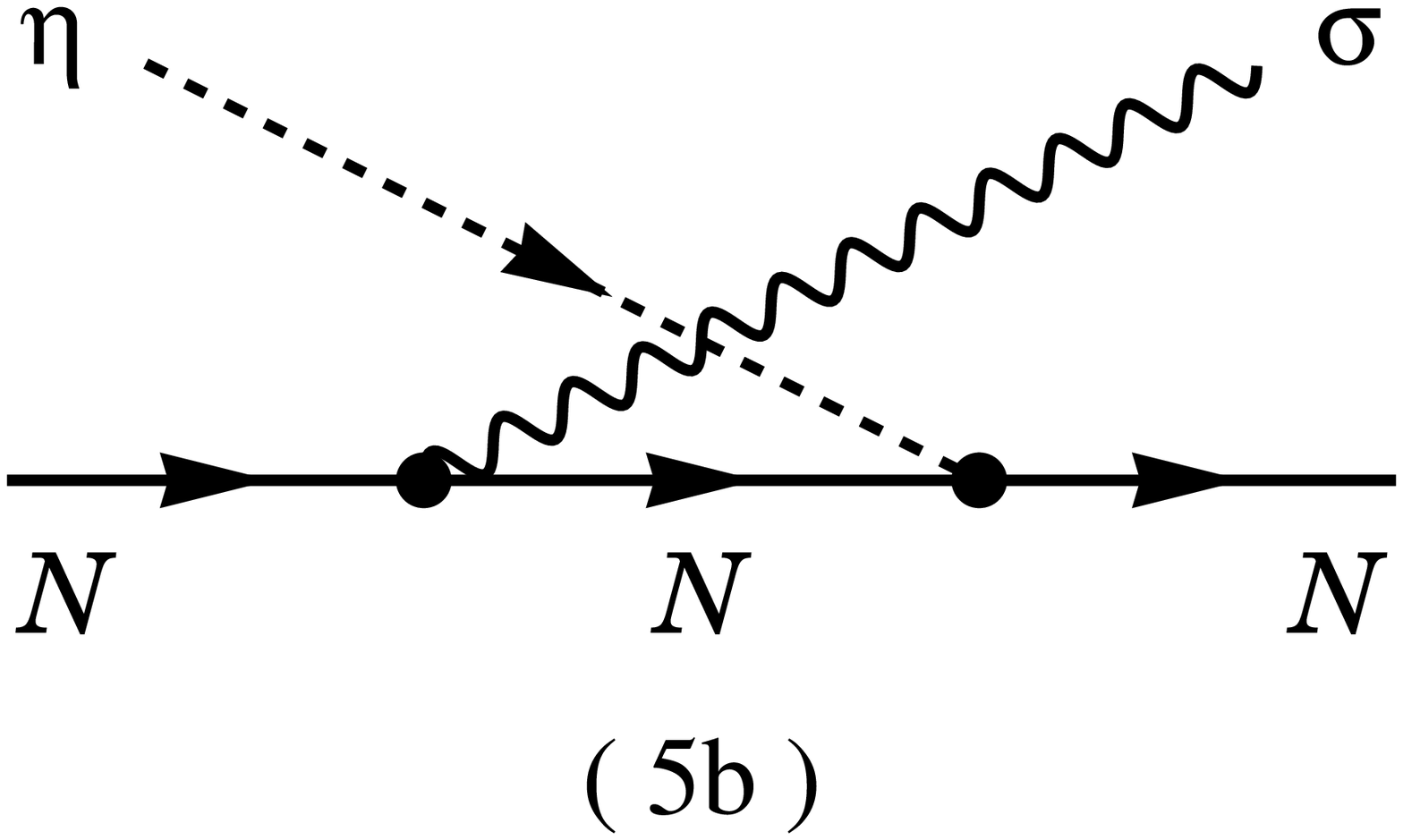} ~
  \PsfigII{0.175}{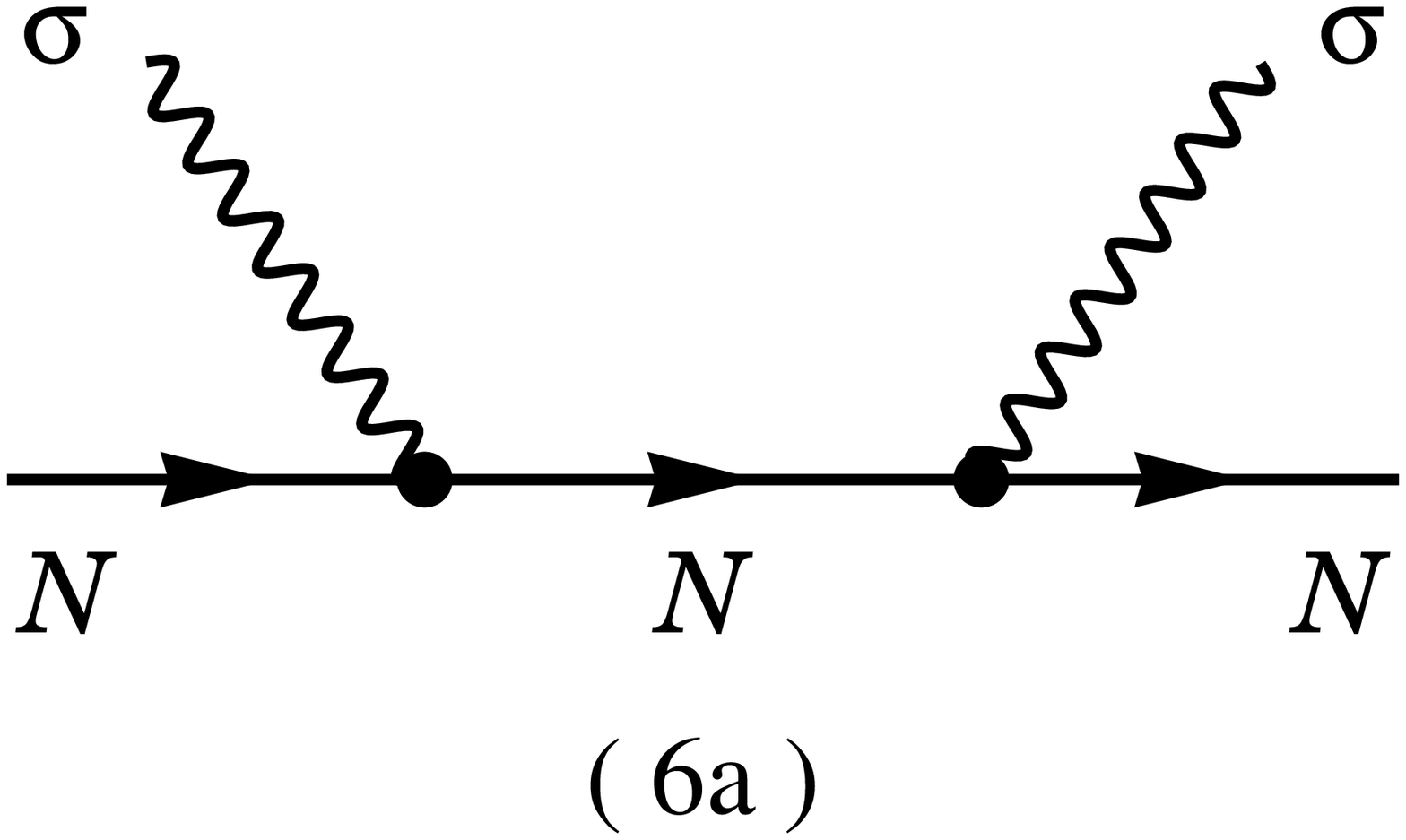} ~
  \PsfigII{0.175}{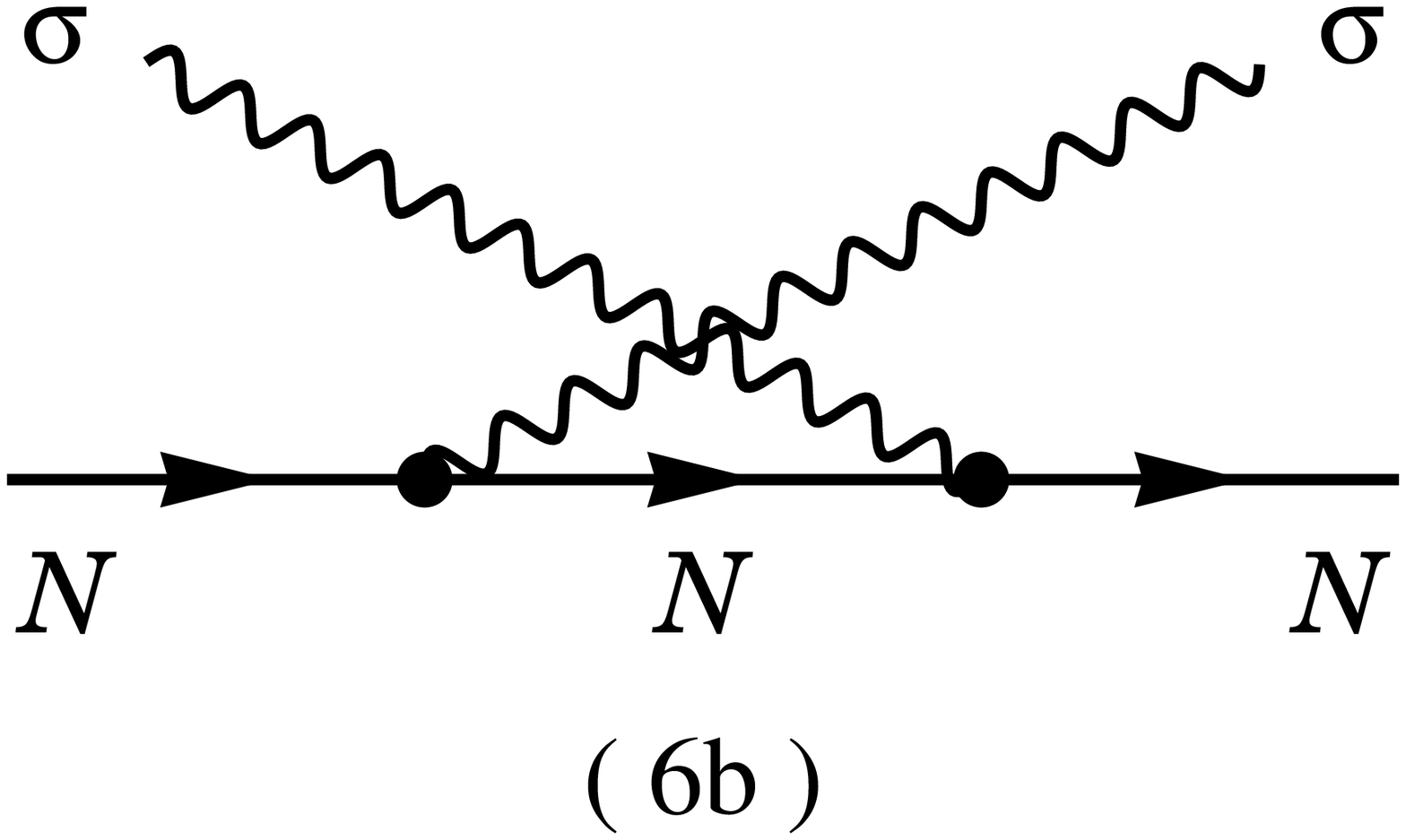} ~
  \PsfigII{0.175}{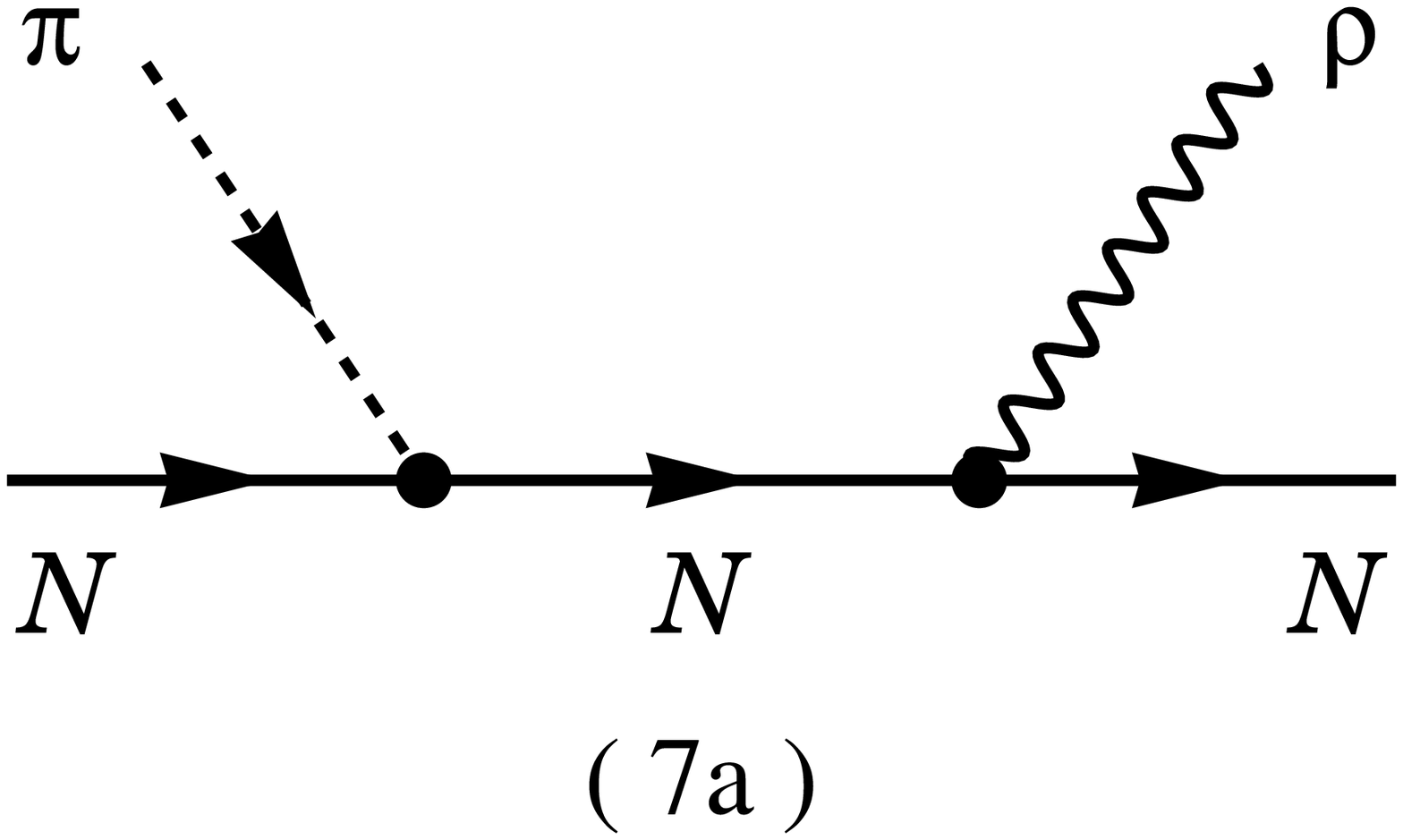} ~
  \PsfigII{0.175}{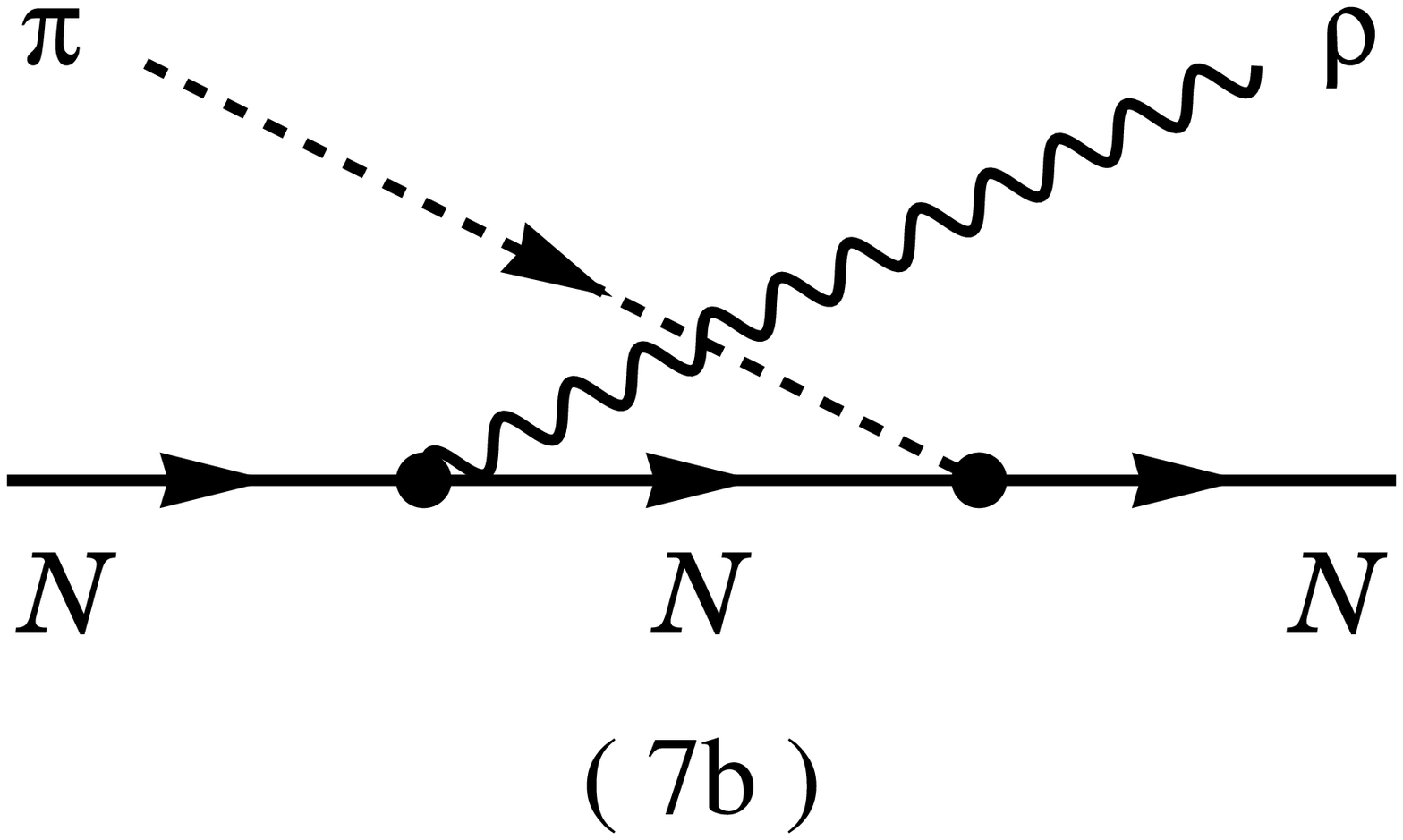} ~
  \PsfigII{0.175}{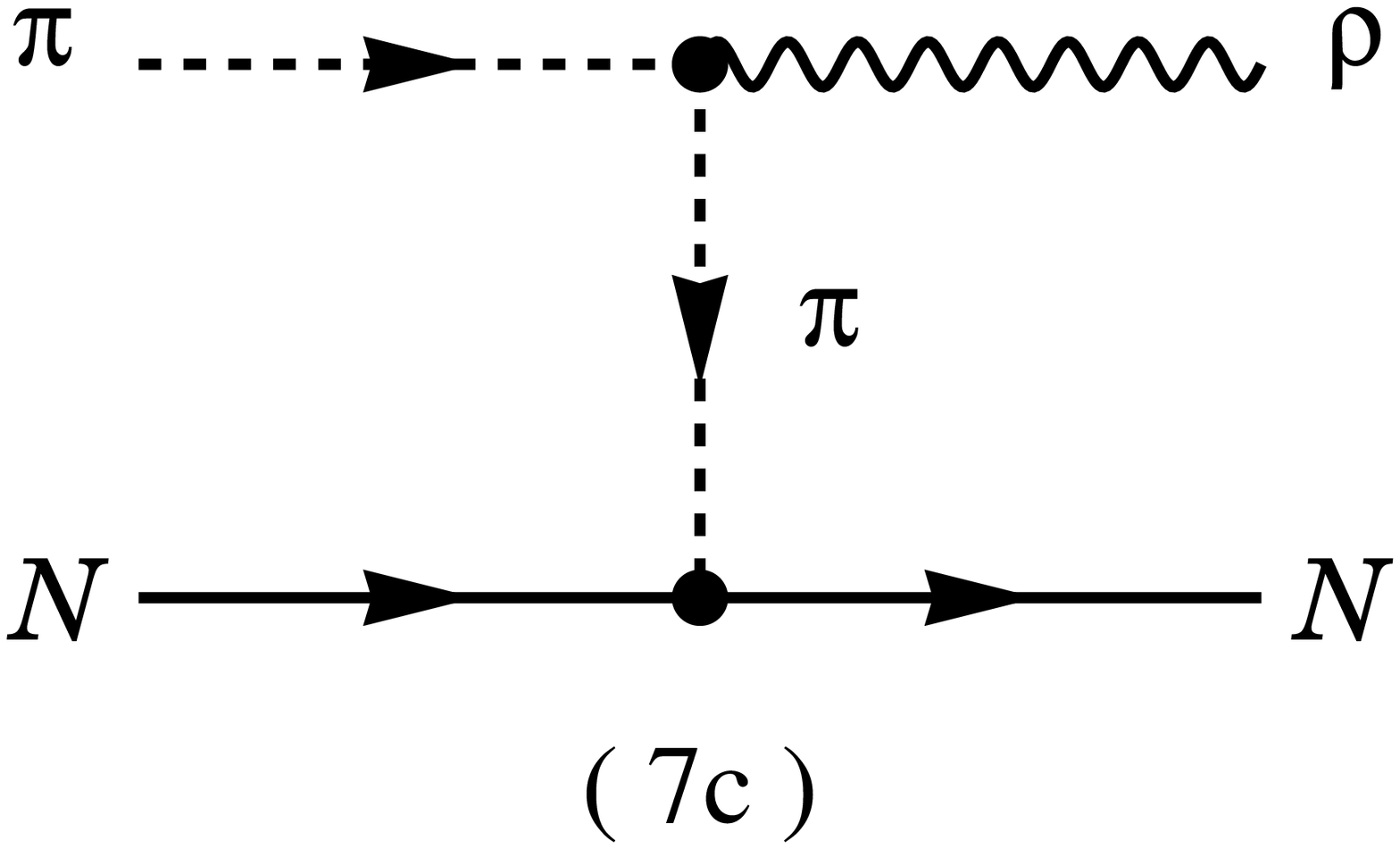} ~
  \PsfigII{0.175}{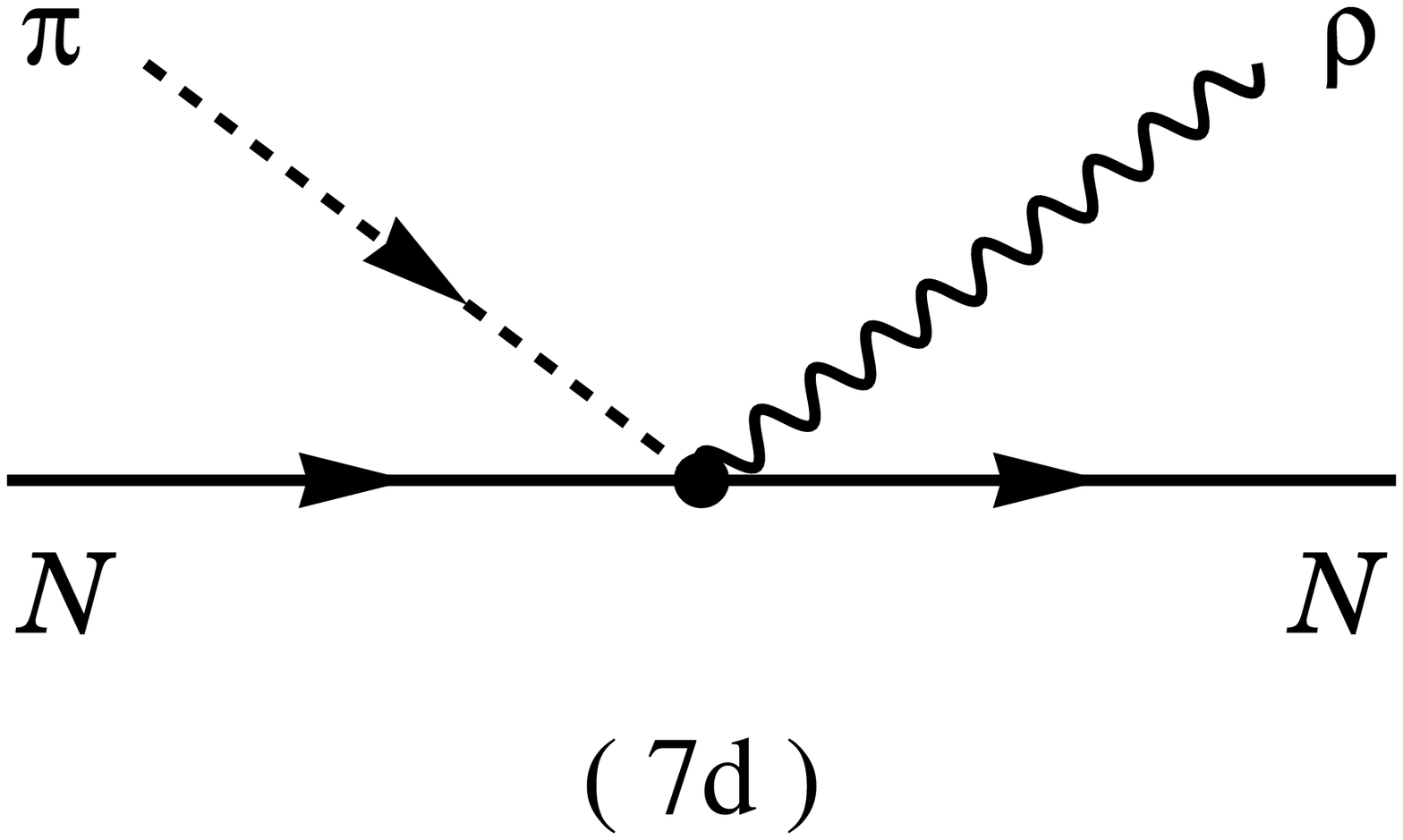} ~
  \PsfigII{0.175}{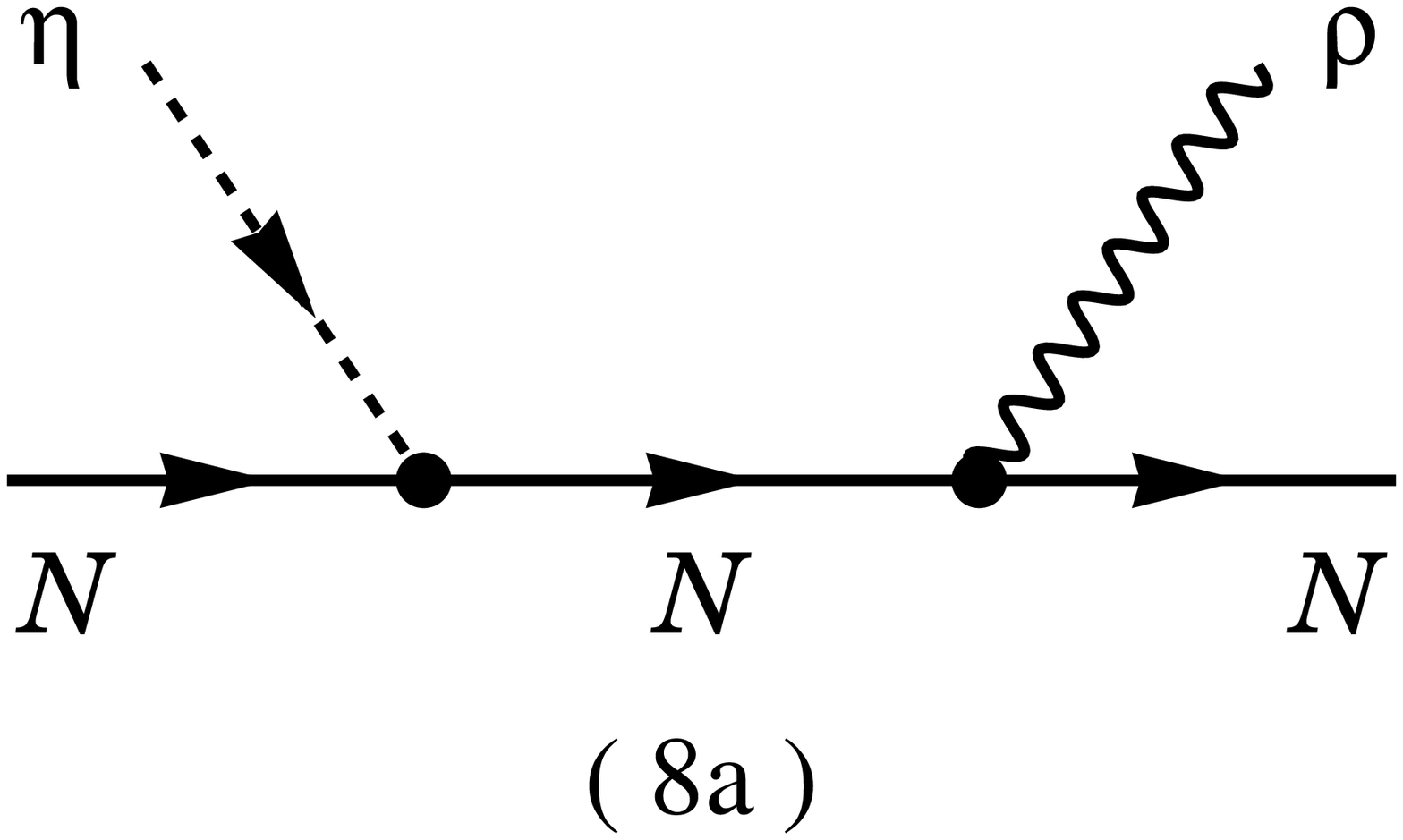} ~
  \PsfigII{0.175}{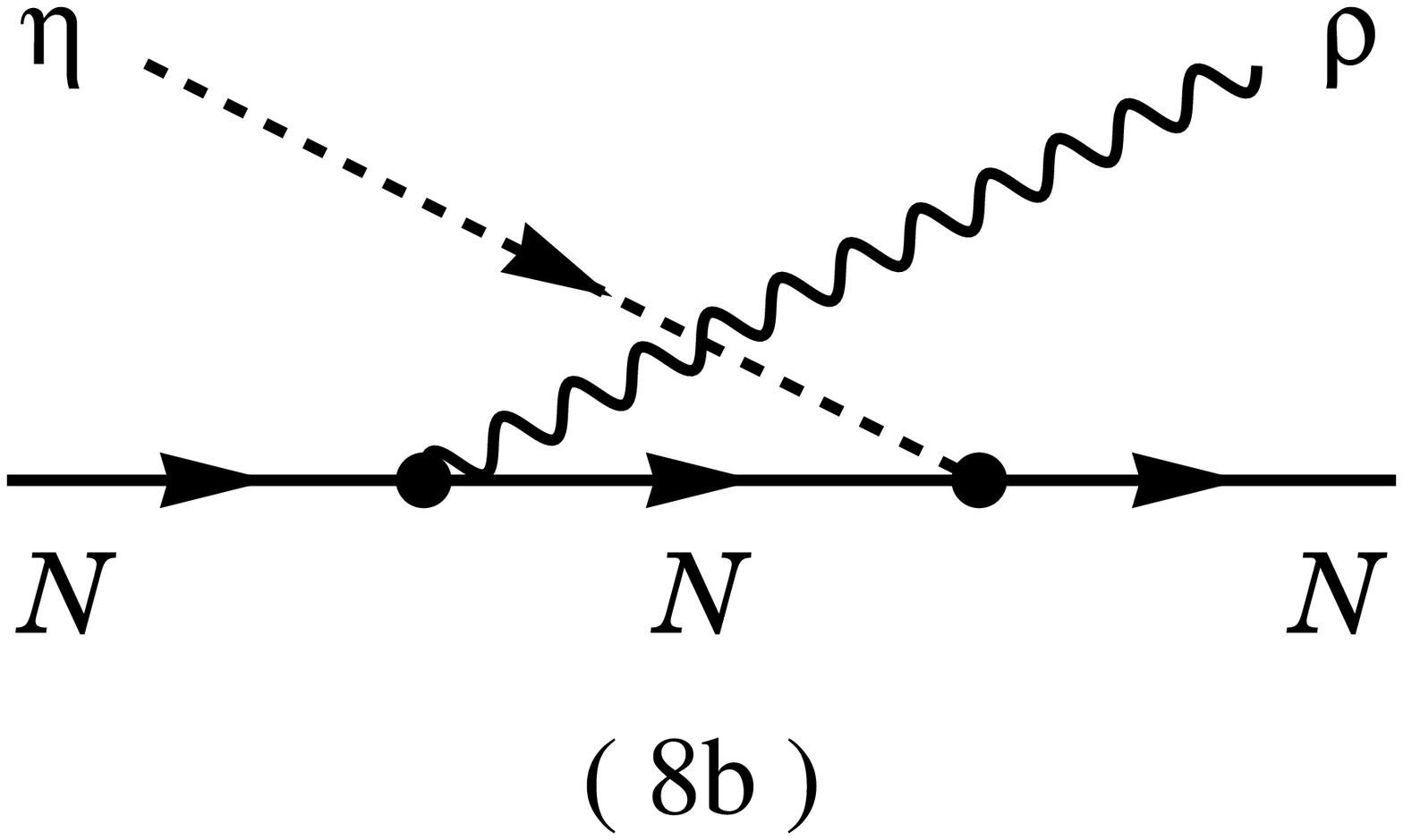} ~
  \PsfigII{0.175}{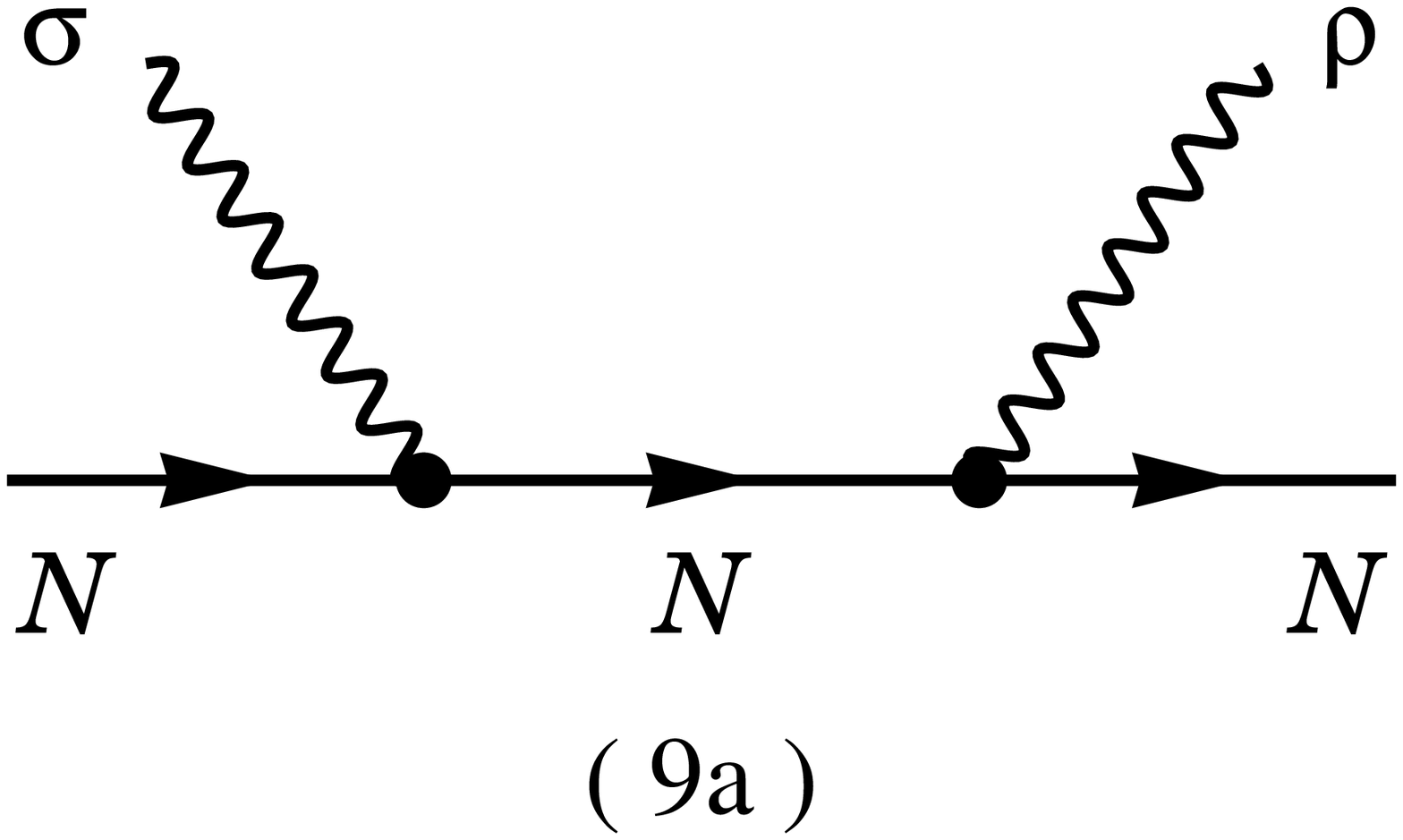} ~
  \PsfigII{0.175}{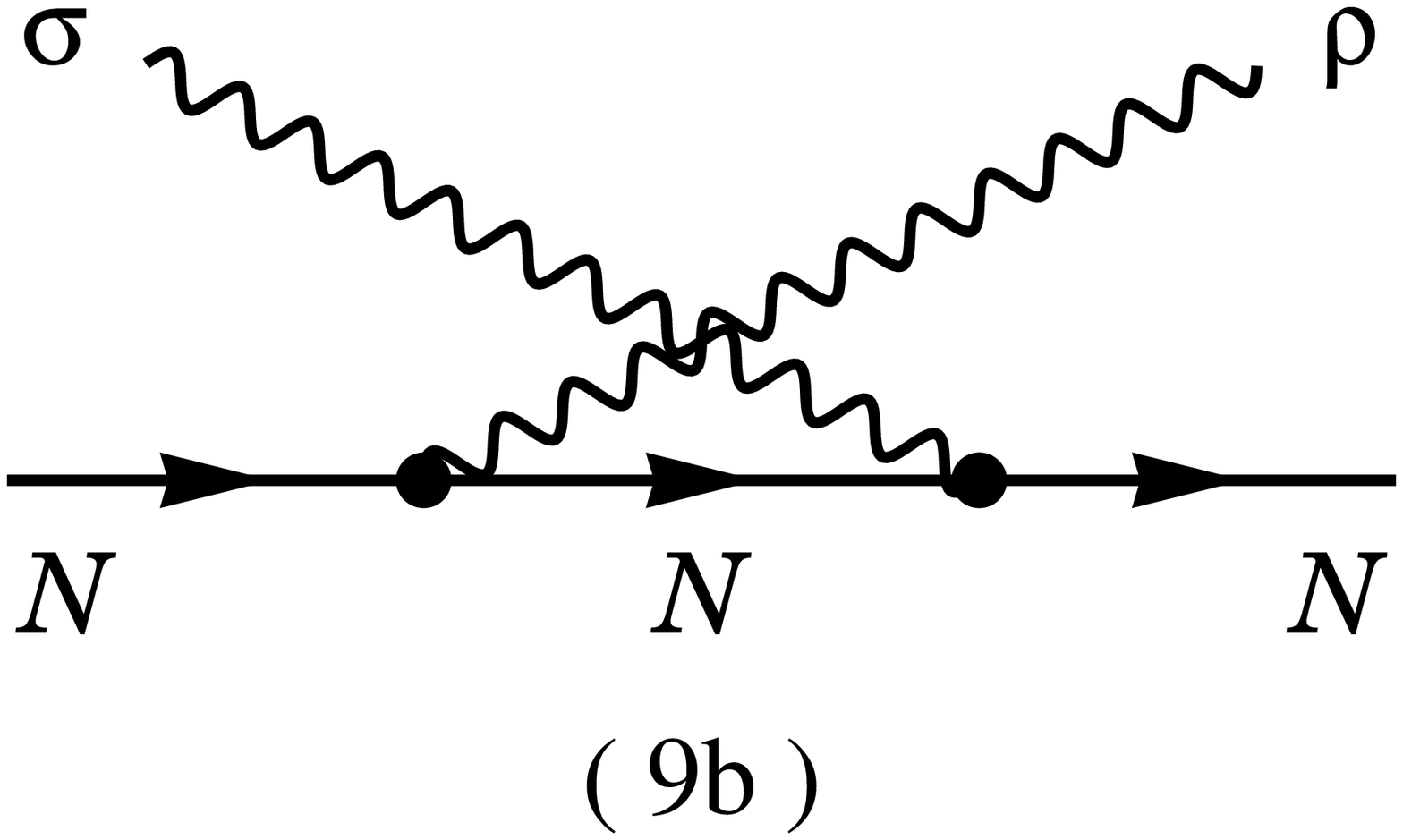} ~
  \PsfigII{0.175}{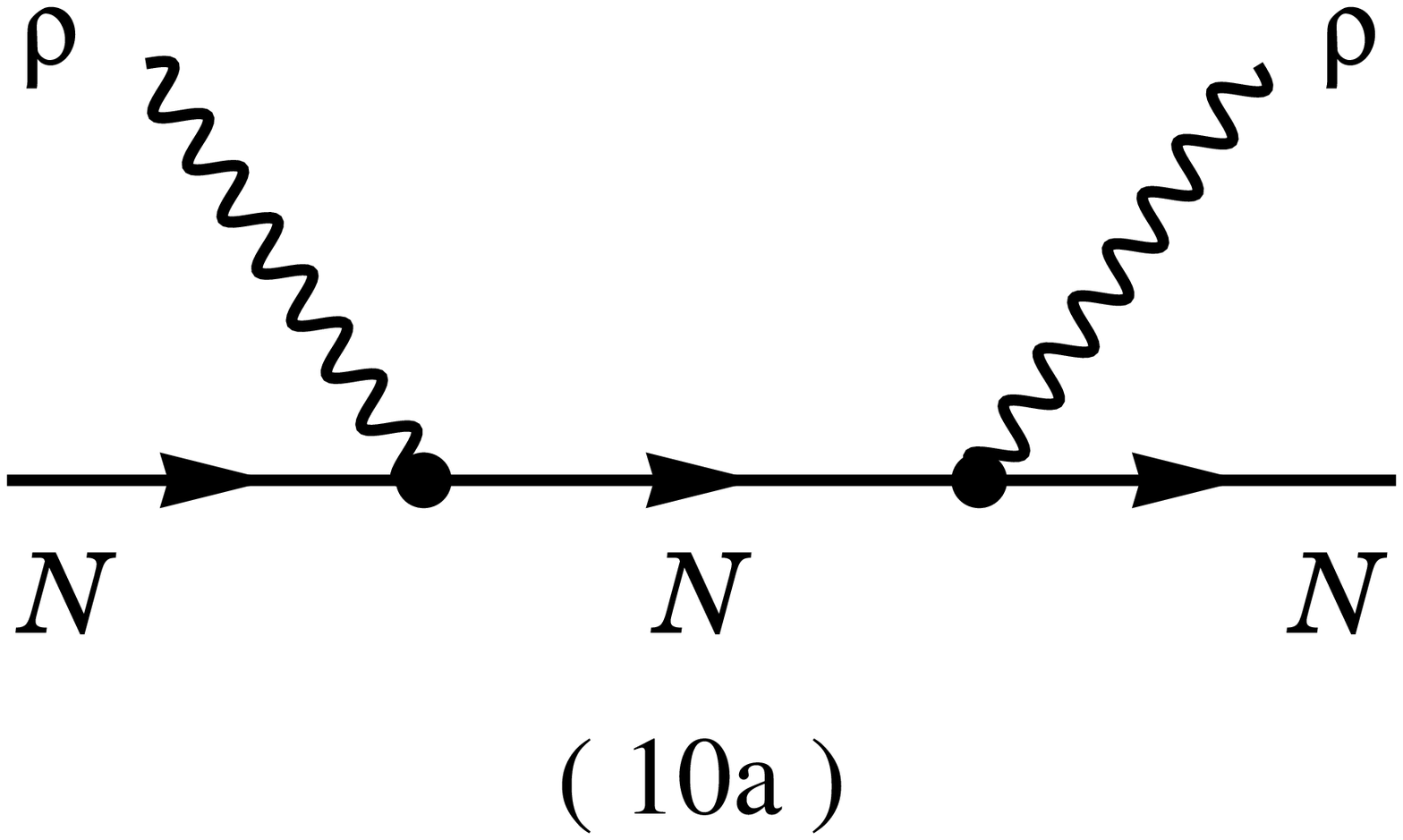} ~
  \PsfigII{0.175}{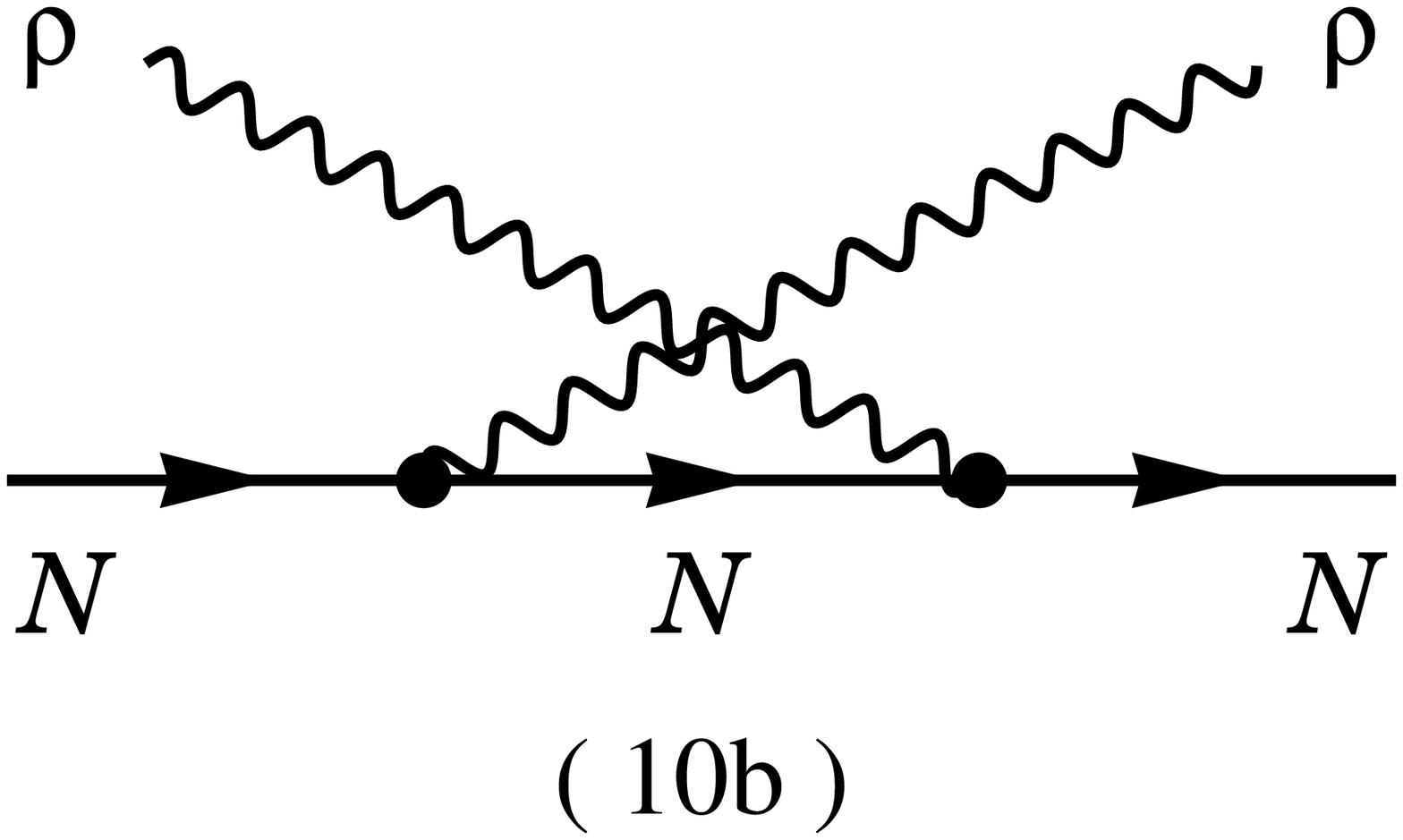} ~
  \PsfigII{0.175}{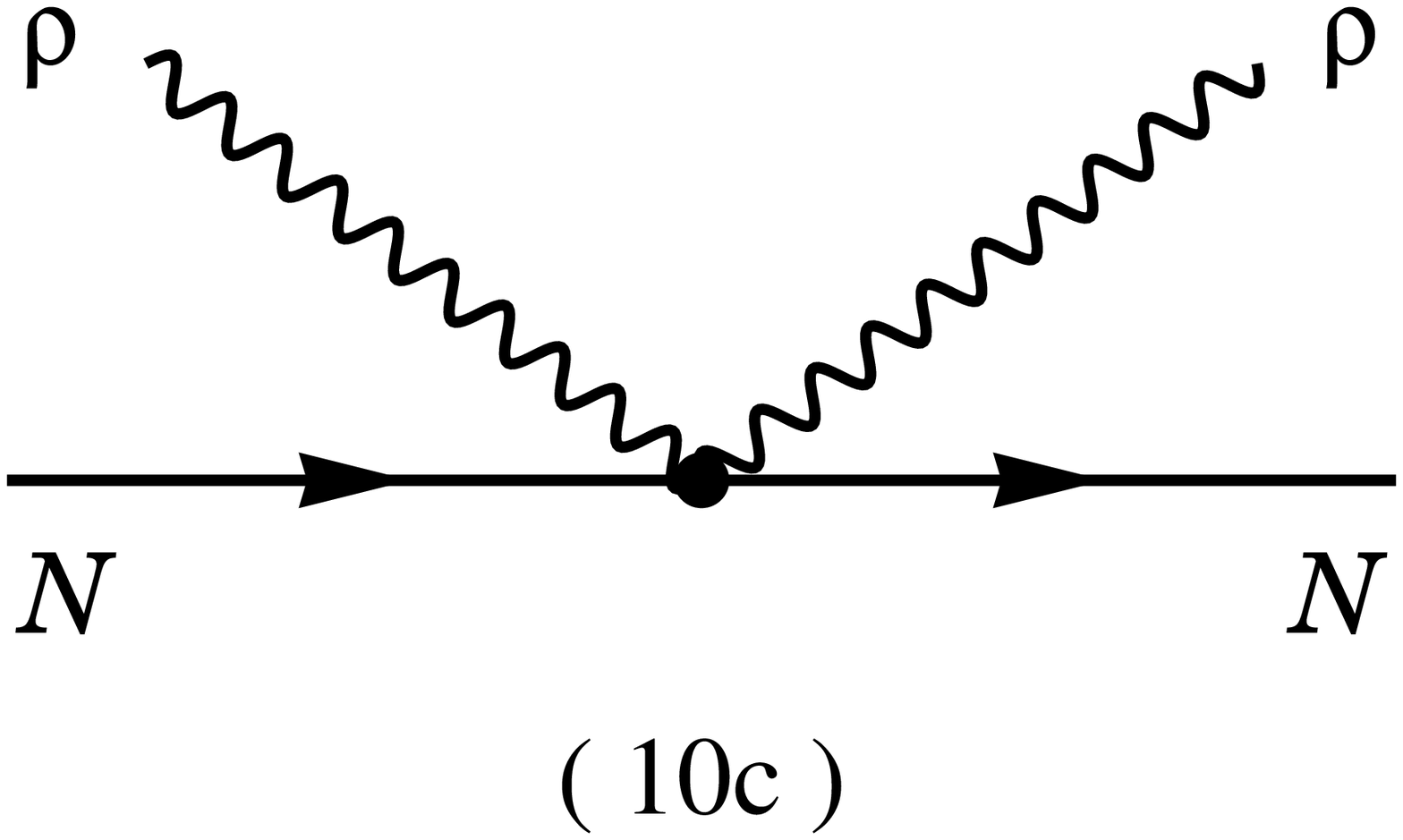} ~
  \PsfigII{0.175}{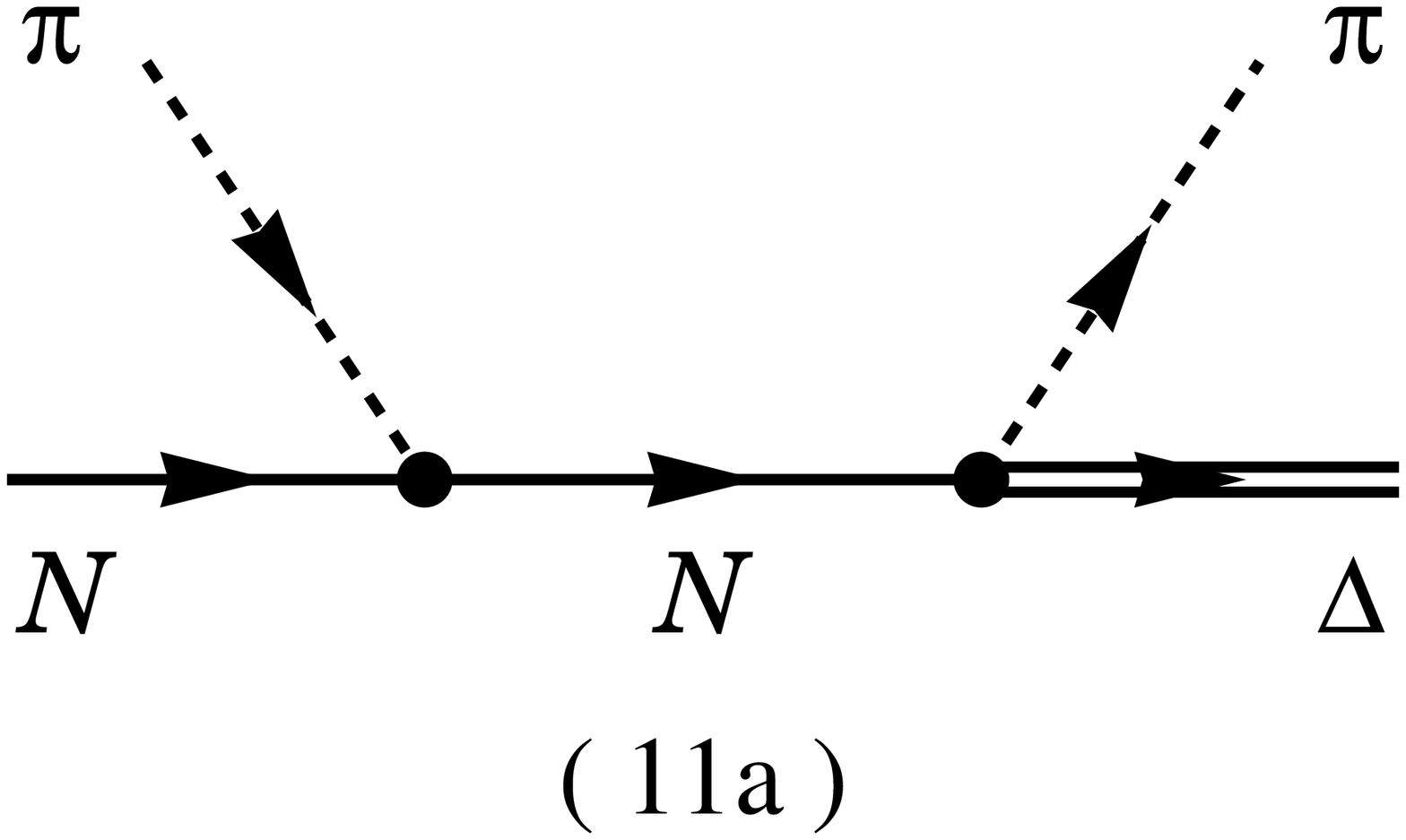} ~
  \PsfigII{0.175}{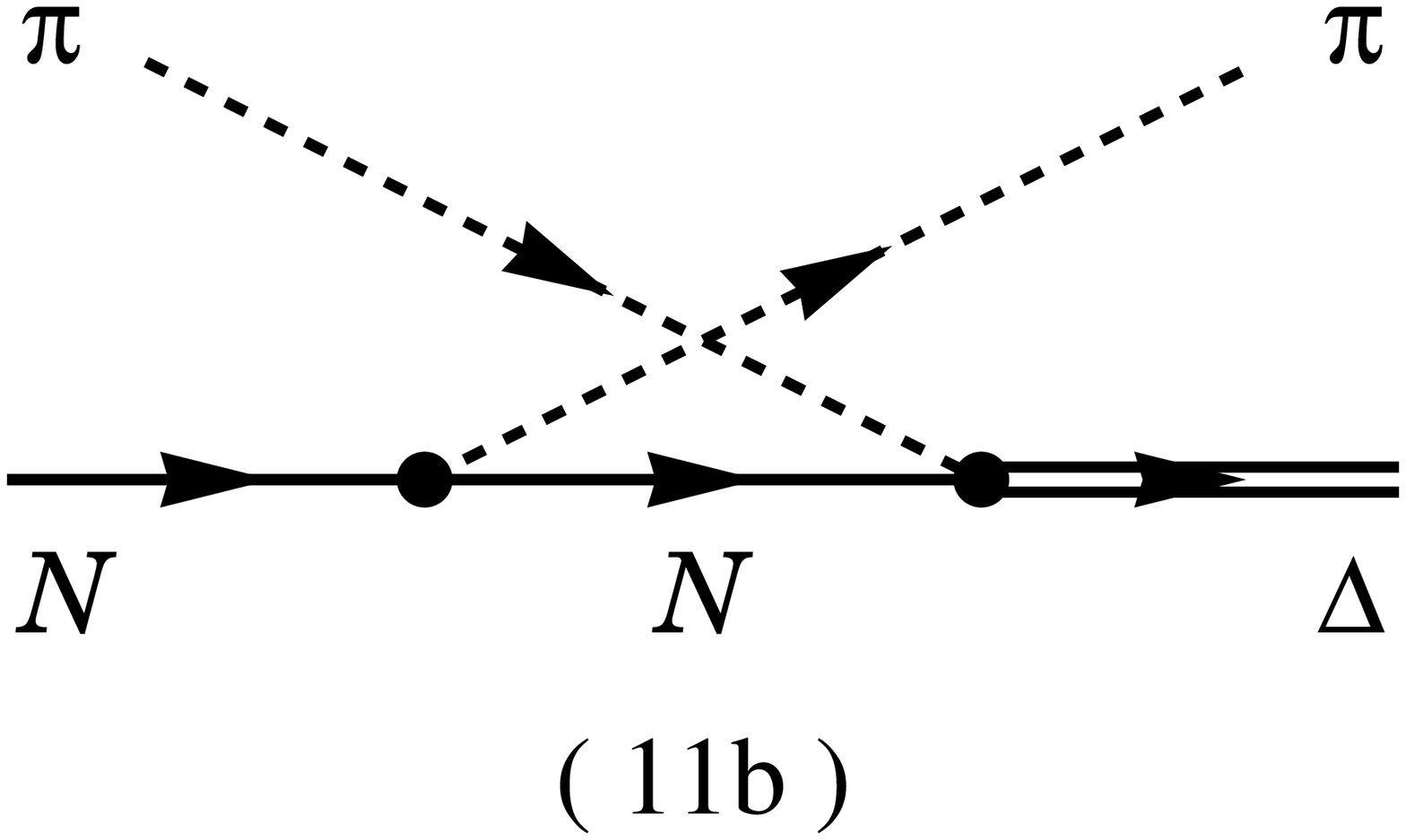} ~
  \PsfigII{0.175}{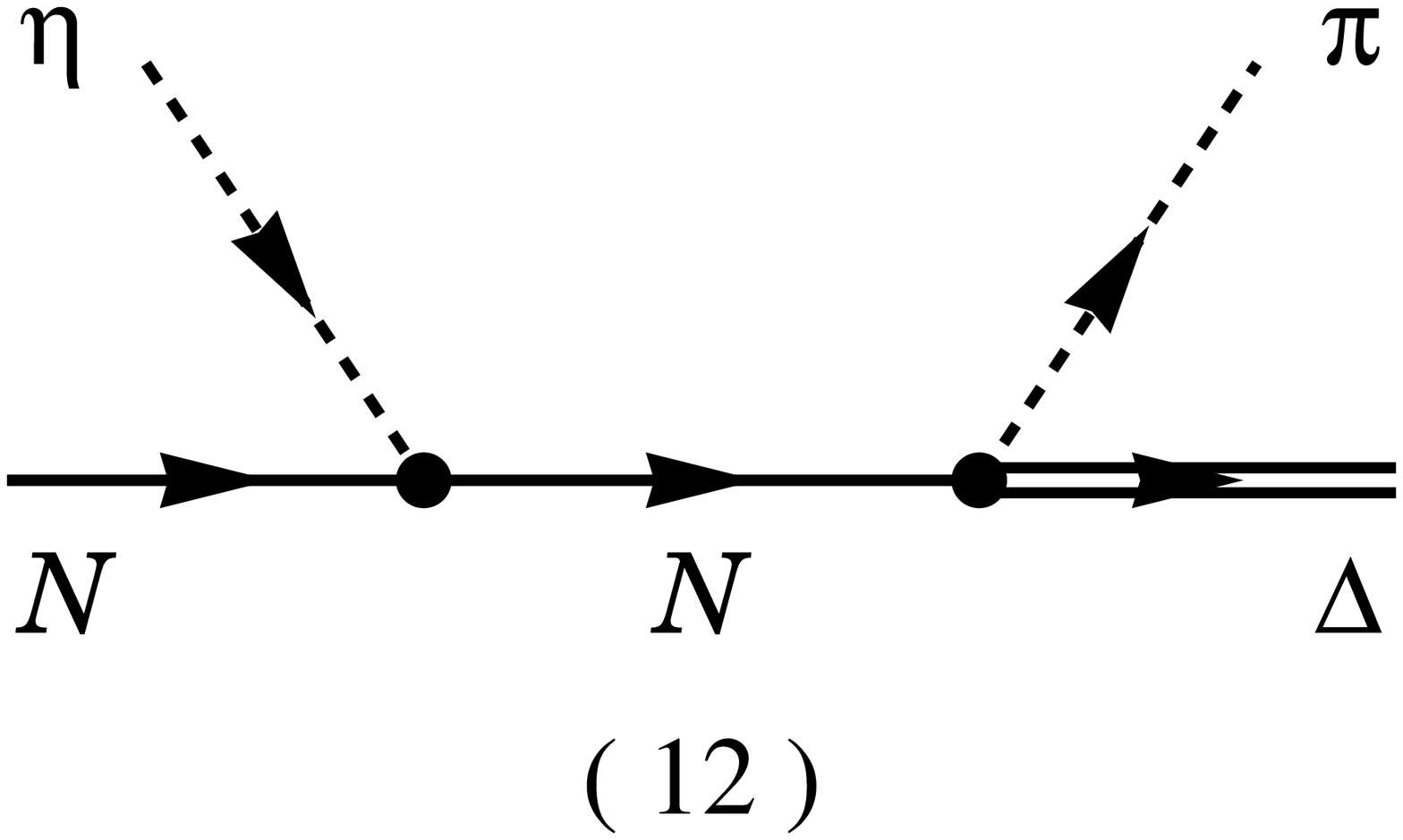} ~
  \PsfigII{0.175}{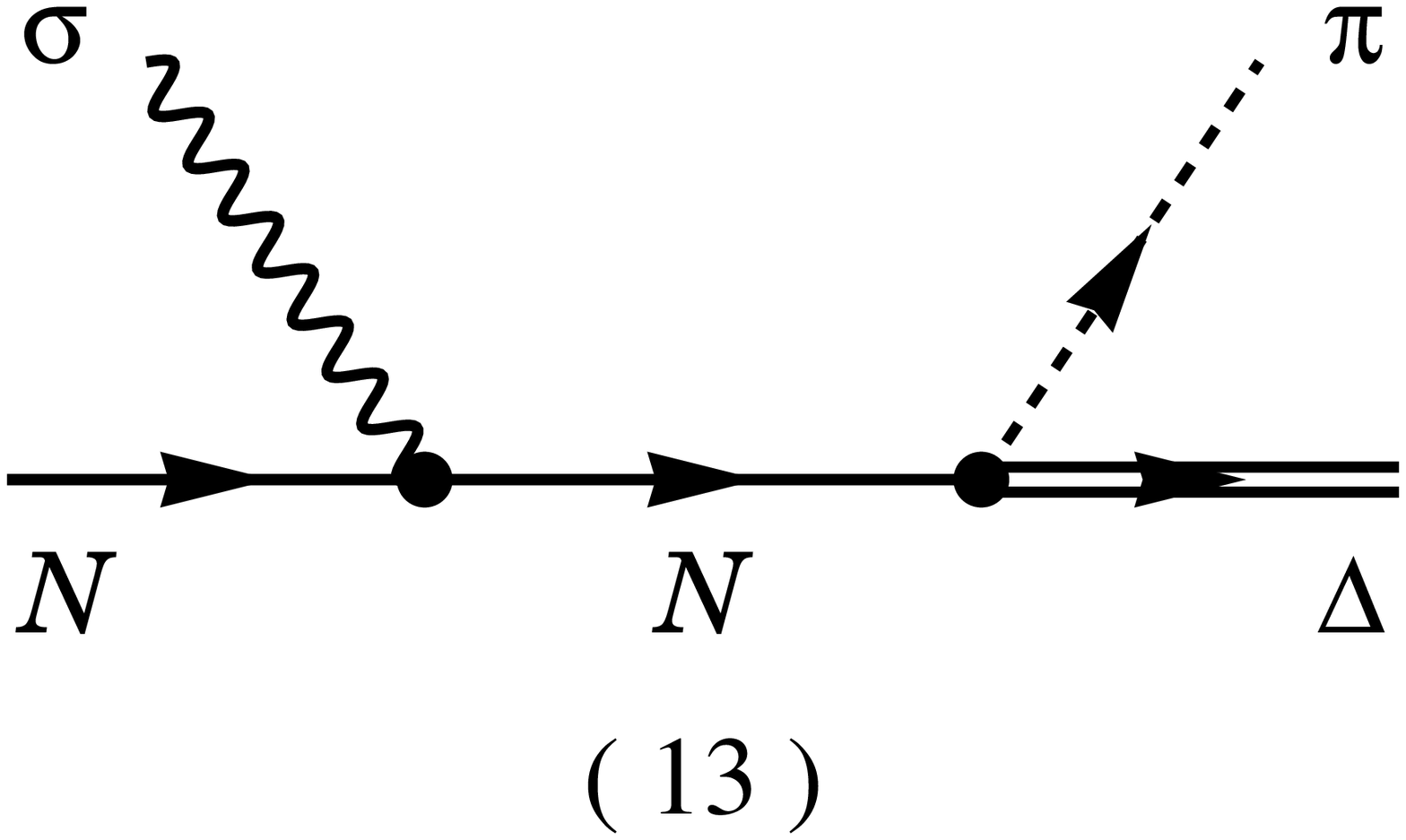} ~
  \PsfigII{0.175}{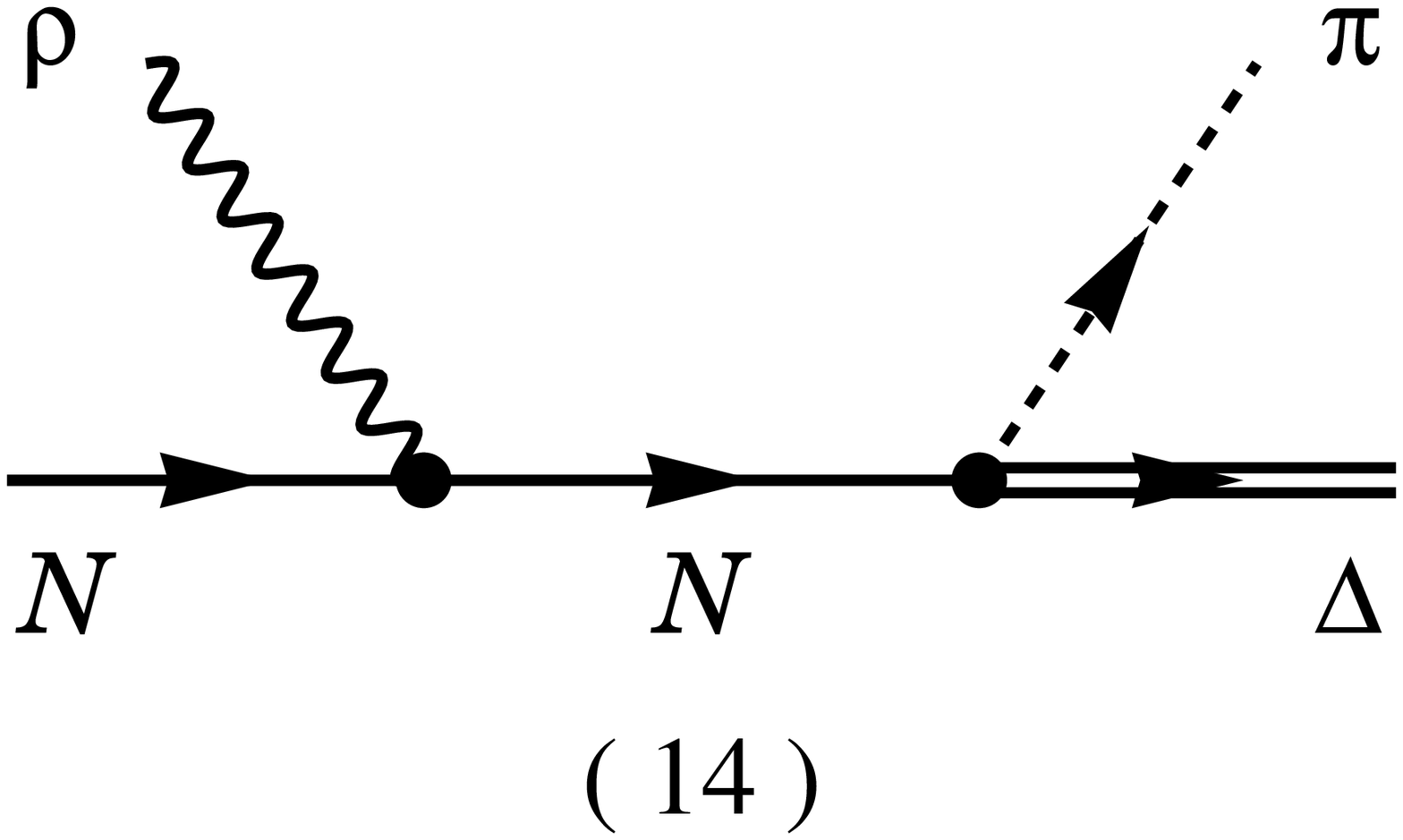} ~
  \PsfigII{0.175}{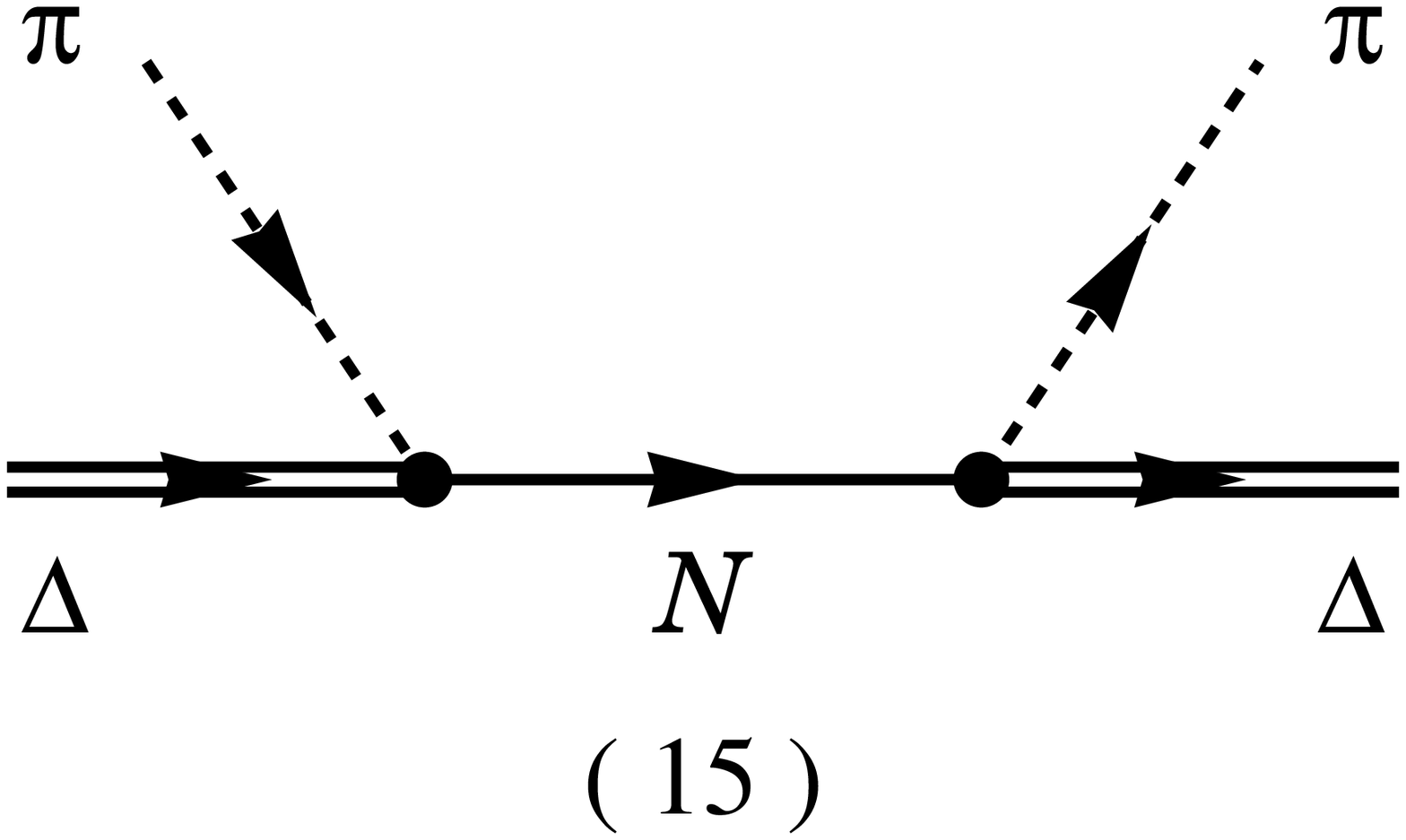} ~
  \caption{Feynman diagrams for the meson--baryon interactions in the
    $\pi N$ coupled-channels scattering amplitudes.}
  \label{fig:diagrams}
\end{figure*}

\begin{table*}[!p]
  \caption{Explicit channels for each quantum number $J^{P}$, $I$
    ($L_{2 I \, 2 J}$ for the elastic $\pi N$) considered in the
    present study.  The meson--baryon states are specified by their
    orbital angular momentum $L$ and total spin $S$ as ($L, S$).}
  \label{tab:exp_chan}
  \begin{ruledtabular}
    \begin{tabular*}{8.6cm}{@{\extracolsep{\fill}}l|llllllll}
      $J^{P}$, $I$ ($L_{2 I \, 2 J}$) & 
      Channel 1 & Channel 2 & Channel 3 &
      Channel 4 & Channel 5 & Channel 6 & Channel 7 & Channel 8 
      \\
      \hline
      $1/2^{-}$, $1/2$ ($S_{11}$) & 
      $\pi N ( 0 , 1/2 )$ &
      $\eta N ( 0 , 1/2 )$ &
      $\sigma N ( 1 , 1/2 )$ &
      $\rho N ( 0 , 1/2 )$ & 
      $\rho N ( 2 , 3/2 )$ &
      $\pi \Delta ( 2 , 3/2 )$ &
      &
      \\
      $1/2^{-}$, $3/2$ ($S_{31}$) & 
      $\pi N ( 0 , 1/2 )$ &
      $\rho N ( 0 , 1/2 )$ & 
      $\rho N ( 2 , 3/2 )$ & 
      $\pi \Delta ( 2 , 3/2 )$ &
      &
      & & \\
      $1/2^{+}$, $1/2$ ($P_{11}$) & 
      $\pi N ( 1 , 1/2 )$ &
      $\eta N ( 1 , 1/2 )$ &
      $\sigma N ( 0 , 1/2 )$ &
      $\rho N ( 1 , 1/2 )$ & 
      $\rho N ( 1 , 3/2 )$ & 
      $\pi \Delta ( 1 , 3/2 )$ &
      &
      \\
      $1/2^{+}$, $3/2$ ($P_{31}$) & 
      $\pi N ( 1 , 1/2 )$ &
      $\rho N ( 1 , 1/2 )$ & 
      $\rho N ( 1 , 3/2 )$ & 
      $\pi \Delta ( 1 , 3/2 )$ &
      &
      & & \\
      $3/2^{+}$, $1/2$ ($P_{13}$) & 
      $\pi N ( 1 , 1/2 )$ &
      $\eta N ( 1 , 1/2 )$ &
      $\sigma N ( 2 , 1/2 )$ &
      $\rho N ( 1 , 1/2 )$ & 
      $\rho N ( 1 , 3/2 )$ & 
      $\pi \Delta ( 1 , 3/2 )$ &
      &
      \\
      $3/2^{+}$, $3/2$ ($P_{33}$) & 
      $\pi N ( 1 , 1/2 )$ &
      $\rho N ( 1 , 1/2 )$ & 
      $\rho N ( 1 , 3/2 )$ & 
      $\pi \Delta ( 1 , 3/2 )$ &
      &
      & & \\
      $3/2^{-}$, $1/2$ ($D_{13}$) & 
      $\pi N ( 2 , 1/2 )$ &
      $\eta N ( 2 , 1/2 )$ &
      $\sigma N ( 1 , 1/2 )$ &
      $\rho N ( 0 , 3/2 )$ & 
      $\rho N ( 2 , 1/2 )$ & 
      $\rho N ( 2 , 3/2 )$ &
      $\pi \Delta ( 0 , 3/2 )$ &
      $\pi \Delta ( 2 , 3/2 )$ 
      \\
      $3/2^{-}$, $3/2$ ($D_{33}$) & 
      $\pi N ( 2 , 1/2 )$ &
      $\rho N ( 0 , 3/2 )$ & 
      $\rho N ( 2 , 1/2 )$ & 
      $\rho N ( 2 , 3/2 )$ &
      $\pi \Delta ( 0 , 3/2 )$ &
      $\pi \Delta ( 2 , 3/2 )$ &
      & 
      \\
      $5/2^{-}$, $1/2$ ($D_{15}$) & 
      $\pi N ( 2 , 1/2 )$ &
      $\eta N ( 2 , 1/2 )$ &
      $\rho N ( 2 , 1/2 )$ & 
      $\rho N ( 2 , 3/2 )$ & 
      $\pi \Delta ( 2 , 3/2 )$ &
      & & \\
      $5/2^{-}$, $3/2$ ($D_{35}$) & 
      $\pi N ( 2 , 1/2 )$ &
      $\rho N ( 2 , 1/2 )$ & 
      $\rho N ( 2 , 3/2 )$ & 
      $\pi \Delta ( 2 , 3/2 )$ &
      & & &
    \end{tabular*}
  \end{ruledtabular}
\end{table*}

\subsection{Complex scaling method for resonances}
\label{sec:2B}

To calculate numerically the resonance wave function from the residue
of the scattering amplitude, one has to perform an analytic
continuation to reach the resonance pole $E_{\rm pole}$ in the complex
energy plane.  One way to do this is the complex scaling
method~\cite{Aoyama:2006, Myo:2020rni}, which I employ in the present
study.

In the complex scaling method, one transforms the relative coordinate
$\bm{r}$ and relative momenta $\bm{q}$ for two-body states into the
complex-scaled values in the following manner:
\begin{equation}
  \bm{r} \to \bm{r} e^{i \theta} ,
  \quad
  \bm{q} \to \bm{q} e^{- i \theta} ,
\end{equation}
with the scaling angle $\theta$.  Then, the equations relevant to my
study become:
\begin{align}
  & T_{\alpha , j k} ( E ; q^{\prime} e^{- i \theta} , q e^{- i \theta} ) 
  = V_{\alpha , j k} ( E ; q^{\prime} e^{- i \theta} , q e^{- i \theta} ) 
  \notag \\
  & + e^{- 3 i \theta}
  \sum _{l = 1}^{N_{\rm chan}} \int _{0}^{\infty} d k \, k^{2}
  \frac{V_{\alpha , j l} ( E ; q^{\prime} e^{- i \theta} , k e^{- i \theta} )}
       {E - \mathcal{E}_{l} ( k e^{- i \theta} )}
  \notag \\
  & \quad \quad \quad \quad \quad \quad \quad 
  \times T_{\alpha , l k} ( E ; k e^{- i \theta} , q e^{- i \theta} )
  ,
\label{eq:LS_CSM}
\end{align}
\begin{align}
  & T_{\alpha , j k} ( E ; q^{\prime} e^{- i \theta} , q e^{- i \theta} )
  \notag \\
  & = 
  \frac{\gamma _{j} ( q^{\prime} e^{- i \theta} ) \gamma _{k} ( q e^{- i \theta} )}
       {E - E_{\rm pole}}
       + (\text{regular at } E = E_{\rm pole}) ,
\end{align}
\begin{equation}
  \gamma _{j} ( q e^{- i \theta} )
  = \frac{1}{( 2 \pi )^{3/2}}
  [ E_{\rm pole} - \mathcal{E}_{j} ( q e^{- i \theta} ) ] 
  \tilde{R}_{j} ( q e^{- i \theta} ) ,
  \label{eq:gamma_CSM}
\end{equation}
\begin{equation}
  X_{j} = \int _{0}^{\infty} d q \, \tilde{\Rho} _{j}^{( \theta )} ( q )
  = \int _{0}^{\infty} d r \, \Rho _{j}^{( \theta )} ( r ) ,
  \label{eq:Xj_CSM}
\end{equation}
\begin{equation}
  \tilde{\Rho} _{j}^{( \theta )} ( q ) = e^{- 3 i \theta} q^{2}
    \left [ \frac{\gamma _{j} ( q e^{- i \theta} )}
      {E_{\rm pole} - \mathcal{E}_{j} ( q e^{- i \theta} )} \right ] ^{2} ,
  \label{eq:Rho_mom_CSM}
\end{equation}
\begin{equation}
  \Rho _{j}^{( \theta )} ( r ) = \frac{2 r^{2} e^{-3 i \theta}}{\pi}
  \left [ \int _{0}^{\infty} d q \, q^{2}
    \frac{\gamma _{j} ( q e^{- i \theta} ) j_{L} ( q r )}
         {E_{\rm pole} - \mathcal{E}_{j} ( q e^{- i \theta} )}
         \right ] ^{2} .
  \label{eq:Rho_CSM}
\end{equation}
The definitions of the missing-channel contribution
$Z$~\eqref{eq:missing}, uncertainties $U$~\eqref{eq:XtildeU} and
$U_{\rm r}$~\eqref{eq:Ur}, and real-valued quantities $\tilde{X}$ and
$\tilde{Z}$~\eqref{eq:XtildeZtilde} are unchanged.

An important property of the complex scaling method is that the pole
position $E_{\rm pole}$ and compositeness $X_{j}$ do not depend on the
scaling angle $\theta$ while density distributions $\Rho _{j}^{(
  \theta )}$ and $\tilde{\Rho}_{j}^{( \theta )}$ depend on $\theta$.

\section{\boldmath The $\pi N$ coupled-channels scattering amplitudes}
\label{sec:3}

As described in the previous section, one can extract normalized
two-body wave functions of resonance states from the residues of the
off-shell scattering amplitudes at the resonance poles.  This fact is
of special important when investigating the internal structure of the
$N^{\ast}$ and $\Delta ^{\ast}$ resonances, because nowadays precise
$\pi N$ scattering amplitudes are available from the partial wave
analysis of the experimental data (see, e.g., the database of
SAID~\cite{SAID}), which allows us to construct sophisticated models
for the $\pi N$ coupled-channels scattering amplitudes as done in
Refs.~\cite{Ronchen:2012eg, Kamano:2013iva}.

In this study I investigate the meson--baryon molecular components of
the $N^{\ast}$ and $\Delta ^{\ast}$ resonances by constructing a
meson--baryon coupled-channels model for the $\pi N$ amplitudes partly
based on Ref.~\cite{JuliaDiaz:2007kz}.  For this purpose I take into
account the $\pi N$, $\eta N$, $\sigma N$, $\rho N$, and $\pi \Delta$
channels.  The interactions are calculated according to the Feynman
diagrams shown in Fig.~\ref{fig:diagrams}.  Model parameters are fixed
so as to reproduce the results of the SAID partial wave analysis for
the on-shell $\pi N$ amplitudes~\cite{SAID} up to $E = \SI{1.9}{GeV}$
with the orbital angular momentum $L \le 2$, i.e., $S_{11}$, $S_{31}$,
$P_{11}$, $P_{31}$, $P_{13}$, $P_{33}$, $D_{13}$, $D_{33}$, $D_{15}$,
and $D_{35}$ partial waves of the elastic $\pi N$ scattering.  Here
and below I use the notation $L_{2 I \, 2 J}$ with the orbital angular
momentum $L$, isospin $I$, and total angular momentum $J$ for the $\pi
N$ system.

I first summarize the notation of the $\pi N$ scattering amplitudes in
Sec.~\ref{sec:3A} and then construct the tree-level interactions in
Sec.~\ref{sec:3B}.  The self-energies for the unstable channels, i.e.,
$\sigma N$, $\rho N$, and $\pi \Delta$, are introduced in
Sec.~\ref{sec:3C}.  In Sec.~\ref{sec:3D} bare $N^{\ast}$ and $\Delta
^{\ast}$ states are introduced.  Finally in Sec.~\ref{sec:3E} the
model parameters are fitted to reproduce the on-shell $\pi N$ partial
wave amplitudes.  Throughout the calculations isospin symmetry is
assumed.

\subsection{Notation of the scattering amplitudes}
\label{sec:3A}

First of all, I fix the quantum number of the $\pi N$ scattering.
When one considers the elastic $\pi N$ scattering, its partial wave
can be uniquely specified by $L_{2 I \, 2 J}$.  In a general
coupled-channels analysis, however, the quantum number of the system
should be specified by the spin/parity $J^{P}$ and isospin $I$,
because the orbital angular momentum $L$ may change in different
channels such as $\pi N$ and $\sigma N$.  Therefore, for the
coupled-channels scattering amplitude in the Lippmann--Schwinger
equation~\eqref{eq:LSeq}, I use the notation $\alpha = ( J^{P} , I)$,
but for the $\pi N$ partial waves I use the notation $L_{2 I \, 2 J}$
as well.

For each quantum number $\alpha = ( J^{P} , I)$, I take into
account the $\pi N$, $\eta N$, $\sigma N$, $\rho N$, and $\pi \Delta$
channels with their orbital angular momenta $L \le 2$.  The explicit
channels considered are summarized in Table~\ref{tab:exp_chan}.  From
the coupled-channels amplitude, I calculate the normalized on-shell
$\pi N$ amplitude $L_{2 I \, 2 J} ( E )$ as
\begin{equation}
  L_{2 I \, 2 J} ( E ) = - \frac{\rho _{\pi N} ( E )}{2}
  T_{( J^{P} , I), 1 1}^{\text{on-shell}} ( E ) ,
\end{equation}
where the phase space $\rho _{\pi N}$ of the $\pi N$ channel is
calculated according to Eq.~\eqref{eq:PSrho}.  This normalized $\pi N$
amplitude satisfies the optical theorem
\begin{equation}
  \text{Im} \, L_{2 I \, 2 J} ( E ) = 
  \left | L_{2 I \, 2 J} ( E ) \right | ^{2} .
\end{equation}
below the inelastic threshold for the $\pi N$ state.

\def\arraystretch{2.0}

\begin{table}[!t]
  \caption{Effective Lagrangians for the $\pi N$ coupled-channels
    interaction.}
  \label{tab:Lag}
  \begin{ruledtabular}
    \begin{tabular*}{8.6cm}{@{\extracolsep{\fill}}ll}
      Vertex & $\mathcal{L}_{\rm int}$
      \\
      \hline
      $\pi N N$ &
      $\displaystyle - \frac{D + F}{2 f_{\pi}}
      \bar{N} \gamma ^{\mu} \gamma _{5} \prt _{\mu} \vec{\pi} \cdot \vec{\tau} N$
      \\
      $\pi N \Delta$ &
      $\displaystyle - \frac{f_{\pi N \Delta}}{m_{\pi}}
      \bar{N} \prt _{\mu} \vec{\pi} \cdot \vec{T} \Delta ^{\mu}
      + \text{h.c.}$
      \\
      $\rho N N$ &
      $\displaystyle \frac{g_{\rho N N}}{2}
      \bar{N} \left ( \gamma ^{\mu} - \frac{\kappa _{\rho}}{2 m_{N}}
      \sigma ^{\mu \nu} \partial _{\nu} \right )
      \vec{\rho}_{\mu} \cdot \vec{\tau} N$
      \\
      $\pi \pi \rho$ &
      $\displaystyle  g_{\pi \pi \rho} ( \vec{\pi} \times \prt ^{\mu} \vec{\pi} )
      \cdot \vec{\rho}_{\mu}$
      \\
      $\sigma N N$ &
      $g_{\sigma N N} \sigma \bar{N} N$
      \\
      $\pi \pi \sigma$ &
      $\displaystyle - \frac{g_{\pi \pi \sigma}}{2 m_{\pi}}
      \prt _{\mu} \vec{\pi} \cdot \prt ^{\mu} \vec{\pi} \sigma$
      \\
      $\eta N N$ &
      $\displaystyle \frac{D - 3 F}{2 \sqrt{3} f_{\eta}}
      \prt _{\mu} \eta \bar{N} \gamma ^{\mu} \gamma _{5} N $
      \\
      $\pi \rho N N$ &
      $\displaystyle \frac{( D + F ) g_{\rho N N}}{2 f_{\pi}} \bar{N}
        \gamma ^{\mu} \gamma _{5} ( \vec{\pi} \times \vec{\rho}_{\mu} )
      \cdot \vec{\tau} N$
      \\
      $\rho \rho N N$ &
      $\displaystyle \frac{g_{\rho N N}^{2} \kappa _{\rho}}{8 m_{N}}
      \bar{N} \sigma ^{\mu \nu} ( \vec{\rho}_{\mu}\times \vec{\rho}_{\nu} )
      \cdot \vec{\tau} N$
    \end{tabular*}
  \end{ruledtabular}
\end{table}

\def\arraystretch{1.0}

\subsection{\boldmath Tree-level interactions}
\label{sec:3B}

Next I formulate the tree-level $\pi N$ coupled-channels interactions
diagrammatically shown in Fig.~\ref{fig:diagrams}.  For this purpose I
employ effective Lagrangians summarized in Table~\ref{tab:Lag}.  Here
the notations for the hadron fields are: $N = ( p , n )^{\rm t}$ and
\begin{equation}
  \vec{\pi} = ( \pi ^{1} , \pi ^{2} , \pi ^{3} ) 
  = \left ( \frac{\pi ^{+} + \pi ^{-}}{\sqrt{2}} , 
    - \frac{\pi ^{+} - \pi ^{-}}{\sqrt{2} i} , 
    \pi ^{0} \right ) ,
    \label{eq:pi}
\end{equation}
hence
\begin{equation}
  \vec{\pi} \cdot \vec{\tau} = 
  \begin{pmatrix}
    \pi ^{0} & \sqrt{2} \pi ^{+} \\
    \sqrt{2} \pi ^{-} & - \pi ^{0} \\
  \end{pmatrix} ,
\end{equation}
where $\vec{\tau} = ( \tau ^{1} , \tau ^{2} , \tau ^{3} )$ is the
Pauli matrices acting on the isospin states.  The $\vec{\rho}$ field
is expressed in the same manner to the $\vec{\pi}$ field.  The
$\Delta$ field $\Delta = ( \Delta ^{++} , \Delta ^{+} , \Delta ^{0} ,
\Delta ^{-} )^{\rm t}$ is tied to the isospin transition operators
$\vec{T} = ( T^{1} , T^{2} , T^{3} )$ from isospin $3/2$ to $1/2$:
\begin{equation}
T^{1} = 
  \begin{pmatrix}
    - 1 / \sqrt{2} & 0 & 1 / \sqrt{6} & 0 \\
    0 & - 1 / \sqrt{6} & 0 & 1 / \sqrt{2} \\
  \end{pmatrix} ,
\end{equation}
\begin{equation}
T^{2} = 
  \begin{pmatrix}
    - i / \sqrt{2} & 0 & - i / \sqrt{6} & 0 \\
    0 & - i / \sqrt{6} & 0 & - i / \sqrt{2} \\
  \end{pmatrix} ,
\end{equation}
\begin{equation}
T^{3} = 
  \begin{pmatrix}
    0 & \sqrt{2/3} & 0 & 0 \\
    0 & 0 & \sqrt{2/3} & 0 \\
  \end{pmatrix} ,
\end{equation}
hence
\begin{equation}
  \vec{\pi} \cdot \vec{T}
  = 
  \begin{pmatrix}
    - \pi ^{-} & \sqrt{2/3} \pi ^{0} & \sqrt{1/3} \pi ^{+} & 0 \\
    0 & - \sqrt{1/3} \pi ^{-} & \sqrt{2/3} \pi ^{0} & \pi ^{+} \\
  \end{pmatrix} .
\end{equation}

For the $\pi N N$ and $\eta N N$ vertices, I employ the chiral
Lagrangian with the meson decay constants, $f_{\pi}$ and $f_{\eta}$,
at their physical values~\cite{Zyla:2020zbs}, $f_{\pi} =
\SI{92.1}{MeV}$ and $f_{\eta} = 1.3 f_{\pi}$, and fix the coupling
constants $D + F = 1.26$ and $D - F = 0.33$ by the weak decay of octet
baryons.  Therefore, both the $\pi N N$ and $\eta N N$ couplings do
not contain free parameters, except for the cutoffs [see
  Eq.~\eqref{eq:FF}].

As for the $\pi N \Delta$ vertex, I treat the coupling constant
$f_{\pi N \Delta}$ as a free parameter.  I use the real-valued
physical $\Delta (1232)$ mass $m_{\Delta} = \SI{1210}{MeV}$ for the
propagator of the diagram (1c) in Fig.~\ref{fig:diagrams}.  I allow
that the parameters for the $\Delta$ in the diagram (1c) may differ
from those for the bare $\Delta$ introduced in Sec.~\ref{sec:3D} for
better reproduction of the experimental data.

In the case of the $\rho$ and $\sigma$ exchanges, the coupling
constants $g_{\rho N N}$, $\kappa _{\rho}$, $g_{\pi \pi \rho}$,
$g_{\sigma N N}$, and $g_{\pi \pi \sigma}$ are free parameters.  For
these $t$-channel $\rho$ and $\sigma$ exchanges, I use the real-valued
physical mass for the $\rho$ meson, $m_{\rho} = \SI{775.3}{MeV}$, but
a real-valued bare mass for the $\sigma$ meson, $m_{\sigma _{0}} =
\SI{700}{MeV}$ (see Sec.~\ref{sec:3C}).

To regularize the divergences from the integrals of the
Lippmann--Schwinger equation~\eqref{eq:LSeq}, I introduce a dipole
form factor
\begin{equation}
  \mathcal{F} ( \Lambda , q ) \equiv 
  \left ( \frac{\Lambda ^{2}}
        {\Lambda ^{2} + q^{2}} \right ) ^{2} ,
        \label{eq:FF}
\end{equation}
with a cutoff $\Lambda$ for each meson--baryon--baryon vertex with $q$
being the three-momentum of the meson.  I also use the dipole form
factor~\eqref{eq:FF} for the meson--meson--meson vertex with $q$ being
the three-momentum of the exchanged meson.  The cutoffs $\Lambda$ are
model parameters and may take different values for different vertices.

The energies of the mesons and baryons in the initial and final states
are respectively fixed to their on-shell values [see
  Eq.~\eqref{eqA:k0p0} in Appendix~\ref{app:int}].  Therefore, the
meson--baryon interactions of the diagrams in Fig.~\ref{fig:diagrams}
do not depend on the center-of-mass energy $E$ but only on the
relative momenta $q$ and $q^{\prime}$.  The explicit forms of the
interaction terms and their partial-wave projections are given in
Appendix~\ref{app:forms}.

\subsection{\boldmath Self-energies for the $\sigma N$, $\rho N$, and
  $\pi \Delta$ channels}
\label{sec:3C}

Let us turn to the self-energies for the $\sigma N$, $\rho N$, and
$\pi \Delta$ channels.

Because the $\sigma$, $\rho$, and $\Delta$ resonances are unstable
particles, one should take into account their self-energies as in
Eq.~\eqref{eq:Ej_SE}.  I take the strategy to calculate the
self-energy developed in Refs.~\cite{Kamano:2008gr, Kamano:2013iva}.
For the $\rho$ and $\Delta$ resonances, I use the same formulae and
parameters in Ref.~\cite{Kamano:2013iva}.  On the other hand, to
describe the $\sigma$ resonance I use the same formula but different
parameters so as to reproduce the $\pi \pi$ ($I = 0$, $L = 0$) phase
shift in particular near the $\pi \pi$ threshold where the $\sigma$
resonance exists.  The effective interaction of the $\pi \pi$ ($I =
0$, $L = 0$) scattering is
\begin{equation}
  \mathcal{V}_{\sigma} ( E_{2} ; q^{\prime} , q )
  = \frac{g_{0}}{E_{2} - m_{\sigma _{0}}} f_{g} ( q^{\prime} ) f_{g} ( q )
  + h_{0} f_{h} ( q^{\prime} ) f_{h} ( q ) ,
\end{equation}
with monopole form factors
\begin{equation}
  f_{g} ( q )
  = \frac{\lambda _{g}^{2}}{q^{2} + \lambda _{g}^{2}} ,
  \quad 
  f_{h} ( q )
  = \frac{\lambda _{h}^{2}}{q^{2} + \lambda _{h}^{2}} .
\end{equation}
Here $E_{2}$ is the total energy of the $\pi \pi$ system and $q^{(
  \prime )}$ is the relative momentum of the initial (final) $\pi \pi$
state.  The bare $\sigma$ mass $m_{\sigma _{0}}$, coupling constants
$g_{0}$ and $h_{0}$, and cutoffs $\lambda _{g}$ and $\lambda _{h}$
are the model parameters for the $\sigma$ resonance.  As a result of
the fit to the $\pi \pi$ ($I = 0$, $L = 0$) phase shift, I obtain
$m_{\sigma _{0}} = \SI{700}{MeV}$, $g_{0} = \SI{616}{GeV^{-1}}$,
$h_{0} = \SI{1189}{GeV^{-2}}$, $\lambda _{g} = \SI{178}{MeV}$, and
$\lambda _{h} = \SI{217}{MeV}$ for the $\sigma$ resonance.

With the parameters for the $\sigma$, $\rho$, and $\Delta$ resonances,
I find resonance poles at $E_{2} = 486 - 213 i \, \si{MeV}$, $765 - 75
i \, \si{MeV}$, and $1210 - 55 i \, \si{MeV}$ for the $\sigma$,
$\rho$, and $\Delta$ resonances, respectively.

\begin{figure}[!t]
  \centering
  \Psfig{8.6cm}{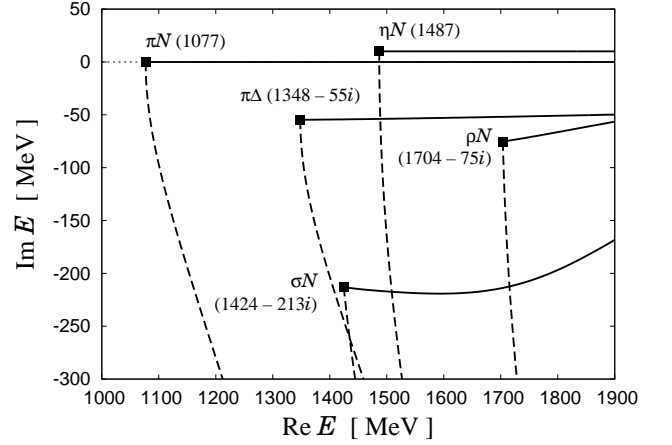}
  \caption{Branch cuts for the $\pi N$, $\eta N$, $\sigma N$, $\rho
    N$, and $\pi \Delta$ channels.  The solid lines indicate the usual
    cuts, i.e., cuts with the complex-scaling angle $\theta =
    0^{\circ}$, while the dashed lines are those with $\theta =
    45^{\circ}$.  The line for the $\eta N$ channel is shifted
    slightly upward for a better visualization.  The boxes denote the
    branch points, whose positions are written in the parentheses in
    units of MeV.}
  \label{fig:branch}
\end{figure}

\begin{table*}[t]
  \caption{Bare $N^{\ast}$ and $\Delta ^{\ast}$ states introduced in
    the present model.  Their fitted bare masses $M_{0}$, cutoffs
    $\Lambda$, and coupling constants $g$ are also shown.  The
    channels are specified by the indices $1$, $2$, $\ldots$, $8$ in
    the same order as in Table~\ref{tab:exp_chan}.}
  \label{tab:bare}
  \begin{ruledtabular}
    \begin{tabular*}{\textwidth}{@{\extracolsep{\fill}}lcc|rrrrrrrr}
      $J^{P}$, $I$ ($L_{2 I \, 2 J}$) & $M_{0}$ [MeV] & $\Lambda$ [MeV]
      & \multicolumn{1}{c}{$g_{1}$}
      & \multicolumn{1}{c}{$g_{2}$}
      & \multicolumn{1}{c}{$g_{3}$}
      & \multicolumn{1}{c}{$g_{4}$}
      & \multicolumn{1}{c}{$g_{5}$}
      & \multicolumn{1}{c}{$g_{6}$}
      & \multicolumn{1}{c}{$g_{7}$}
      & \multicolumn{1}{c}{$g_{8}$} 
      \\
      \hline
      $1/2^{-}$, $1/2$ ($S_{1 1}$) &
      $1919$ & $855$
      & $2.054$ & $-0.570$ & $0.415$ & $-1.817$
      & $-0.031$ & $-0.043$ & &
      \\
      '' &
      $2150$ & $575$
      & $1.095$ & $-0.761$ & $-0.227$ & $-2.651$
      & $-0.386$ & $0.181$ & &
      \\
      $1/2^{+}$, $1/2$ ($P_{1 1}$) & 
      $1912$ & $634$
      & $-1.116$ & $0.299$ & $2.221$ & $0.157$
      & $0.719$ & $0.610$ & & 
      \\
      $3/2^{+}$, $3/2$ ($P_{3 3}$) & 
      $1314$ & $476$
      & $1.089$ & $-0.310$ & $-0.288$ & $0.046$
      & & & & 
      \\
      $3/2^{-}$, $1/2$ ($D_{1 3}$) & 
      $2043$ & $608$
      & $0.148$ & $0.033$ & $-1.028$ & $-1.935$
      & $-0.152$ & $-0.213$ & $-0.149$ & $-0.068$
      \\
      $3/2^{-}$, $3/2$ ($D_{3 3}$) & 
      $1824$ & $446$
      & $0.168$ & $-2.488$ & $0.142$ & $0.262$
      & $2.762$ & $0.150$ & &
      \\
      $5/2^{-}$, $1/2$ ($D_{1 5}$) & 
      $1821$ & $494$
      & $0.142$ & $0.189$ & $0.060$ & $0.096$
      & $0.167$ & & &
    \end{tabular*}
  \end{ruledtabular}
\end{table*}

The $\sigma N$, $\rho N$, and $\pi \Delta$ self-energies, $\Sigma
_{\sigma N}$, $\Sigma _{\rho N}$, and $\Sigma _{\pi \Delta}$,
respectively, are calculated in the same manner as in
Ref.~\cite{Kamano:2013iva}, and then I obtain the kinetic
energy~\eqref{eq:Ej_SE} in these channels:
\begin{equation}
  \mathcal{E}_{\sigma N} ( E ; q ) = \sqrt{m_{\sigma _{0}}^{2} + q^{2}}
  + \sqrt{m_{N}^{2} + q^{2}} + \Sigma _{\sigma N} ( E ; q ) ,
  \label{eq:EsigmaN}
\end{equation}
\begin{equation}
  \mathcal{E}_{\rho N} ( E ; q ) = \sqrt{m_{\rho _{0}}^{2} + q^{2}}
    + \sqrt{m_{N}^{2} + q^{2}} + \Sigma _{\rho N} ( E ; q ) ,
  \label{eq:ErhoN}
\end{equation}
\begin{equation}
  \mathcal{E}_{\pi \Delta} ( E ; q ) = \sqrt{m_{\pi}^{2} + q^{2}}
    + \sqrt{m_{\Delta _{0}}^{2} + q^{2}} + \Sigma _{\pi \Delta} ( E ; q ) ,
  \label{eq:EpiDelta}
\end{equation}
with the bare $\rho$ mass $m_{\rho _{0}} = \SI{812}{MeV}$ and bare
$\Delta$ mass $m_{\Delta _{0}} =
\SI{1280}{MeV}$~\cite{Kamano:2013iva}.  I note that the kinetic
energies for the unstable channels as well as the self-energies depend
on the energy $E$ because they implicitly involve the three-body $\pi
\pi N$ state.  Therefore, this $E$ dependence in the kinetic energy
may give a deviation of the compositeness from unity and hence a
nonzero missing-channel contribution $Z$~\eqref{eq:missing}
corresponding to the implicit $\pi \pi N$ state, as discussed in
Ref.~\cite{Sekihara:2016xnq}.

By calculating the energy $E$ which satisfies $E = \mathcal{E}_{\sigma
  N, \rho N, \pi \Delta} ( E ; q )$ with the momentum $q$ from $0$ to
$+ \infty$, I obtain the branch cuts for these channels, which are
plotted in Fig.~\ref{fig:branch} as solid lines together with the
branch cuts for the stable channels $\pi N$ and $\eta N$.  I note that
for the energy satisfying $\text{Re} \, E > 2 m_{\pi} + m_{N} =
\SI{1215}{MeV}$ and $\text{Im} \, E < 0$ I perform the analytic
continuation to the second Riemann sheet of the $\pi \pi N$ channel by
deforming appropriately the momentum integral paths in the formulae of
the self-energies.

Here it is instructive to see how these branch cuts behave with a
finite value of the complex-scaling angle $\theta$ in the complex
scaling method.  For this purpose, in Fig.~\ref{fig:branch} I also
plot the branch cuts with $\theta = 45^{\circ}$ in the complex scaling
method as the dashed lines.  As one can see, each branch cut rotates
clockwise with a finite value of the scaling angle.  In practical
calculations, one can reach the second Riemann sheet in each channel
for the complex energy $E$ in the region sandwiched between the solid
and dashed lines of Fig.~\ref{fig:branch}.

\begin{figure}[!t]
  \centering
  \PsfigII{0.175}{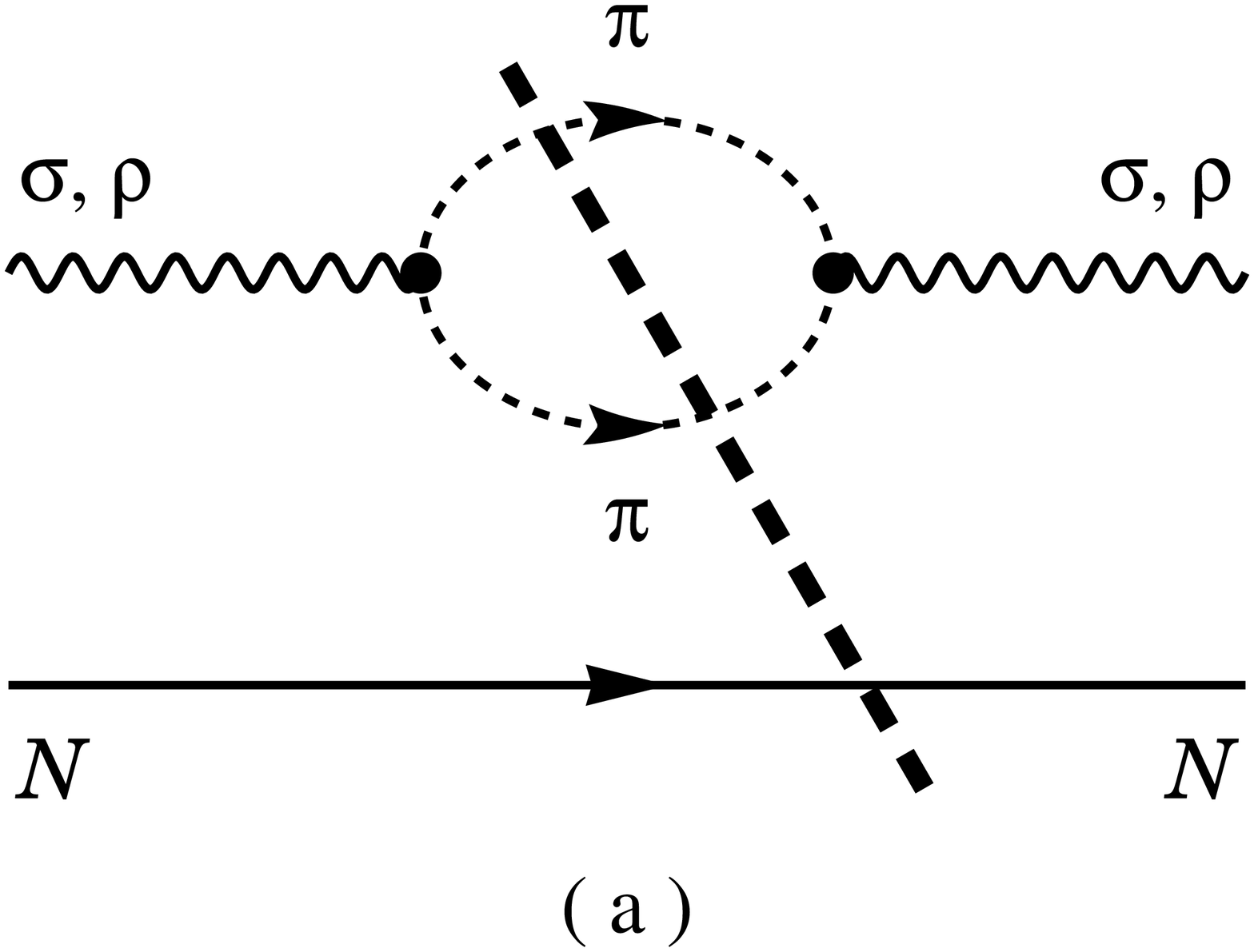} 
  \PsfigII{0.175}{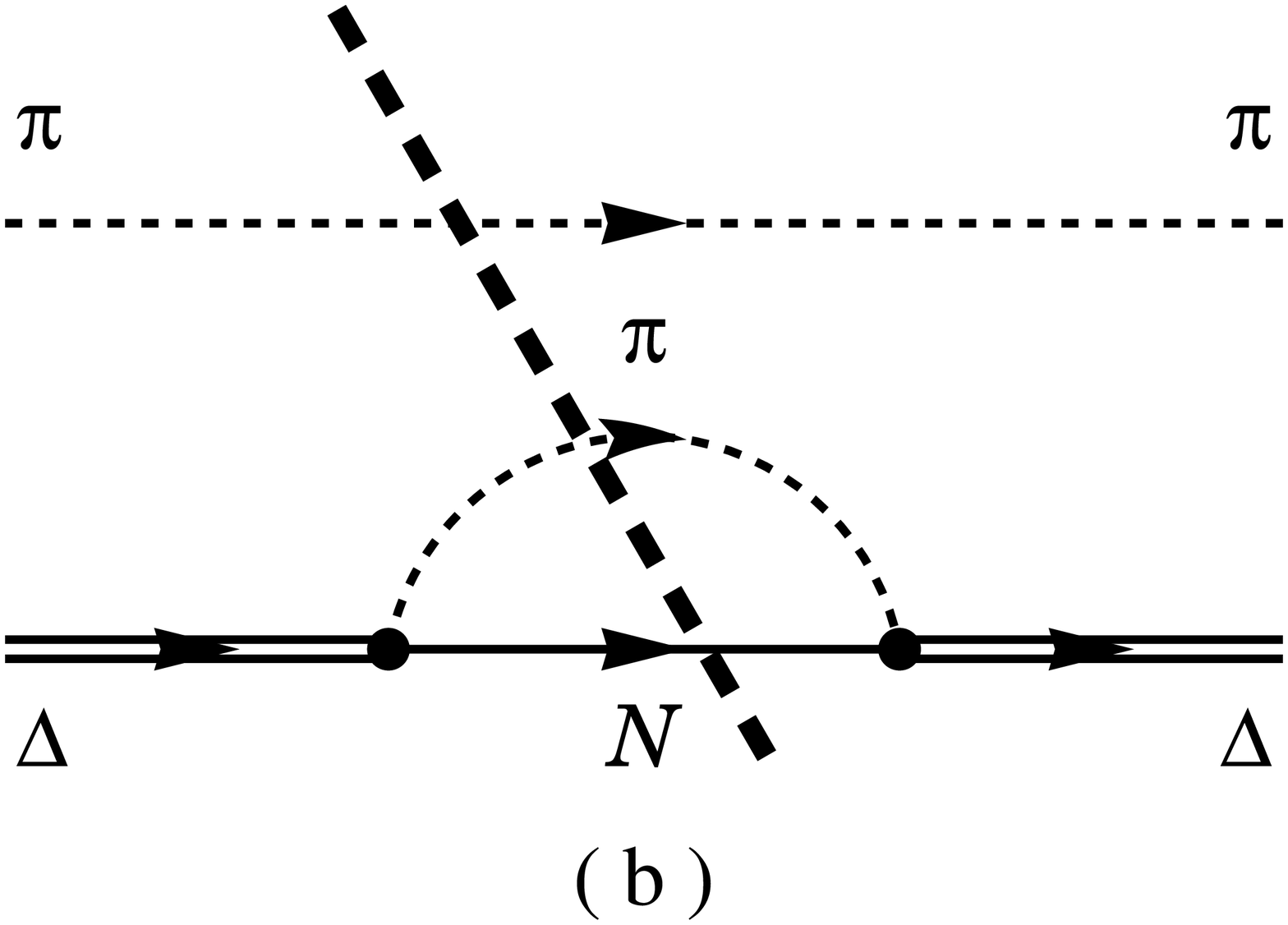} 
  \PsfigII{0.175}{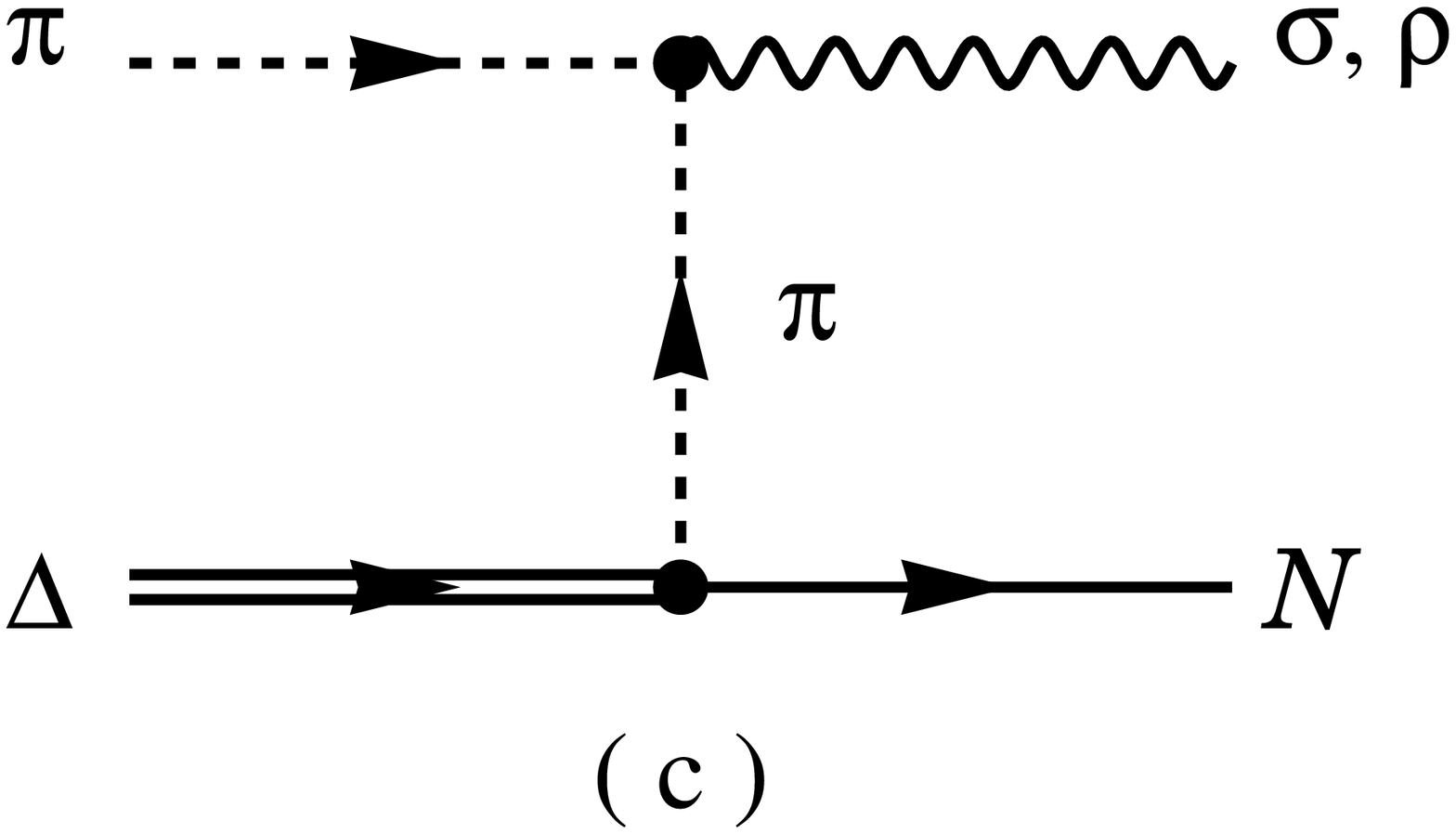}
  \PsfigII{0.175}{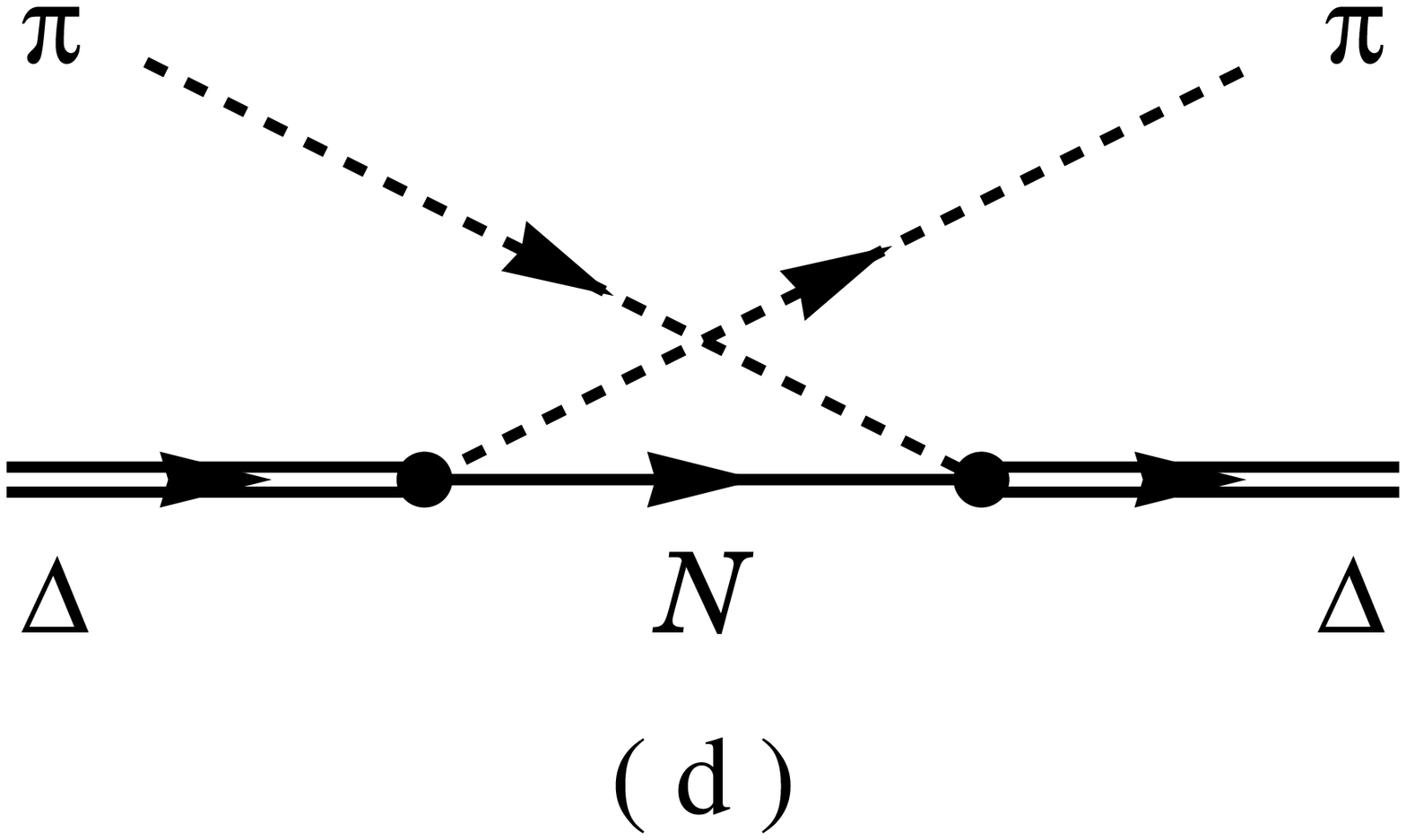}
  \caption{Diagrams for the $\pi \pi N$ three-body unitary cut.  (a)
    and (b) are taken into account in the present model, while (c) and
    (d) are not included.}
  \label{fig:cut}
\end{figure}

Finally I comment on the three-body unitarity.  In the $\pi N$
scattering the $\pi \pi N$ three-body channel opens at the $\pi \pi N$
threshold.  In the present model I implicitly include the $\pi \pi N$
discontinuities arising from the $s$-channel propagation of the
three-body states in the self-energies for the $\sigma N$, $\rho N$,
and $\pi \Delta$ channels.  The three-body cut for this process is
depicted in Figs.~\ref{fig:cut}(a) and \ref{fig:cut}(b) as the thick
dashed lines.  On the other hand, I do not include the $\pi \pi N$
discontinuities induced by the $t$-channel $\pi$ exchanges in the $\pi
\Delta \to \sigma N$, $\rho N$ transitions and by the $u$-channel $N$
exchange in the $\pi \Delta \to \pi \Delta$ transition, which are
depicted in Figs.~\ref{fig:cut}(c) and \ref{fig:cut}(d), respectively.
In this sense, I partly take into account the $\pi \pi N$ three-body
unitarity.  To satisfy the three-body $\pi \pi N$ unitarity fully, it
is necessary to include the so-called $Z$ diagrams, which corresponds
to Figs.~\ref{fig:cut}(c) and \ref{fig:cut}(d), in addition to the
usual two-body to two-body interaction terms~\cite{Matsuyama:2006rp}.

\subsection{\boldmath Bare $N^{\ast}$ and $\Delta ^{\ast}$ states}
\label{sec:3D}

Now I introduce bare $N^{\ast}$ and $\Delta ^{\ast}$ states, which are
embedded into the $\pi N$ coupled-channels interactions.  They are
described as $s$-channel interactions
\begin{align}
  V_{j k}^{\rm bare} ( E ; q^{\prime} , q )
  = & 
  \frac{g_{j} g_{k}}
       {2 m_{\pi} ( E - M_{0} )}
  \left ( \frac{q^{\prime}}{m_{\pi}} \right )^{L^{\prime}}
  \left ( \frac{q}{m_{\pi}} \right )^{L}
  \notag \\
  & \times
  \mathcal{F} ( \Lambda , q^{\prime} )
  \mathcal{F} ( \Lambda , q ) .
  \label{eq:Vbare}
\end{align}
These interactions are added to each partial-wave component of the
$\pi N$ coupled-channels interaction $V$ in Eq.~\eqref{eq:LSeq}.  Here
$M_{0}$ is the bare mass, $g_{j}$ is the coupling constant of the bare
state to the channel $j$, form factor $\mathcal{F}$ is the same as in
Eq.~\eqref{eq:FF}, and $L^{( \prime )}$ is the orbital angular
momentum of the initial (final) meson--baryon state.  The bare mass
$M_{0}$, coupling constant $g_{j}$, and cutoff $\Lambda$ are free
parameters and are fixed in the fits.  I use a single cutoff $\Lambda$
for each $N^{\ast}$ or $\Delta ^{\ast}$ state to reduce the number of
parameters.

Because the bare-state contributions are implemented into the two-body
coupled-channels interactions as in Eq.~\eqref{eq:Vbare}, they depend
on the energy $E$, which will bring nonzero missing-channel
contribution $Z$~\eqref{eq:missing} corresponding to the implicit bare
states~\cite{Sekihara:2016xnq}.

A problem is to specify the number of bare $N^{\ast}$ and $\Delta
^{\ast}$ states in each partial wave.  In the present study I take
into account the bare states only if the bare states can significantly
improve the fitting and reproduce well resonance-like behaviors of the
on-shell $\pi N$ amplitudes.  In this strategy I introduce the bare
$N^{\ast}$ and $\Delta ^{\ast}$ states listed in Table~\ref{tab:bare},
in which the fitted values of the parameters are shown as well.

\begin{table}[!b]
  \caption{Fitted values of coupling constants and cutoff parameters.
    Cutoff parameters are in units of MeV.}
  \label{tab:para}
  \begin{ruledtabular}
    \begin{tabular*}{8.6cm}{@{\extracolsep{\fill}}lrc|lr}
      $f_{\pi N \Delta}$ & $-0.439$ &
      & $\Lambda _{\pi N N}$ & $566$
      \\
      $g_{\pi \pi \rho}$ & $2.876$ &
      & $\Lambda _{\pi N \Delta}$ & $510$
      \\
      $g_{\rho N N}$ & $8.783$ &
      & $\Lambda _{\pi \pi \rho}$ & $1000$
      \\
      $\kappa _{\rho}$ & $4.806$ &
      & $\Lambda _{\rho N N}$ & $564$
      \\
      $g_{\pi \pi \sigma}$ & $3.188$ &
      & $\Lambda _{\pi \pi \sigma}$ & $1000$
      \\
      $g_{\sigma N N}$ & $21.571$ &
      & $\Lambda _{\sigma N N}$ & $564$ 
      \\
      & & &
      $\Lambda _{\eta N N}$ & $843$
    \end{tabular*}
  \end{ruledtabular}
\end{table}

\subsection{\boldmath Fit to the experimental $\pi N$ amplitudes}
\label{sec:3E}

\begin{figure*}[!p]
  \centering
  \Psfig{\textwidth}{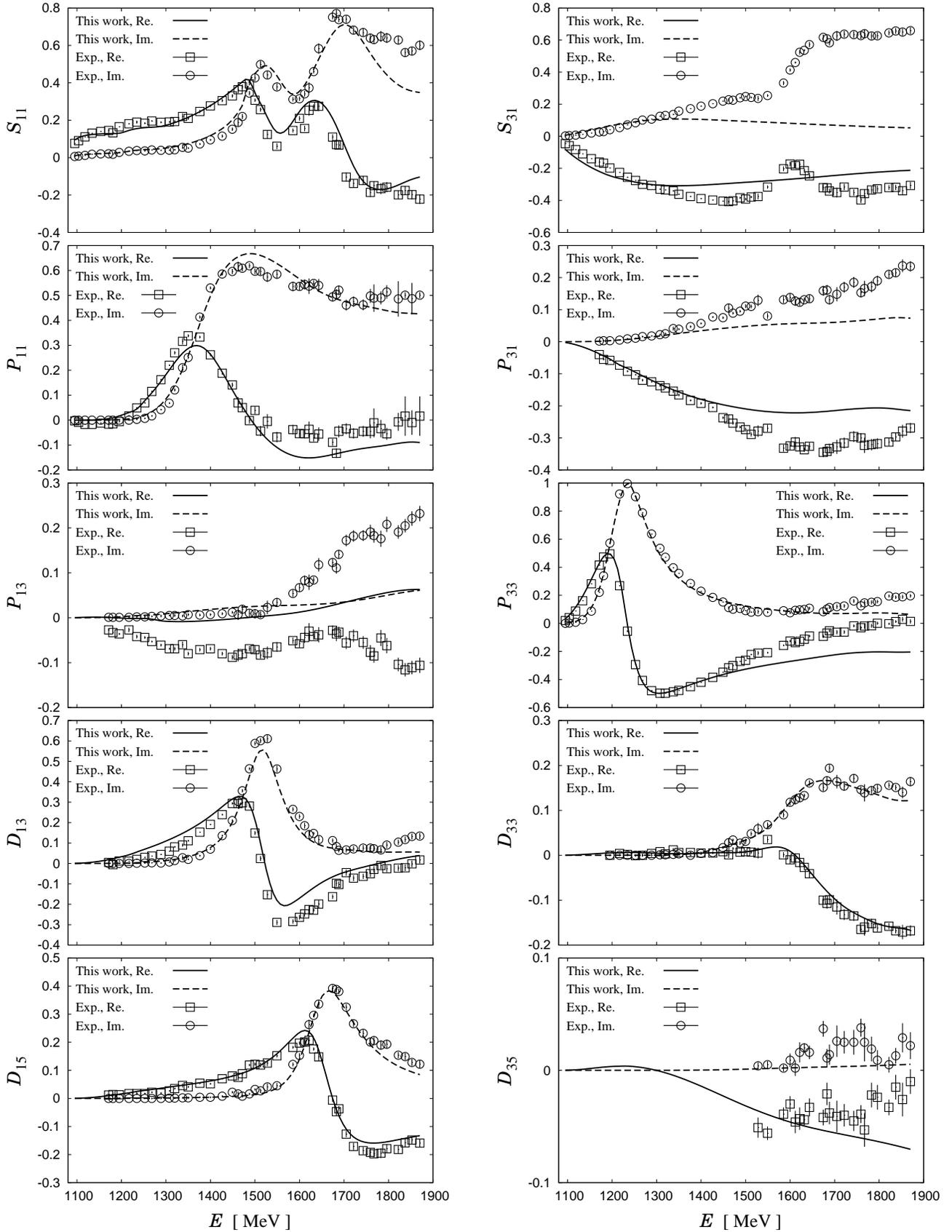}
  \caption{Fitted $\pi N$ scattering amplitudes $L_{2 I \, 2 J}$.
    Lines (points) represent results of the present calculation (SAID
    analysis~\cite{SAID}).}
  \label{fig:amp}
\end{figure*}

\begin{table*}[!t]
  \caption{Nucleon resonances obtained in this study.  I here show
    their pole positions in the present model $E_{\rm pole}$ and their
    locations in terms of the Riemann sheets, pole positions listed in
    Particle Data Group (PDG)~\cite{Zyla:2020zbs}, compositeness
    $X$~\eqref{eq:Xj_CSM}, missing-channel contributions
    $Z$~\eqref{eq:missing}, uncertainties $U$~\eqref{eq:XtildeU} and
    $U_{\rm r}$~\eqref{eq:Ur}, and real-valued quantities $\tilde{X}$
    and $\tilde{Z}$~\eqref{eq:XtildeZtilde}.  The notation of the pole
    locations is explained in the text.  The indices $1$, $2$, $3$
    ($1$, $2$) for the $\rho N$ ($\pi \Delta$) channel represent the
    same order as in Table~\ref{tab:exp_chan}.}
  \label{tab:pole}
  \begin{ruledtabular}
    \begin{tabular*}{\textwidth}{@{\extracolsep{\fill}}ccccc}
      & $N (1535)$ $1/2^{-}$
      & $N (1650)$ $1/2^{-}$
      & $N (1440)_{1}$ $1/2^{+}$
      & $N (1440)_{2}$ $1/2^{+}$
      \\
      \hline
      $E_{\rm pole}$ [MeV]
      & $1527 - 45 i$
      & $1699 - 70 i$
      & $1362 - 106 i$
      & $1361 - 114 i$
      \\
      Location
      & (22111)
      & (22112)
      & (21112)
      & (21111)
      \\
      $E_{\rm pole}$(PDG) [MeV]
      & $(1500\text{--}1520) - (55\text{--}75) i$
      & $(1640\text{--}1670) - (50\text{--}85) i$
      & $(1360\text{--}1380) - (80\text{--}95) i$
      & $(1360\text{--}1380) - (80\text{--}95) i$
      \\
      \hline
      $X_{\pi N}$
      & $-0.09 + 0.01 i \phantom{-}$
      & $-0.05 - 0.02 i \phantom{-}$
      & $0.47 + 0.35 i$
      & $0.55 + 0.28 i$
      \\
      $X_{\eta N}$
      & $-0.42 - 0.19 i \phantom{-}$
      & $-0.02 - 0.01 i \phantom{-}$
      & $-0.00 - 0.01 i \phantom{-}$
      & $-0.00 - 0.01 i \phantom{-}$
      \\
      $X_{\sigma N}$
      & $-0.00 + 0.03 i \phantom{-}$
      & $-0.00 + 0.09 i \phantom{-}$
      & $0.40 + 0.15 i$
      & $0.46 + 0.03 i$
      \\
      $X_{\rho N (1)}$
      & $0.37 - 0.09 i$
      & $0.21 - 0.60 i$
      & $-0.00 + 0.00 i \phantom{-}$
      & $-0.00 + 0.00 i \phantom{-}$
      \\
      $X_{\rho N (2)}$
      & $0.31 - 0.00 i$
      & $0.09 - 0.01 i$
      & $-0.00 - 0.03 i \phantom{-}$
      & $-0.01 - 0.03 i \phantom{-}$
      \\
      $X_{\pi \Delta (1)}$
      & $0.06 - 0.01 i$
      & $0.07 + 0.12 i$
      & $0.03 - 0.24 i$
      & $0.02 - 0.06 i$
      \\
      $Z$
      & $0.78 + 0.25 i$
      & $0.71 + 0.43 i$
      & $0.11 - 0.23 i$
      & $-0.01 - 0.22 i \phantom{-}$
      \\
      \hline
      $U$
      & $1.15$
      & $0.87$
      & $0.55$
      & $0.40$
      \\
      $U_{\rm r}$
      & $0.16$
      & $0.12$
      & $0.08$
      & $0.06$
      \\
      $\tilde{X}_{\pi N}$
      & $0.04$
      & $0.03$
      & $0.38$
      & $0.44$
      \\
      $\tilde{X}_{\eta N}$
      & $0.22$
      & $0.01$
      & $0.00$
      & $0.00$
      \\
      $\tilde{X}_{\sigma N}$
      & $0.01$
      & $0.05$
      & $0.28$
      & $0.33$
      \\
      $\tilde{X}_{\rho N (1)}$
      & $0.18$
      & $0.34$
      & $0.00$
      & $0.00$
      \\
      $\tilde{X}_{\rho N (2)}$
      & $0.14$
      & $0.05$
      & $0.02$
      & $0.02$
      \\
      $\tilde{X}_{\pi \Delta (1)}$
      & $0.03$
      & $0.08$
      & $0.15$
      & $0.05$
      \\
      $\tilde{Z}$
      & $0.38$
      & $0.44$
      & $0.17$
      & $0.16$
    \end{tabular*}
  \end{ruledtabular}

  ~

  \begin{ruledtabular}
    \begin{tabular*}{\textwidth}{@{\extracolsep{\fill}}ccccc}
      & $N (1520)$ $3/2^{-}$
      & $N (1675)$ $5/2^{-}$
      & $\Delta (1232)$ $3/2^{+}$
      & $\Delta (1700)$ $3/2^{-}$
      \\
      \hline
      $E_{\rm pole}$ [MeV]
      & $1506 - 53 i$
      & $1652 - 57 i$
      & $1216 - 54 i$
      & $1666 - 84 i$
      \\
      Location
      & (22112)
      & (22 - 12)
      & (2 - - 11)
      & (2 - - 12)
      \\
      $E_{\rm pole}$(PDG) [MeV]
      & $(1505\text{--}1515) - (52.5\text{--}60) i$
      & $(1655\text{--}1665) - (62.5\text{--}75) i$
      & $(1209\text{--}1211) - (49\text{--}51) i$
      & $(1640\text{--}1690) - (100\text{--}150) i$
      \\
      \hline
      $X_{\pi N}$
      & $0.05 + 0.16 i$
      & $-0.03 + 0.05 i \phantom{-}$
      & $-0.03 + 0.59 i \phantom{-}$
      & $-0.03 + 0.00 i \phantom{-}$
      \\
      $X_{\eta N}$
      & $0.01 - 0.00 i$
      & $0.06 + 0.24 i$
      & &
      \\
      $X_{\sigma N}$
      & $0.15 + 0.11 i$
      & 
      & &
      \\
      $X_{\rho N (1)}$
      & $0.01 - 0.06 i$
      & $0.01 - 0.00 i$
      & $0.00 - 0.00 i$
      & $0.48 - 0.09 i$
      \\
      $X_{\rho N (2)}$
      & $0.04 - 0.02 i$
      & $0.02 - 0.01 i$
      & $0.01 + 0.00 i$
      & $0.05 - 0.02 i$
      \\
      $X_{\rho N (3)}$
      & $0.08 - 0.04 i$
      & & & $0.09 - 0.02 i$
      \\
      $X_{\pi \Delta (1)}$
      & $- 0.00 + 0.00 i \phantom{-}$
      & $0.08 + 0.07 i$
      & $0.00 - 0.00 i$
      & $-0.08 - 0.23 i \phantom{-}$
      \\
      $X_{\pi \Delta (2)}$
      & $0.02 - 0.00 i$
      & & & $-0.01 + 0.04 i \phantom{-}$
      \\
      $Z$
      & $0.62 - 0.15 i$
      & $0.86 - 0.35 i$
      & $1.02 - 0.59 i$
      & $0.50 + 0.31 i$
      \\
      \hline
      $U$
      & $0.24$
      & $0.37$
      & $0.77$
      & $0.54$
      \\
      $U_{\rm r}$
      & $0.03$
      & $0.06$
      & $0.15$
      & $0.08$
      \\
      $\tilde{X}_{\pi N}$
      & $0.14$
      & $0.04$
      & $0.33$
      & $0.02$
      \\
      $\tilde{X}_{\eta N}$
      & $0.01$
      & $0.18$
      & &
      \\
      $\tilde{X}_{\sigma N}$
      & $0.15$
      & & &
      \\
      $\tilde{X}_{\rho N (1)}$
      & $0.05$ 
      & $0.01$
      & $0.00$
      & $0.32$
      \\
      $\tilde{X}_{\rho N (2)}$
      & $0.04$
      & $0.02$
      & $0.01$
      & $0.03$
      \\
      $\tilde{X}_{\rho N (3)}$
      & $0.08$
      & & & $0.06$
      \\
      $\tilde{X}_{\pi \Delta (1)}$
      & $0.00$
      & $0.08$
      & $0.00$
      & $0.16$
      \\
      $\tilde{X}_{\pi \Delta (2)}$
      & $0.02$
      & & & $0.03$
      \\
      $\tilde{Z}$
      & $0.51$
      & $0.67$
      & $0.66$
      & $0.39$
    \end{tabular*}
  \end{ruledtabular}  
\end{table*}

The present model for the $\pi N$ coupled-channels amplitudes contains
the model parameters: coupling constants, cutoffs, and bare-state
masses.  They can be determined by fitting the experimental data of
the on-shell $\pi N$ scattering amplitudes.  In the present study,
while I fix the cutoffs for the meson--meson--meson vertices as
$\Lambda _{\pi \pi \rho} = \Lambda _{\pi \pi \sigma} =
\SI{1000}{MeV}$, I allow the other model parameters to vary freely so
as to reproduce the on-shell $\pi N$ scattering amplitudes.  The fixed
value $\Lambda _{\pi \pi \rho} = \Lambda _{\pi \pi \sigma} =
\SI{1000}{MeV}$ is a typical value of the hadronic scale, and I
checked that the on-shell $\pi N$ scattering amplitudes only weakly
depend on the meson--meson--meson cutoffs $\Lambda _{\pi \pi \rho}$,
$\Lambda _{\pi \pi \sigma}$.

For the on-shell $\pi N$ scattering amplitudes, I employ the database
of SAID~\cite{SAID}.  Restricting the center-of-mass energy $E \le
\SI{1.9}{GeV}$ and orbital angular momentum $L \le 2$, I obtain the
model parameters listed in Tables~\ref{tab:bare} and \ref{tab:para} by
the fitting.

The results of the on-shell $\pi N$ amplitudes are shown in
Fig.~\ref{fig:amp}.  As one can see, the present model reproduces the
on-shell $\pi N$ amplitudes fairly well except for the $S_{31}$,
$P_{31}$, and $P_{13}$ amplitudes.  In particular, the resonance-like
behavior in $S_{11}$, $P_{11}$, $P_{33}$, $D_{13}$, $D_{33}$, and
$D_{15}$ is well reproduced.  Indeed, fittings to these amplitudes are
significantly improved by introducing two bare $N^{\ast}$ states in
the $S_{11}$ and one bare $N^{\ast}$ or $\Delta ^{\ast}$ state in the
$P_{11}$, $P_{33}$, $D_{13}$, $D_{33}$, and $D_{15}$, respectively.

On the other hand, I can quantitatively reproduce the $S_{31}$ and
$P_{31}$ amplitudes only in the low-energy region $E \lesssim
\SI{1250}{MeV}$.  For the $P_{13}$ amplitude, I can only reproduce the
smallness of the absolute value of the amplitude ($\lesssim 0.1$).  I
expect that one could cure these discrepancies in the $S_{31}$,
$P_{31}$, and $P_{13}$ amplitudes by employing further diagrams of the
meson--baryon interactions or phenomenological contact potentials as
done in Ref.~\cite{Kamano:2013iva} for the $S_{31}$ partial wave, and
by taking into account other meson--baryon channels such as $K
\Lambda$ and $K \Sigma$.  I note that, although the experimental data
imply resonances around $\SI{1600}{MeV}$ and $\SI{1700}{MeV}$ in the
$S_{31}$ and $P_{13}$, respectively, I do not include bare states in
these partial waves because bare states will not significantly improve
the fittings.

\begin{figure*}[!t]
  \centering
  \Psfig{8.6cm}{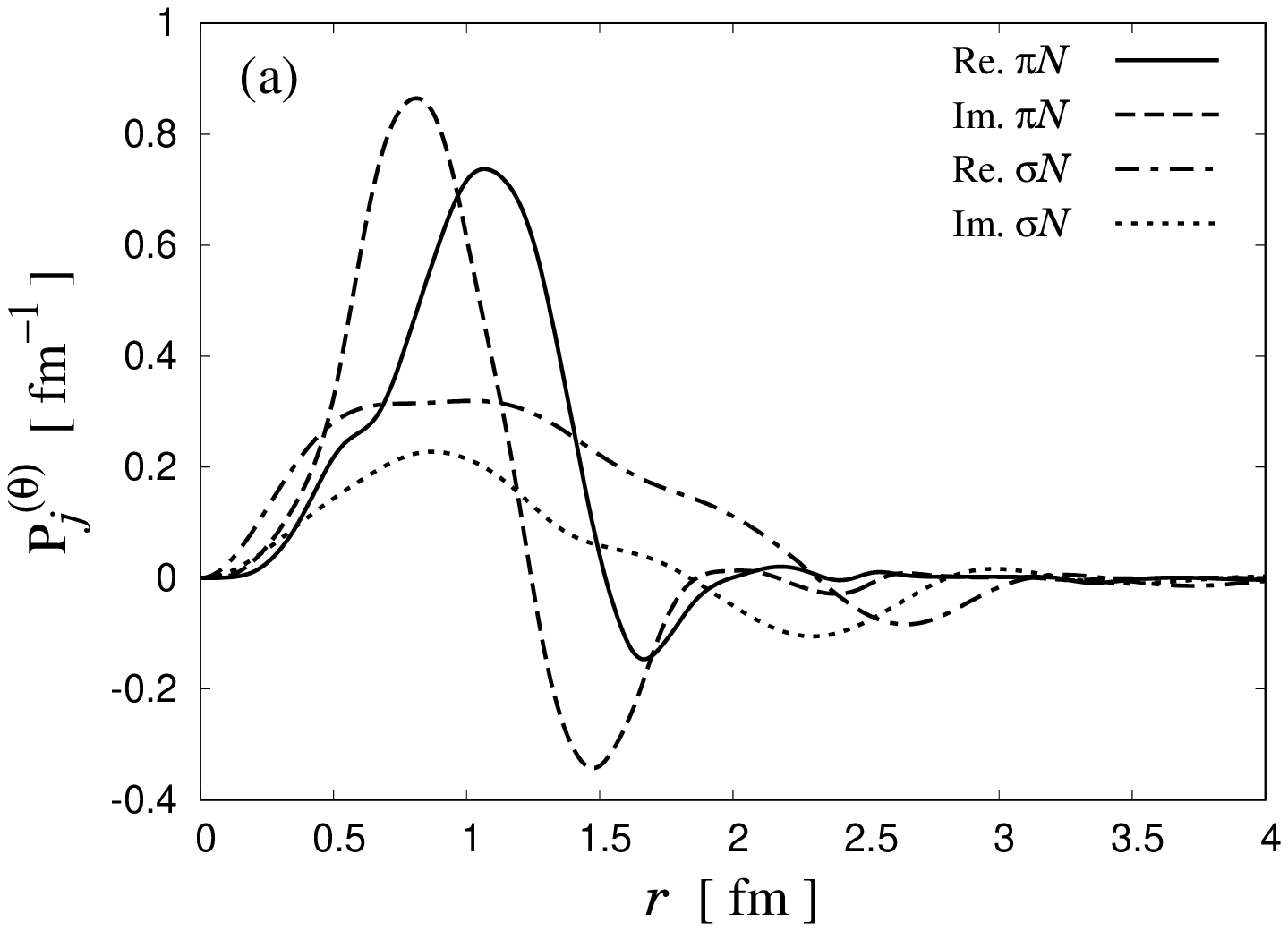} ~ \Psfig{8.6cm}{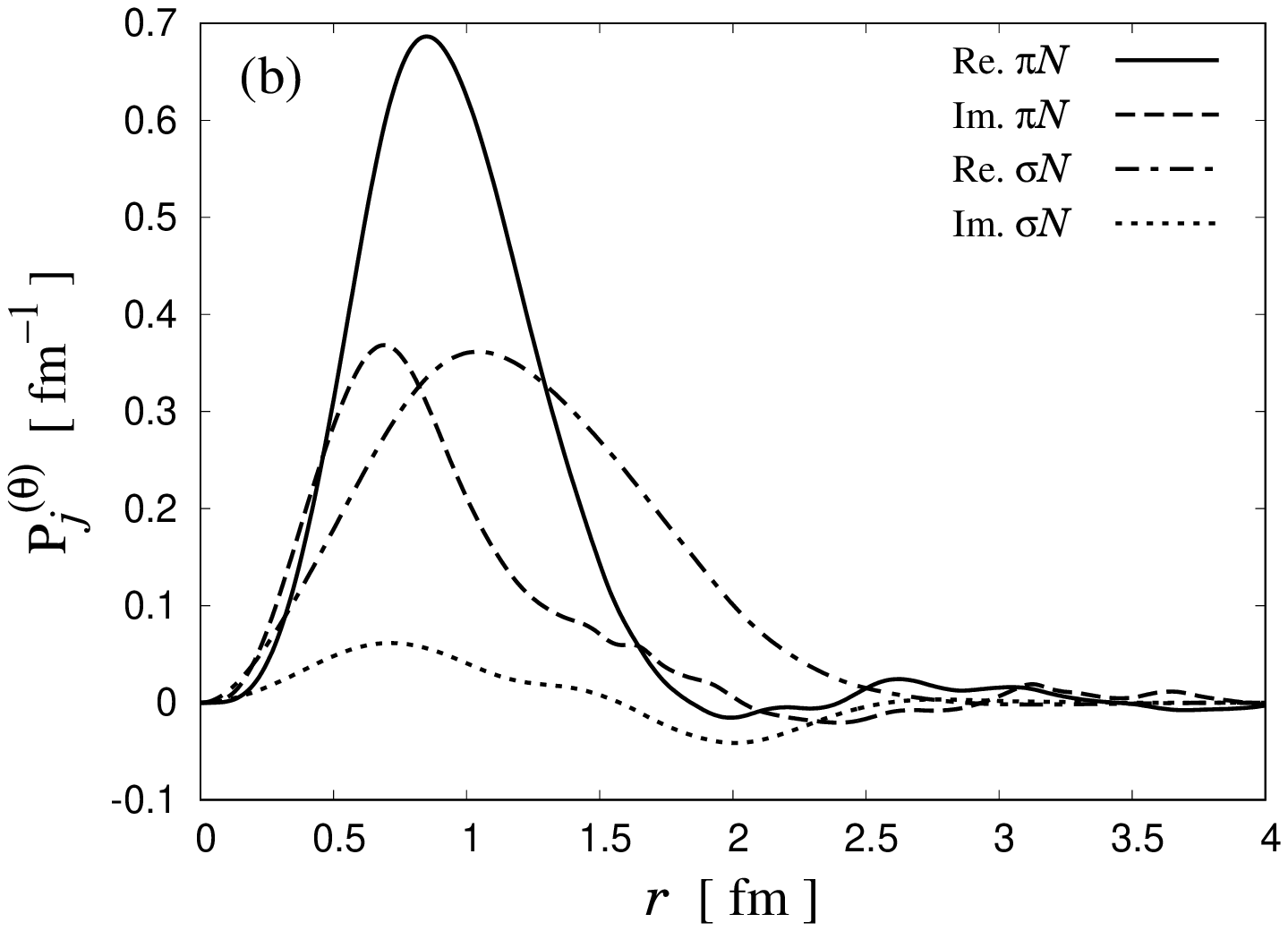}
  \caption{Density distributions of the $\pi N$ and $\sigma N$
    components for the $N (1440)$ resonance in coordinate space.  (a)
    For the $N (1440)_{1}$ pole with the complex-scaling angle $\theta
    = 50^{\circ}$.  (b) For the $N (1440)_{2}$ pole with the
    complex-scaling angle $\theta = 35^{\circ}$.}
  \label{fig:Roper}
\end{figure*}

\section{\boldmath The meson--baryon compositeness for the $N^{\ast}$
and $\Delta ^{\ast}$ resonances}
\label{sec:4}

In this section I calculate the compositeness of the $\pi N$, $\eta
N$, $\sigma N$, $\rho N$, and $\pi \Delta$ channels for the $N^{\ast}$
and $\Delta ^{\ast}$ resonances.

One of the most interesting features in hadron physics is the
competition between hadron degrees of freedom and quark degrees of
freedom.  In the present $N^{\ast}$/$\Delta ^{\ast}$ case, bare states
which are expected to originate from quark degrees of freedom are
embedded into the $\pi N$ coupled channels.  As a consequence, even if
a meson--baryon interaction is strongly attractive enough to make a
bound state, the bound state is in general contaminated by bare
$N^{\ast}$/$\Delta ^{\ast}$ states which couple to the meson--baryon
system.  Conversely, it is inevitable that a bare $N^{\ast}$/$\Delta
^{\ast}$ state is dressed in meson--baryon clouds.  The compositeness
is applicable to evaluating both the dominance of the meson--baryon
molecular components and the fractions of the meson--baryon clouds for
physical nucleon resonances.

In the present study, I employ the interaction diagrams in
Fig.~\ref{fig:diagrams} and several bare $N^{\ast}$ and $\Delta
^{\ast}$ states for the $\pi N$ coupled-channels scattering, and fix
the model parameters so as to reproduce the on-shell $\pi N$
amplitude, as explained in the previous section.  I perform the
analytic continuations of the scattering amplitudes to the complex
energy plane in the complex scaling method, and find resonance poles
corresponding to the $N (1535)$ and $N (1650)$ in the spin/parity
$J^{P} = 1/2^{-}$, $N (1440)$ in $1/2^{+}$, $N (1520)$ in $3/2^{-}$,
$N (1675)$ in $5/2^{-}$, $\Delta (1232)$ in $3/2^{+}$, and $\Delta
(1700)$ in $3/2^{-}$, where the names are taken from Particle Data
Group (PDG)~\cite{Zyla:2020zbs}.  Their pole positions are listed in
Table~\ref{tab:pole} together with the results of the compositeness
$X$~\eqref{eq:Xj_CSM}, missing-channel contributions
$Z$~\eqref{eq:missing}, uncertainties $U$~\eqref{eq:XtildeU} and
$U_{\rm r}$~\eqref{eq:Ur}, and real-valued quantities $\tilde{X}$ and
$\tilde{Z}$~\eqref{eq:XtildeZtilde}.

Below I discuss the internal structure of the $N^{\ast}$ and $\Delta
^{\ast}$ resonances on their resonance poles.

\subsection{\boldmath $N (1440)$}

The $N (1440)$ resonance in $J^{P} = 1/2^{+}$, also known as the Roper
resonance, is one of the most interesting states among the nucleon
resonances.  The Roper resonance is lighter than the lowest
negative-parity nucleon excitations, i.e., $N (1535)$ in $J^{P} =
1/2^{-}$ and $N (1520)$ in $J^{P} = 3/2^{-}$, which cannot be easily
explained if one assumes that the Roper resonance is a radial
excitation of nucleon as a three-quark system.  A promising physical
interpretation is that the Roper resonance is the first radial
excitation of nucleon but consists of a dressed-quark core augmented
by a meson cloud~\cite{Burkert:2017djo}.  Importance of the
contribution from the $\pi N$ coupled-channels dynamics was also
pointed out in, e.g., Refs.~\cite{Krehl:1999km, Lang:2016hnn,
  Liu:2016uzk, Wu:2017qve, Suzuki:2009nj, Golli:2017nid}.  In this
sense, the use of the two-body wave functions and compositeness in my
approach is quite suitable for studying the internal structure of the
Roper resonance.

In the present $\pi N$ coupled-channels model, I find two poles of the
scattering amplitude corresponding to the Roper resonance at $E_{\rm
  pole} = 1362 - 106 i \, \si{MeV}$ and $1361 - 114 i \, \si{MeV}$.
The pole positions deviate only slightly from the value reported by
PDG.  The former pole is found with the complex-scaling angle $\theta
\ge 45^{\circ}$ in the complex scaling method.  This indicates that
the former pole exists in the second Riemann sheets of the $\pi N$ and
$\pi \Delta$ channels but in the first Riemann sheets of the $\eta N$,
$\sigma N$, and $\rho N$ channels (see Fig.~\ref{fig:branch}), to
which I refer as $(2 1 1 1 2)$ in the order $\pi N$, $\eta N$, $\sigma
N$, $\rho N$, and $\pi \Delta$.  On the other hand, the latter pole is
found with the scaling angle $25^{\circ} \le \theta \le 40^{\circ}$,
and hence it exists in the sheet $(2 1 1 1 1)$.  Because the former
pole is closer to the physical region, i.e., the real energy axis, the
former is the resonance pole for the Roper resonance and write it as
$N (1440)_{1}$.  On the other hand, the latter pole is the shadow pole
for the Roper resonance~\cite{Suzuki:2009nj} and write it as $N
(1440)_{2}$.  I note that the pole positions $E_{\rm pole}$ do not
depend on the scaling angle $\theta$ but one can switch the Riemann
sheets at a certain energy by varying $\theta$.

From the residues of the off-shell scattering amplitudes at the poles,
I can calculate the two-body wave functions and compositeness as their
norms both for the $N (1440)_{1}$ and $N (1440)_{2}$, according to the
method in Sec.~\ref{sec:2}.  The results of the compositeness
$X_{j}$~\eqref{eq:Xj_CSM} of the $j$th meson--baryon channel and
missing-channel contribution $Z$~\eqref{eq:missing} are listed in
Table~\ref{tab:pole}.  I checked that the compositeness does not
depend on the scaling angle $\theta$.  As one can see from the Table,
although the values of the compositeness are complex due to the
resonance nature, the real parts of $X_{\pi N}$ and $X_{\sigma N}$ for
the both poles are as large as $0.5$, which should be compared with
unity, and their imaginary parts are smaller than the real parts.  On
the other hand, the missing-channel contribution $Z$ is close to zero.
Because the nonzero value of $Z$ comes from the bare state in the
meson--baryon interaction~\eqref{eq:Vbare} as well as the $\pi \pi N$
state in the self-energies of
Eqs.~\eqref{eq:EsigmaN}--\eqref{eq:EpiDelta}, the present result
strongly implies that, while the $\pi N$ and $\sigma N$ molecular
components dominate both the $N (1440)_{1}$ and $N (1440)_{2}$ states
as ``thick meson clouds'', the bare-state contribution is small.

The complex-valued compositeness, however, cannot be interpreted as
probabilities of finding meson--baryon components.  To draw a more
definite conclusion, I calculate the real-valued quantities
$\tilde{X}$ and $\tilde{Z}$~\eqref{eq:XtildeZtilde} together with
$U$~\eqref{eq:XtildeU} and $U_{\rm r}$~\eqref{eq:Ur} for the $N
(1440)_{1}$ and $N (1440)_{2}$ states.  The results are listed in
Table~\ref{tab:pole}.  The value of the reduced uncertainty $U_{\rm
  r}$ is smaller than $0.1$ for both poles.  Therefore, according to
the discussions in the end of Sec.~\ref{sec:2A}, I can interpret
$\tilde{X}_{j}$ ($\tilde{Z}$) as the probability of finding the
composite (missing) part with small uncertainties.  From the values in
Table~\ref{tab:pole}, I can conclude more definitely that the $\pi N$
and $\sigma N$ molecular components, whose contributions are almost
the same as each other, dominate the $N (1440)_{1}$ and $N (1440)_{2}$
states while the bare-state contribution is about less than $20 \%$ in
the present model.  These finding are consistent with the previous
studies in, e.g., Refs.~\cite{Krehl:1999km, Lang:2016hnn, Liu:2016uzk,
  Wu:2017qve, Suzuki:2009nj, Golli:2017nid, Burkert:2017djo}, in which
a significant contribution from meson--baryon coupled channels was
reported.  In particular, the present study supports the scenario
drawn in Ref.~\cite{Krehl:1999km} that the $\pi N \to \sigma N$
transition potential and treatment of the $\pi \pi N$ components in
the unstable $\sigma N$ channel are important for the description of
the Roper resonance.

The squared wave functions in coordinate space~\eqref{eq:Rho_CSM}
represent the behavior of relative motions between the mesons and
baryons as the density distributions.  The density distributions $\Rho
_{j}^{( \theta )}$ for the $N (1440)_{1}$ and $N (1440)_{2}$ states
are plotted in Fig.~\ref{fig:Roper} as functions of the relative
distance $r$ between the mesons and baryons.  The scaling angle is
fixed as $\theta = 50^{\circ}$ for the $N (1440)_{1}$ and $\theta =
35^{\circ}$ for the $N (1440)_{2}$.  Although the density
distributions are complex for resonances in general and depend on
$\theta$, they provide information on the typical distance between the
mesons and baryons.  The results of the density distributions imply
that both in the $\pi N$ and $\sigma N$ channels the meson--baryon
separation is about more than $\SI{1}{fm}$ for the $N (1440)_{1}$ and
$N (1440)_{2}$ states.

\subsection{\boldmath $\Delta (1232)$}

The $\Delta (1232)$ resonance in $J^{P} = 3/2^{+}$ is also
interesting, because there are several suggestions of its large $\pi N$
component.  Historically, it was pointed out in
Ref.~\cite{Chew:1955zz} that the $\Delta (1232)$ resonance can occur
by the attractive $p$-wave $\pi N$ interaction.  A hint of the large
effect of the meson cloud is seen, for instance, in the $M1$
transition form factor for $\gamma ^{\ast} N \to \Delta$ at $Q^{2} =
0$~\cite{Sato:2009de}.  Further studies on the dynamical generation of
the $\Delta (1232)$ resonance can be seen, e.g., in
Refs.~\cite{Aceti:2014ala, Sekihara:2015gvw, Golli:2019vxw}.  I can
examine the picture of a large $\pi N$ component in terms of the
compositeness.

In the present model, I observe the resonance pole for the $\Delta
(1232)$ at $1216 - 54 i \, \si{MeV}$ in the Riemann sheet
$(2\text{~-~-~}11)$, where hyphen represents a decoupled channel.  The
compositeness calculated from the residue and pole position for the
$\Delta (1232)$ is listed in Table~\ref{tab:pole}.  The $\pi N$
compositeness $X_{\pi N}$ has nonnegligible imaginary part but its
real part is small.  The other meson--baryon channels, i.e., the $\rho
N ( L = 1, S = 1/2 )$, $\rho N ( L = 1, S = 3/2 )$, and $\pi \Delta$
channels give negligible contributions to the compositeness, and hence
the missing-channel contribution $Z$ is almost unity in the real part
and negatively large in the imaginary part.  Because the bare $\Delta
^{\ast}$ state exists near the physical pole position, I can expect
that $Z$ is dominated by the bare state.  Therefore, the results imply
that the bare-state contribution is the most essential for the
physical $\Delta (1232)$ resonance.  Nevertheless, I expect that the
large absolute value $| X_{\pi N} |$ reflects the importance of the
$\pi N$ channel in the $\Delta (1232)$ and affects the properties of
the $\Delta (1232)$ as the meson clouds.

Besides, the reduced uncertainty $U_{\rm r}$ is as large as $0.15$
owing to the large imaginary part in the $\pi N$ channel.  Therefore,
although $\tilde{X}_{\pi N}$ takes a nonnegligible value $\sim 0.3$, I
cannot definitely interpret it as the probability of finding $\pi N$
component inside the $\Delta (1232)$ resonance.

\subsection{\boldmath $N(1535)$, $N(1650)$,
  $N(1520)$, $N(1675)$, and $\Delta (1700)$}

Next I consider the other resonances: $N(1535)$ and $N(1650)$ in
$J^{P} = 1/2^{-}$, $N(1520)$ in $3/2^{-}$, $N(1675)$ in $5/2^{-}$, and
$\Delta (1700)$ in $3/2^{-}$.  The results of their pole positions and
compositeness are listed in Table~\ref{tab:pole}.

The $N(1535)$ and $N(1650)$ resonances exist in the Riemann sheets
$(22111)$ and $(22112)$, respectively.  The results of the
compositeness for the $N (1535)$ imply that the bare-state
contribution is dominant but the coupling to the $\eta N$ channel,
whose branch point is the closest to the $N (1535)$ pole position,
would be large.  However, the real part of the $\eta N$ compositeness
$X_{\eta N}$ is negatively large, which is canceled with the real
parts of the $\rho N$ compositeness.  As a consequence, the reduced
uncertainty $U_{\rm r}$ is as large as $\tilde{X}_{\eta N, \rho N (1),
  \rho N (2)}$ for the $N (1535)$ and I cannot interpret
$\tilde{X}_{\eta N, \rho N (1), \rho N (2)}$ as the probabilities of
finding the $\eta N$ and $\rho N$ components, respectively.

On the other hand, for the $N (1650)$, because $U_{\rm r} = 0.12$, I
can interpret $\tilde{X}_{\rho N (1)}$ and $\tilde{Z}$ as the
probabilities with uncertainties $\sim 0.1$.  The results of
$\tilde{X}_{j}$ and $\tilde{Z}$ indicate that about half of the $N
(1650)$ comes from missing channels, in the present case the bare
state, while it has a certain fraction of the $\rho N ( L = 0 , S =
1/2 )$ cloud.

Similarly, the poles of the $N (1520)$, $N (1675)$, and $\Delta
(1700)$ resonances are found in the $(22112)$, $(22\text{~-~}12)$, and
$(2\text{~-~-~}12)$ Riemann sheets, respectively.  The results of the
compositeness indicate that they are dominated by missing channels,
i.e., the bare states.  The $\Delta (1700)$ resonance has a certain
fraction of the $\rho N ( L = 0, S = 3/2 )$ cloud, while the $N
(1520)$ and $N (1675)$ resonances have only small fractions of
meson--baryon clouds.

These results indicate that, although the $N(1535)$, $N(1650)$,
$N(1520)$, $N(1675)$, and $\Delta (1700)$ resonances have some
fractions of the meson--baryon clouds, they do not have dominant
meson--baryon molecular components.  The largest fractions of the
meson--baryon clouds are the $\rho N$ for the $N (1650)$ and $\Delta
(1700)$ resonances, which amount to $\sim 0.3$ with uncertainties
$\sim 0.1$.  These are because the resonance pole positions are close
to the $\rho N$ branch point and coupling constants of the
$N^{\ast}$/$\Delta ^{\ast}$ bare states to the $\rho N$ channel are
large, as seen in Table~\ref{tab:bare}.

\section{Conclusion}
\label{sec:5}

In this study I have investigated the internal structure of the
nucleon resonances $N^{\ast}$ and $\Delta ^{\ast}$ in terms of the
meson--baryon two-body wave functions and compositeness.  One of the
most essential parts in my approach is to extract the meson--baryon
two-body wave functions by using the pole positions for the nucleon
resonances in the complex energy plane of the $\pi N$ coupled-channels
scattering amplitudes and residues for them.  Here the scattering
amplitudes are solutions of the Lippmann--Schwinger equation.  In this
strategy, for each resonance pole I can obtain wave functions of the
$\pi N$ and coupled channels which are automatically scaled, thanks to
the inhomogeneous property of the Lippmann--Schwinger equation.  As a
consequence, by calculating the compositeness, which is defined as the
norm of the two-body wave function from the meson--baryon scattering
amplitudes, and by comparing the compositeness with unity, I can
evaluate the dominance of the meson--baryon molecular components as
well as the fractions of meson--baryon clouds for physical nucleon
resonances.

For this purpose I have constructed a meson exchange model in a $\pi
N$-$\eta N$-$\sigma N$-$\rho N$-$\pi \Delta$ coupled-channels problem
and involve several bare $N^{\ast}$ and $\Delta ^{\ast}$ states.  The
coupling constants, cutoffs, and bare-state masses as the model
parameters were fixed so as to reproduce the experimental data of the
on-shell $\pi N$ scattering amplitudes in the center-of-mass energy $E
\le \SI{1.9}{GeV}$ and orbital angular momentum $L \le 2$.  The
constructed model reproduced the on-shell $\pi N$ amplitudes fairly
well and generated resonance poles corresponding to the $N (1535)$ and
$N (1650)$ in the spin/parity $J^{P} = 1/2^{-}$, $N (1440)$ in
$1/2^{+}$, $N (1520)$ in $3/2^{-}$, $N (1675)$ in $5/2^{-}$, $\Delta
(1232)$ in $3/2^{+}$, and $\Delta (1700)$ in $3/2^{-}$ in the complex
energy plane of the scattering amplitudes.  In the present model the
Roper resonance $N (1440)$ is composed of two poles in different $\pi
\Delta$ sheets.

Then I have calculated the meson--baryon wave functions and
compositeness from the scattering amplitudes for these nucleon
resonances.  As a result, the Roper resonance $N (1440)$, for both the
two poles, was found to be dominated by the $\pi N$ and $\sigma N$
molecular components, whose contributions are almost the same as each
other, while the bare-state contribution is about less than $20 \%$ in
the present model.  The squared wave functions in coordinate space
imply that both in the $\pi N$ and $\sigma N$ channels the separation
between the meson and baryon is about more than $\SI{1}{fm}$ for the
$N (1440)$ resonance.  On the other hand, dominant meson--baryon
molecular components were not observed in any other $N^{\ast}$ and
$\Delta ^{\ast}$ resonances in the present model, although they have
some fractions of the meson--baryon clouds.

Here I emphasize that the present strategy to calculate the
compositeness is in general applicable as long as the
Lippmann--Schwinger equation is fully solved for hadron--hadron
scatterings.  In this sense, more definitely conclusion about the
composite nature of nucleon resonances will be drawn with more
sophisticated models such as Refs.~\cite{Ronchen:2012eg,
  Kamano:2013iva, Ronchen:2018ury}, in which they precisely reproduced
experimental data of not only the on-shell $\pi N$ amplitudes but also
the pion- and photon-induced reactions.  Furthermore, excited baryons
with non-zero strangeness, i.e., $\Lambda ^{\ast}$, $\Sigma ^{\ast}$,
$\Xi ^{\ast}$, and $\Omega ^{\ast}$ states, will be important as a
next target, because they will be discovered and be investigated
extensively in near future experiments at J-PARC, JLab, etc., as well
as in the relativistic heavy ion collisions.  To construct scattering
amplitudes for these resonances, approaches in, e.g.,
Refs.~\cite{Kamano:2014zba, Kamano:2015hxa} will be helpful.

Finally I comment on the model dependence of the two-body wave
functions and compositeness.  Because the compositeness as well as the
wave functions is not observable, the compositeness is in general a
model dependent quantity.  This fact has a special meaning when one
discusses a hadron--hadron molecular component in a hadron resonance.
The strong interactions between hadrons emerge as a nonperturbative
phenomenon of the underlying theory, QCD, and hence one cannot obtain
the strong interactions between hadrons by analytically solving QCD.
This is in contrast to the electromagnetic case, in which the
electromagnetic interactions can be directly obtained by the
fundamental theory, quantum electrodynamics (QED).  Therefore, to pin
down the hadron--hadron interaction and calculate its off-shell part,
which plays an essential role in the two-body wave functions and
compositeness, we have to fix a scheme based on a certain principle
such as meson exchanges and employ effective Lagrangians which govern
the hadron--hadron interaction.  The present article indeed suggests a
strategy in this line to elucidate the internal structure of hadron
resonances in terms of the hadron--hadron molecular components.

\begin{acknowledgments}
  The author acknowledges H.~Kamano, S.~X.~Nakamura, and T.~Sato for
  helpful discussions on the $\pi N$ partial-wave analysis and on the
  $N^{\ast}$ and $\Delta ^{\ast}$ physics.  He is also very grateful
  to D.~Jido and T.~Hyodo for fruitful discussions on the
  compositeness.
  This work is partly supported by JSPS KAKENHI Grant No.~JP15K17649.
\end{acknowledgments}

\appendix

\section{Masses of hadrons}
\label{app:mass}

In this study I employ isospin symmetric masses for hadrons: $m_{\pi}
= \SI{138.0}{MeV}$, $m_{\eta} = \SI{547.9}{MeV}$, and $m_{N} =
\SI{938.9}{MeV}$.  These values are used in the $t$-channel pion
propagators and in the $s$- and $u$-channel nucleon propagators as
well as in the initial and final states.  The masses in the unstable
initial and final states are taken from the bare states in the
self-energies: $m_{\sigma _{0}} = \SI{700}{MeV}$, $m_{\rho _{0}} =
\SI{812}{MeV}$, and $m_{\Delta _{0}} = \SI{1280}{MeV}$.  The value
$m_{\sigma _{0}} = \SI{700}{MeV}$ is used for the $t$-channel $\sigma$
exchange in the $\pi N \to \pi N$ interaction, while physical values
$m_{\rho} = \SI{775.3}{MeV}$ and $m_{\Delta} = \SI{1210}{MeV}$ are
used for the $t$-channel $\rho$ exchange and $u$-channel $\Delta$
exchange in the $\pi N \to \pi N$ interaction, respectively.

\section{Partial waves of the meson--baryon interactions}
\label{app:int}

In this Appendix I summarize the notations and partial-wave
projections of the meson--baryon ($M B$) interactions.

\subsection{Notations}

The meson--baryon scatterings are denoted by $M ( k^{\mu} , \lambda
_{M} ) B ( p^{\mu} , \lambda _{B} ) \to M^{\prime} ( k^{\prime \mu}
, \lambda _{M^{\prime}}) B^{\prime} ( p^{\prime \mu} , \lambda
_{B^{\prime}} )$, where $k^{( \prime ) \mu}$ and $p^{( \prime ) \mu}$
are momenta and $\lambda _{M, B}$ are helicities of the meson and
baryon, respectively.  Since I consider scatterings in the
center-of-mass frame, I can write the three-momenta as $\bm{q} \equiv
\bm{k} = - \bm{p}$ and $\bm{q}^{\prime} \equiv \bm{k}^{\prime} = -
\bm{p}^{\prime}$.  Without loss of generality, I can choose the
coordinates such that
\begin{equation}
  \bm{q} = ( 0 , 0 , q ) ,
  \quad 
  \bm{q}^{\prime} = ( q^{\prime} \sin \theta , 0 ,
  q^{\prime} \cos \theta ) ,
\end{equation}
with the scattering angle $\theta$.  The masses of the meson $M^{(
  \prime )}$ and baryon $B^{( \prime )}$ are expressed as $m_{M^{(
    \prime )}}$ and $m_{B^{( \prime )}}$, respectively.  Throughout
this study I fix the energies of the four-momenta $k^{\mu}$ and
$p^{\mu}$ to their on-shell values as
\begin{equation}
  k^{0} = \sqrt{m_{M}^{2} + q^{2}} ,
  \quad
  p^{0} = \sqrt{m_{B}^{2} + q^{2}} ,
  \label{eqA:k0p0}
\end{equation}
and similarly for $k^{\prime 0}$ and $p^{\prime 0}$.

\subsection{Partial-wave projections}

I calculate the partial-wave matrix elements of the interaction
$V_{\alpha}$ by following the Jacob--Wick
formulation~\cite{Jacob:1959at}, where $\alpha$ specifies the quantum
numbers of the system.  In the $\pi N$ coupled-channels scattering
case, I take $\alpha = ( J^{P} , I )$ with the total angular momentum
$J$, parity $P$, and isospin $I$.

First, according to Feynman diagrams, I calculate the interactions in
terms of the helicity eigenstates
\begin{equation}
  V_{M B \to M^{\prime} B^{\prime}}
  = V_{M B \to M^{\prime} B^{\prime}} ( \bm{q}^{\prime} , \lambda
  _{M^{\prime}} , \lambda _{B^{\prime}} , \bm{q} , \lambda _{M}
  , \lambda _{B}) .
  \label{eqA:Vhelicity}
\end{equation}

Then, the interactions are projected to the total angular
momentum $J$ as
\begin{align}
  & V^{J} ( q^{\prime} ,  \lambda _{M^{\prime}} , \lambda
  _{B^{\prime}} , q , \lambda _{M} , \lambda _{B} )
  \notag \\
  & = 2 \pi 
  \int _{-1}^{1} d \cos \theta \,
  d^{J}_{\lambda _{M} - \lambda _{B} \, \lambda _{M^{\prime}} - \lambda _{B^{\prime}}}
  ( \theta )
  \notag \\
  & \phantom{=}
  \times V ( \bm{q}^{\prime} , \lambda _{M^{\prime}} ,
  \lambda _{B^{\prime}} , \bm{q} , \lambda _{M} , \lambda _{B} ) ,
\end{align}
where I omitted the subscript $M B \to M^{\prime} B^{\prime}$ of $V$,
and $d_{m^{\prime} \, m}^{j}$ is the Wigner $d$-matrix.

The interaction of the total angular momentum $J$ is projected to the
states with definite orbital angular momenta and spins for the
meson--baryon channels as
\begin{align}
  V_{\alpha} ( q^{\prime} , q )
  & = \kappa ( q^{\prime} , q ) \sum _{\lambda _{M} , \lambda _{B} ,
    \lambda _{M^{\prime}} , \lambda _{B^{\prime}}}
  \frac{\sqrt{( 2 L + 1 )( 2 L^{\prime} + 1 )}}{2 J + 1}
  \notag \\
  & \quad \times
  \langle j_{M^{\prime}} \, j_{B^{\prime}} \, \lambda _{M^{\prime}} \,
  - \lambda _{B^{\prime}} | S^{\prime} \, S_{z}^{\prime} \rangle
  \langle L^{\prime} \, S^{\prime} \, 0 \, S_{z}^{\prime} | J \, S_{z}^{\prime} \rangle
  \notag \\
  & \quad \times
  \langle j_{M} \, j_{B} \, \lambda _{M} \, - \lambda _{B} | S S_{z} \rangle
  \langle L \, S \, 0 \, S_{z} | J \, S_{z} \rangle
  \notag \\
  & \quad \times
  V^{J} ( q^{\prime} , \lambda _{M^{\prime}} , \lambda _{B^{\prime}} ,
  q , \lambda _{M} , \lambda _{B}) ,
  \label{eqA:Vfinal}
\end{align}
where $j_{M^{( \prime )}}$ and $j_{B^{( \prime )}}$ are the spins of
the meson $M^{( \prime )}$ and baryon $B^{( \prime )}$, respectively,
$L^{( \prime )}$ and $S^{( \prime )}$ are the orbital angular momentum
and spin in the initial (final) state, respectively, $S_{z} \equiv
\lambda _{M} - \lambda _{B}$, and $S_{z}^{\prime} \equiv \lambda
_{M^{\prime}} - \lambda _{B^{\prime}}$.  $\langle j_{M} \, j_{B} \,
\lambda _{M} \, - \lambda _{B} | S \, S_{z} \rangle$ is the
Clebsch--Gordan coefficient.  The factor $\kappa ( q^{\prime} , q
)$, which is defined as
\begin{equation}
  \kappa ( q^{\prime} , q ) \equiv
  \frac{1}{( 2 \pi )^{3}}
  \sqrt{\frac{m_{B} m_{B^{\prime}}}
    {4 \omega _{M} ( q ) E_{B} ( q )
      \omega _{M^{\prime}} ( q^{\prime} ) E_{B^{\prime}} ( q^{\prime} )}} ,
\end{equation}
with $\omega _{M^{( \prime )}} ( q ) \equiv \sqrt{q^{2} + m_{M^{(
      \prime )}}}$ and $E_{B^{( \prime )}} ( q ) \equiv \sqrt{q^{2} +
  m_{B^{( \prime )}}^{2}}$, was introduced so as to satisfy the
optical theorem with the correct coefficients.  The
interaction~\eqref{eqA:Vfinal} is applicable to the
Lippmann--Schwinger equation~\eqref{eq:LSeq}.

\section{Explicit forms of the meson--baryon interactions}
\label{app:forms}

In this Appendix I show the explicit forms of the interactions $M (
k^{\mu} , \lambda _{M} ) B ( p^{\mu} , \lambda _{B} ) \to M^{\prime} (
k^{\prime \mu} , \lambda _{M^{\prime}}) B^{\prime} ( p^{\prime \mu} ,
\lambda _{B^{\prime}} )$ used in the present study.  The interactions
are written in terms of the helicity eigenstates, i.e., those in
Eq.~\eqref{eqA:Vhelicity}, except for the $s$-channel contributions of
the bare $N^{\ast}$ and $\Delta ^{\ast}$ states in
Appendix~\ref{app:bare}.

\begin{table}[!t]
  \caption{Isospin factors for the interactions.}
  \label{tab:iso}
  \begin{ruledtabular}
    \begin{tabular*}{8.6cm}{@{\extracolsep{\fill}}lcc}
      & $I = 1/2$ & $I = 3/2$ 
      \\
      \hline
      $\tau ^{j} \tau ^{i}$ & $3$ & $0$
      \\
      $\tau ^{i} \tau ^{j}$ & $-1$ & $2$
      \\
      $\delta _{i j}$ & $1$ & $1$
      \\
      $i \epsilon _{j i k} \tau ^{k}$ & $2$ & $-1$
      \\
      $T^{i} T^{\dagger j}$ & $4/3$ & $1/3$
      \\
      $\tau ^{i}$ & $-\sqrt{3}$ & $0$
      \\
      $\tau ^{j}$ & $-\sqrt{3}$ & $0$
      \\
      $T^{\dagger j} \tau ^{i}$ & $\sqrt{6}$ & $0$
      \\
      $T^{\dagger i} \tau ^{j}$ & $\sqrt{8/3}$ & $- \sqrt{5/3}$
      \\
      $T^{\dagger j}$ & $- \sqrt{2}$ & $0$
      \\
      $T^{\dagger j} T^{i}$ & $2$ & $0$
    \end{tabular*}
  \end{ruledtabular}
\end{table}%

For the incoming and outgoing nucleons, I express the helicity
eigenstates by the the Dirac spinors $u_{N} ( - \bm{q} , \lambda _{N}
)$ and $\bar{u}_{N} ( - \bm{q}^{\prime} , \lambda _{N^{\prime}} )$,
respectively.  The spinors for the incoming and outgoing $\rho$ meson
are $e_{\rho}^{\mu} ( \bm{q} , \lambda _{M} )$ and $e_{\rho}^{\mu
  \ast} ( \bm{q}^{\prime} , \lambda _{M^{\prime}} )$, respectively.
The helicity eigenstates of the $\Delta$ baryon as the
Rarita-Schwinger spinor are $u_{\Delta}^{\mu} ( - \bm{q} , \lambda
_{\Delta} )$ for the incoming and $\bar{u}_{\Delta}^{\mu} ( -
\bm{q}^{\prime} , \lambda _{\Delta ^{\prime}} )$ for the outgoing
states.  The explicit forms of the spinors are given in
Ref.~\cite{Sekihara:2018tsb}.

In this study I multiply a factor $i$ for the incoming $\sigma$ and
$\rho$ mesons and accordingly a factor $- i$ for the outgoing $\sigma$
and $\rho$ mesons to obtain a real-valued interaction.

\begin{widetext}

\subsection{\boldmath $\pi N \to \pi N$}

The $\pi ^{i} N \to \pi ^{j} N$ interactions, where the isospin
indices for mesons $i, j = 1, 2, 3$ correspond to those in
Eq.~\eqref{eq:pi}, are given as
\begin{equation}
  V_{\pi N \to \pi N} =
  \bar{u}_{N} ( - \bm{q}^{\prime} , \lambda _{N^{\prime}} )
  ( \bar{V}_{\text{1a}} + \bar{V}_{\text{1b}}
  + \bar{V}_{\text{1c}} + \bar{V}_{\text{1d}} + \bar{V}_{\text{1e}} )
  u_{N} ( - \bm{q} , \lambda _{N} )
  ,
\end{equation}
with
\begin{equation}
  \bar{V}_{\text{1a}} = ( \tau ^{j} \tau ^{i} ) 
  \left ( \frac{D + F}{2 f_{\pi}} \right ) ^{2}
  \Slash{k}^{\prime} \gamma _{5}
  \frac{S_{N} ( p + k ) + S_{N} ( p^{\prime} + k^{\prime} )}{2}
  \Slash{k} \gamma _{5}
  \mathcal{F} ( \Lambda _{\pi N N} , q^{\prime} )
  \mathcal{F} ( \Lambda _{\pi N N} , q )
  ,
\end{equation}
\begin{equation}
  \bar{V}_{\text{1b}} = ( \tau ^{i} \tau ^{j} ) 
  \left ( \frac{D + F}{2 f_{\pi}} \right ) ^{2}
  \Slash{k} \gamma _{5} 
  \frac{S_{N} ( p - k^{\prime} ) + S_{N} ( p^{\prime} - k )}{2}
  \Slash{k}^{\prime} \gamma _{5}
  \mathcal{F} ( \Lambda _{\pi N N} , q^{\prime} )
  \mathcal{F} ( \Lambda _{\pi N N} , q )
  ,
\end{equation}
\begin{equation}
  \bar{V}_{\text{1c}} = ( T^{i} T^{\dagger j} ) 
  \left ( \frac{f_{\pi N \Delta}}{m_{\pi}} \right ) ^{2}
  k_{\mu} 
  \left [ \frac{S_{\Delta}^{\mu \nu} ( p - k^{\prime} )
      + S_{\Delta}^{\mu \nu} ( p^{\prime} - k )}{2} \right ]
  k_{\nu}^{\prime}
  \mathcal{F} ( \Lambda _{\pi N \Delta} , q^{\prime} )
  \mathcal{F} ( \Lambda _{\pi N \Delta} , q )
  ,
\end{equation}
\begin{align}
  \bar{V}_{\text{1d}} = & ( i \epsilon _{j i k} \tau ^{k} )
  \frac{g_{\pi \pi \rho} g_{\rho N N}}{2} 
  \left [ \frac{S_{\rho} ( k - k^{\prime} )
      + S_{\rho} ( p^{\prime} - p )}{2} \right ]
  \notag \\ & \times
  \left \{ \Slash{k} + \Slash{k}^{\prime} + \frac{\kappa _{\rho}}{4 m_{N}}
    [ ( \Slash{k} + \Slash{k}^{\prime} ) ( \Slash{p} - \Slash{p}^{\prime} )
    - ( \Slash{p} - \Slash{p}^{\prime} ) ( \Slash{k} + \Slash{k}^{\prime} ) ]
    \right \}
  \mathcal{F} ( \Lambda _{\pi \pi \rho} , | \bm{q} - \bm{q}^{\prime} | )
  \mathcal{F} ( \Lambda _{\rho N N} , | \bm{q} - \bm{q}^{\prime} | )
  ,
\end{align}
\begin{equation}
  \bar{V}_{\text{1e}} = ( \delta _{i j} )
  \left ( - \frac{g_{\pi \pi \sigma} g_{\sigma N N}}{m_{\pi}} \right )
  k^{\mu} k_{\mu}^{\prime} \left [ \frac{S_{\sigma} ( k - k^{\prime} )
      + S_{\sigma} ( p^{\prime} - p )}{2} \right ]
  \mathcal{F} ( \Lambda _{\pi \pi \sigma} , | \bm{q} - \bm{q}^{\prime} | )
  \mathcal{F} ( \Lambda _{\sigma N N} , | \bm{q} - \bm{q}^{\prime} | )
  .
\end{equation}
The explicit values of the isospin factors are listed in
Table~\ref{tab:iso}.  Propagators $S_{N}$, $S_{\Delta}$, $S_{\rho}$,
and $S_{\sigma}$ are respectively:
\begin{equation}
  \begin{split}
    & S_{N} ( p )
    = \frac{\Slash{p} + m_{N}}{( p^{\mu} )^{2} - m_{N}^{2}} 
    ,
    \quad
    S_{\Delta}^{\mu \nu} ( p )
    = \frac{\Slash{p} + m_{\Delta}}{( p^{\mu} )^{2} - m_{\Delta}^{2}}
    \left [ - g^{\mu \nu} + \frac{1}{3} \gamma ^{\mu} \gamma ^{\nu}
      + \frac{2 p^{\mu} p^{\nu}}{3 m_{\Delta}^{2}}
      - \frac{p^{\mu} \gamma ^{\nu} - p^{\nu} \gamma ^{\mu}}{3 m_{\Delta}} \right ]
    ,
    \\
    & S_{\rho} ( p )
    = \frac{1}{( p^{\mu} )^{2} - m_{\rho}^{2}} 
    ,
    \quad
    S_{\sigma} ( p )
    = \frac{1}{( p^{\mu} )^{2} - m_{\sigma _{0}}^{2}} 
    ,
  \end{split}
\end{equation}
where the physical masses $m_{N}$, $m_{\Delta}$, $m_{\rho}$ are used
for the $N$, $\Delta$, and $\rho$ exchanges, while the bare mass
$m_{\sigma _{0}}$ is used for the $\sigma$ exchange.  The dipole form
factor $\mathcal{F}$ was defined in Eq.~\eqref{eq:FF}.  I note that I
use an idea of the unitary transformation method~\cite{Sato:1996gk,
  Sato:2009de} to calculate denominators of the propagators.  Owing to
the treatment of the energies of the four-momenta in
Eq.~\eqref{eqA:k0p0}, these interaction terms are independent of the
center-of-mass energy $E$.

\subsection{\boldmath $\pi N \to \eta N$}

The $\pi ^{i} N \to \eta N$ interactions, where the $\eta N$ state is
purely isospin $I=1/2$, are given as
\begin{equation}
  V_{\pi N \to \eta N} =
  \bar{u}_{N} ( - \bm{q}^{\prime} , \lambda _{N^{\prime}} )
  ( \bar{V}_{\text{2a}} + \bar{V}_{\text{2b}} )
  u_{N} ( - \bm{q} , \lambda _{N} )
  , 
\end{equation}
with
\begin{equation}
  \bar{V}_{\text{2a}} = ( \tau ^{i} )
  \left [ - \frac{( D + F ) ( D - 3 F )}{4 \sqrt{3} f_{\pi} f_{\eta}} \right ]
  \Slash{k}^{\prime} \gamma _{5}
  \frac{S_{N} ( p + k ) + S_{N} ( p^{\prime} + k^{\prime} )}{2}
  \Slash{k} \gamma _{5}
  \mathcal{F} ( \Lambda _{\eta N N} , q^{\prime} )
  \mathcal{F} ( \Lambda _{\pi N N} , q ) ,
\end{equation}
\begin{equation}
  \bar{V}_{\text{2b}} = ( \tau ^{i} )
  \left [ - \frac{( D + F ) ( D - 3 F )}{4 \sqrt{3} f_{\pi} f_{\eta}} \right ]
  \Slash{k} \gamma _{5}
  \frac{S_{N} ( p - k^{\prime} ) + S_{N} ( p^{\prime} - k )}{2}
  \Slash{k}^{\prime} \gamma _{5} 
  \mathcal{F} ( \Lambda _{\eta N N} , q^{\prime} )
  \mathcal{F} ( \Lambda _{\pi N N} , q ) .
\end{equation}

\subsection{\boldmath $\eta N \to \eta N$}

The $\eta N \to \eta N$ interactions are given as
\begin{equation}
  V_{\eta N \to \eta N} =
  \bar{u}_{N} ( - \bm{q}^{\prime} , \lambda _{N^{\prime}} )
  ( \bar{V}_{\text{3a}} + \bar{V}_{\text{3b}} )
  u_{N} ( - \bm{q} , \lambda _{N} )
  , 
\end{equation}
with
\begin{equation}
  \bar{V}_{\text{3a}} =
  \left ( \frac{D - 3 F}{2 \sqrt{3} f_{\eta}} \right ) ^{2}
  \Slash{k}^{\prime}
  \gamma _{5}
  \frac{S_{N} ( p + k ) + S_{N} ( p^{\prime} + k^{\prime} )}{2}
  \Slash{k} \gamma _{5}
  \mathcal{F} ( \Lambda _{\eta N N} , q^{\prime} )
  \mathcal{F} ( \Lambda _{\eta N N} , q ) ,
\end{equation}
\begin{equation}
  \bar{V}_{\text{3b}} = 
  \left ( \frac{D - 3 F}{2 \sqrt{3} f_{\eta}} \right ) ^{2}
  \Slash{k} \gamma _{5}
  \frac{S_{N} ( p - k^{\prime} ) + S_{N} ( p^{\prime} - k )}{2}
  \Slash{k}^{\prime} \gamma _{5}
  \mathcal{F} ( \Lambda _{\eta N N} , q^{\prime} )
  \mathcal{F} ( \Lambda _{\eta N N} , q )  
  .
\end{equation}

\subsection{\boldmath $\pi N \to \sigma N$}

The $\pi ^{i} N \to \sigma N$ interactions, where the $\sigma
N$ state is purely $I = 1/2$, are given as
\begin{equation}
  V_{\pi N \to \sigma N} =
  \bar{u}_{N} ( - \bm{q}^{\prime} , \lambda _{N^{\prime}} )
  ( \bar{V}_{\text{4a}} + \bar{V}_{\text{4b}} + \bar{V}_{\text{4c}} )
  u_{N} ( - \bm{q} , \lambda _{N} ) ,
\end{equation}
with
\begin{equation}
  \bar{V}_{\text{4a}} = ( \tau ^{i} )
  \frac{g_{\sigma N N} ( D + F )}{2 f_{\pi}}
  \frac{S_{N} ( p + k ) + S_{N} ( p^{\prime} + k^{\prime} )}{2}
  \Slash{k} \gamma _{5}
  \mathcal{F} ( \Lambda _{\sigma N N} , q^{\prime} )
  \mathcal{F} ( \Lambda _{\pi N N} , q )
  ,
\end{equation}
\begin{equation}
  \bar{V}_{\text{4b}} = ( \tau ^{i} )
  \frac{g_{\sigma N N} ( D + F )}{2 f_{\pi}}
  \Slash{k} \gamma _{5}
  \frac{S_{N} ( p - k^{\prime} ) + S_{N} ( p^{\prime} - k )}{2}
  \mathcal{F} ( \Lambda _{\sigma N N} , q^{\prime} )
  \mathcal{F} ( \Lambda _{\pi N N} , q ) ,
\end{equation}
\begin{equation}
  \bar{V}_{\text{4c}} = ( \tau ^{i} )
    \left [ - \frac{g_{\pi \pi \sigma} ( D + F )}{2 m_{\pi} f_{\pi}} \right ]
    k_{\mu} ( k - k^{\prime} )^{\mu}
    ( \Slash{k} - \Slash{k}^{\prime} ) \gamma _{5}
    S_{\pi} ( p^{\prime} - p )
    \mathcal{F} ( \Lambda _{\pi \pi \sigma} , | \bm{q} - \bm{q}^{\prime} | )
    \mathcal{F} ( \Lambda _{\pi N N} , | \bm{q} - \bm{q}^{\prime} | ) 
    .
\end{equation}
The $\pi$ propagator is
\begin{equation}
  S_{\pi} ( p ) = \frac{1}{( p^{\mu} )^{2} - m_{\pi}^{2}} ,
\end{equation}
with the physical pion mass $m_{\pi}$.  I do not include the pion
propagator of $S_{\pi} ( k - k^{\prime} )$ for $\bar{V}_{\text{4c}}$
because the $\pi \pi \sigma$ vertex interaction is ``real'' and hence
it causes divergence.  Similarly, I will omit propagators of momenta
associated with the ``real'' vertex interactions.

\subsection{\boldmath $\eta N \to \sigma N$}

The $\eta N \to \sigma N$ interactions are given as
\begin{equation}
  V_{\eta N \to \sigma N} =
  \bar{u}_{N} ( - \bm{q}^{\prime} , \lambda _{N^{\prime}} )
  ( \bar{V}_{\text{5a}} + \bar{V}_{\text{5b}} )
  u_{N} ( - \bm{q} , \lambda _{N} )
  ,
\end{equation}
with
\begin{equation}
  \bar{V}_{\text{5a}} =
  - \frac{g_{\sigma N N} ( D - 3 F )}{2 \sqrt{3} f_{\eta}} 
  \frac{S_{N} ( p + k ) + S_{N} ( p^{\prime} + k^{\prime} )}{2}
  \Slash{k} \gamma _{5}
  \mathcal{F} ( \Lambda _{\sigma N N} , q^{\prime} )
  \mathcal{F} ( \Lambda _{\eta N N} , q )
    ,  
\end{equation}
\begin{equation}
  \bar{V}_{\text{5b}} =
  - \frac{g_{\sigma N N} ( D - 3 F )}{2 \sqrt{3} f_{\eta}} 
  \Slash{k} \gamma _{5}
  \frac{S_{N} ( p - k^{\prime} ) + S_{N} ( p^{\prime} - k )}{2}
  \mathcal{F} ( \Lambda _{\sigma N N} , q^{\prime} )
  \mathcal{F} ( \Lambda _{\eta N N} , q )
  .
\end{equation}

\subsection{\boldmath $\sigma N \to \sigma N$}

The $\sigma N \to \sigma N$ interactions are given as
\begin{equation}
  V_{\sigma N \to \sigma N} =
  \bar{u}_{N} ( - \bm{q}^{\prime} , \lambda _{N^{\prime}} )
  ( \bar{V}_{\text{6a}} + \bar{V}_{\text{6b}} )
  u_{N} ( - \bm{q} , \lambda _{N} )
  , 
\end{equation}
with
\begin{equation}
  \bar{V}_{\text{6a}} = g_{\sigma N N}^{2}
  \frac{S_{N} ( p + k ) + S_{N} ( p^{\prime} + k^{\prime} )}{2} 
  \mathcal{F} ( \Lambda _{\sigma N N} , q^{\prime} )
  \mathcal{F} ( \Lambda _{\sigma N N} , q ) ,
\end{equation}
\begin{equation}
  \bar{V}_{\text{6b}} = g_{\sigma N N}^{2}
  \frac{S_{N} ( p - k^{\prime} ) + S_{N} ( p^{\prime} - k )}{2}
  \mathcal{F} ( \Lambda _{\sigma N N} , q^{\prime} )
  \mathcal{F} ( \Lambda _{\sigma N N} , q )
  .
\end{equation}

\subsection{\boldmath $\pi N \to \rho N$}

The $\pi ^{i} N \to \rho ^{j} N$ interactions are given as
\begin{equation}
  V_{\pi N \to \rho N}
  =
  e_{\rho \mu}^{\ast} ( \bm{q}^{\prime} , \lambda _{M^{\prime}} )
  \bar{u}_{N} ( - \bm{q}^{\prime} , \lambda _{N^{\prime}} )
  ( \bar{V}_{\text{7a}}^{\mu} + \bar{V}_{\text{7b}}^{\mu}
  + \bar{V}_{\text{7c}}^{\mu} + \bar{V}_{\text{7d}}^{\mu} )
  u_{N} ( - \bm{q} , \lambda _{N} )
  , 
\end{equation}
with
\begin{equation}
  \bar{V}_{\text{7a}}^{\mu} = ( \tau ^{j} \tau ^{i} )
  \frac{g_{\rho N N} ( D + F )}{4 f_{\pi}}
  \left [
    \gamma ^{\mu}
    + \frac{\kappa _{\rho}}{4 m_{N}}
    ( \gamma ^{\mu} \Slash{k}^{\prime} - \Slash{k}^{\prime} \gamma ^{\mu} )
  \right ]
  \frac{S_{N} ( p + k ) + S_{N} ( p^{\prime} + k^{\prime} )}{2}
  \Slash{k} \gamma _{5}
  \mathcal{F} ( \Lambda _{\rho N N} , q^{\prime} )
  \mathcal{F} ( \Lambda _{\pi N N} , q )
  ,
\end{equation}
\begin{equation}
  \bar{V}_{\text{7b}}^{\mu} = ( \tau ^{i} \tau ^{j} )
  \frac{g_{\rho N N} ( D + F )}{4 f_{\pi}}
  \Slash{k} \gamma _{5}
  \frac{S_{N} ( p - k^{\prime} ) + S_{N} ( p^{\prime} - k )}{2}
  \left [ \gamma ^{\mu}
    + \frac{\kappa _{\rho}}{4 m_{N}}
    ( \gamma ^{\mu} \Slash{k}^{\prime} - \Slash{k}^{\prime} \gamma ^{\mu} )
    \right ]
  \mathcal{F} ( \Lambda _{\rho N N} , q^{\prime} )
  \mathcal{F} ( \Lambda _{\pi N N} , q )
,
\end{equation}
\begin{equation}
  \bar{V}_{\text{7c}}^{\mu} = ( i \epsilon _{j i k} \tau ^{k} )
  \frac{g_{\pi \pi \rho} ( D + F )}{2 f_{\pi}}
  ( 2 k^{\mu} - k^{\prime \mu} )
  ( \Slash{k} - \Slash{k}^{\prime} ) \gamma _{5} 
  S_{\pi} ( p^{\prime} - p )
  \mathcal{F} ( \Lambda _{\pi \pi \rho} , | \bm{q} - \bm{q}^{\prime} | )
  \mathcal{F} ( \Lambda _{\pi N N} , | \bm{q} - \bm{q}^{\prime} | ) ,
\end{equation}
\begin{equation}
  \bar{V}_{\text{7d}}^{\mu} = ( i \epsilon _{j i k} \tau ^{k} )
  \left [ - \frac{g_{\rho N N} ( D + F )}{2 f_{\pi}} \right ]
  \gamma ^{\mu}
  \gamma _{5}
  \mathcal{F} ( \Lambda _{\rho N N} , q^{\prime} )
  \mathcal{F} ( \Lambda _{\pi N N} , q )  .
\end{equation}

\subsection{\boldmath $\eta N \to \rho N$}

The $\eta N \to \rho ^{j} N$ interactions are given as
\begin{equation}
  V_{\eta N \to \rho N} =
  e_{\rho \mu}^{\ast} ( \bm{q}^{\prime} , \lambda _{M^{\prime}} )
  \bar{u}_{N} ( - \bm{q}^{\prime} , \lambda _{N^{\prime}} )
  ( \bar{V}_{\text{8a}}^{\mu} + \bar{V}_{\text{8b}}^{\mu} )
  u_{N} ( - \bm{q} , \lambda _{N} )
  , 
\end{equation}
with
\begin{equation}
  \bar{V}_{\text{8a}}^{\mu} = ( \tau ^{j} )
  \left [ - \frac{g_{\rho N N} ( D - 3 F )}{4 \sqrt{3} f_{\eta}} \right ]
  \left [ \gamma ^{\mu}
    + \frac{\kappa _{\rho}}{4 m_{N}}
    ( \gamma ^{\mu} \Slash{k}^{\prime} - \Slash{k}^{\prime} \gamma ^{\mu} )
    \right ]    
  \frac{S_{N} ( p + k ) + S_{N} ( p^{\prime} + k^{\prime} )}{2}
  \Slash{k} \gamma _{5}
  \mathcal{F} ( \Lambda _{\rho N N} , q^{\prime} )
  \mathcal{F} ( \Lambda _{\eta N N} , q )
  ,
\end{equation}
\begin{equation}
  \bar{V}_{\text{8b}}^{\mu} = ( \tau ^{j} )
  \left [ - \frac{g_{\rho N N} ( D - 3 F )}{4 \sqrt{3} f_{\eta}} \right ]
  \Slash{k} \gamma _{5}
  \frac{S_{N} ( p - k^{\prime} ) + S_{N} ( p^{\prime} - k )}{2}
  \left [ \gamma ^{\mu}
    + \frac{\kappa _{\rho}}{4 m_{N}}
    ( \gamma ^{\mu} \Slash{k}^{\prime} - \Slash{k}^{\prime} \gamma ^{\mu} )
    \right ]    
  \mathcal{F} ( \Lambda _{\rho N N} , q^{\prime} )
  \mathcal{F} ( \Lambda _{\eta N N} , q )
  .
\end{equation}

\subsection{\boldmath $\sigma N \to \rho N$}

The $\sigma N \to \rho ^{j} N$ interactions are given as
\begin{equation}
  V_{\sigma N \to \rho N} =
  e_{\rho \mu}^{\ast} ( \bm{q}^{\prime} , \lambda _{M^{\prime}} )
  \bar{u}_{N} ( - \bm{q}^{\prime} , \lambda _{N^{\prime}} )
  ( \bar{V}_{\text{9a}}^{\mu} + \bar{V}_{\text{9b}}^{\mu} )
  u_{N} ( - \bm{q} , \lambda _{N} )
  , 
\end{equation}
with
\begin{equation}
  \bar{V}_{\text{9a}}^{\mu} = ( \tau ^{j} )
  \frac{g_{\rho N N} g_{\sigma N N}}{2}
  \left [ \gamma ^{\mu}
    + \frac{\kappa _{\rho}}{4 m_{N}}
    ( \gamma ^{\mu} \Slash{k}^{\prime} - \Slash{k}^{\prime} \gamma ^{\mu} )
    \right ]
  \frac{S_{N} ( p + k ) + S_{N} ( p^{\prime} + k^{\prime} )}{2}
  \mathcal{F} ( \Lambda _{\rho N N} , q^{\prime} )
  \mathcal{F} ( \Lambda _{\sigma N N} , q ) ,
\end{equation}
\begin{equation}
  \bar{V}_{\text{9b}}^{\mu} = ( \tau ^{j} )
  \frac{g_{\rho N N} g_{\sigma N N}}{2}
  \frac{S_{N} ( p - k^{\prime} ) + S_{N} ( p^{\prime} - k )}{2}
  \left [ \gamma ^{\mu}
    + \frac{\kappa _{\rho}}{4 m_{N}}
    ( \gamma ^{\mu} \Slash{k}^{\prime} - \Slash{k}^{\prime} \gamma ^{\mu} )
    \right ]
  \mathcal{F} ( \Lambda _{\rho N N} , q^{\prime} )
  \mathcal{F} ( \Lambda _{\sigma N N} , q ) .
\end{equation}

\subsection{\boldmath $\rho N \to \rho N$}

The $\rho ^{i} N \to \rho ^{j} N$ interactions are given as
\begin{equation}
  V_{\rho N \to \rho N} =
  e_{\rho \mu}^{\ast} ( \bm{q}^{\prime} , \lambda _{M^{\prime}} )
  \bar{u}_{N} ( - \bm{q}^{\prime} , \lambda _{N^{\prime}} )
  ( \bar{V}_{\text{10a}}^{\mu \nu} + \bar{V}_{\text{10b}}^{\mu \nu}
  + \bar{V}_{\text{10c}}^{\mu \nu} )
  e_{\rho \nu} ( \bm{q} , \lambda _{M} )
  u_{N} ( - \bm{q} , \lambda _{N} )
  , 
\end{equation}
with
\begin{align}
  \bar{V}_{\text{10a}}^{\mu \nu} = & ( \tau ^{j} \tau ^{i} )
  \frac{g_{\rho N N}^{2}}{4}
  \left [
    \gamma ^{\mu}
    + \frac{\kappa _{\rho}}{4 m_{N}}
    ( \gamma ^{\mu} \Slash{k}^{\prime} - \Slash{k}^{\prime} \gamma ^{\mu} )
  \right ]
  \frac{S_{N} ( p + k ) + S_{N} ( p^{\prime} + k^{\prime} )}{2}
  \left [ \gamma ^{\nu}
    - \frac{\kappa _{\rho}}{4 m_{N}}
    ( \gamma ^{\nu} \Slash{k} - \Slash{k} \gamma ^{\nu} ) \right ]
  \notag \\ & \times
  \mathcal{F} ( \Lambda _{\rho N N} , q^{\prime} )
  \mathcal{F} ( \Lambda _{\rho N N} , q )
  ,
\end{align}
\begin{align}
  \bar{V}_{\text{10b}}^{\mu \nu} = & ( \tau ^{i} \tau ^{j} )
  \frac{g_{\rho N N}^{2}}{4}
  \left [ \gamma ^{\nu}
    - \frac{\kappa _{\rho}}{4 m_{N}}
    ( \gamma ^{\nu} \Slash{k} - \Slash{k} \gamma ^{\nu} ) \right ]
  \frac{S_{N} ( p - k^{\prime} ) + S_{N} ( p^{\prime} - k )}{2}
  \left [
    \gamma ^{\mu}
    + \frac{\kappa _{\rho}}{4 m_{N}}
    ( \gamma ^{\mu} \Slash{k}^{\prime} - \Slash{k}^{\prime} \gamma ^{\mu} )
  \right ]
  \notag \\ & \times
  \mathcal{F} ( \Lambda _{\rho N N} , q^{\prime} )
  \mathcal{F} ( \Lambda _{\rho N N} , q )
  ,
\end{align}
\begin{equation}
  \bar{V}_{\text{10c}}^{\mu \nu} = ( i \epsilon _{j i k} \tau ^{k} )
  \frac{g_{\rho N N}^{2} \kappa _{\rho}}{8 m_{N}}
  ( \gamma ^{\nu} \gamma ^{\mu} - \gamma ^{\mu} \gamma ^{\nu} )
  \mathcal{F} ( \Lambda _{\rho N N} , q^{\prime} )
  \mathcal{F} ( \Lambda _{\rho N N} , q ) .
\end{equation}

\subsection{\boldmath $\pi N \to \pi \Delta$}

The $\pi ^{i} N \to \pi ^{j} \Delta$ interaction is given as
\begin{equation}
  V_{\pi N \to \pi \Delta}
  = \bar{u}_{\Delta \mu} ( - \bm{q}^{\prime} , \lambda _{\Delta ^{\prime}} )
  ( \bar{V}_{\text{11a}}^{\mu} + \bar{V}_{\text{11b}}^{\mu} )
  u_{N} ( - \bm{q} , \lambda _{N} )
  , 
\end{equation}
with
\begin{equation}
  \bar{V}_{\text{11a}}^{\mu} = ( T^{\dagger j} \tau ^{i} )
  \frac{f_{\pi N \Delta} ( D + F )}{2 m_{\pi} f_{\pi}}
  k^{\prime \mu}
  \frac{S_{N} ( p + k ) + S_{N} ( p^{\prime} + k^{\prime} )}{2} 
  \Slash{k} \gamma _{5}
  \mathcal{F} ( \Lambda _{\pi N \Delta} , q^{\prime} )
  \mathcal{F} ( \Lambda _{\pi N N} , q ) ,
\end{equation}
\begin{equation}
  \bar{V}_{\text{11b}}^{\mu} = ( T^{\dagger i} \tau ^{j} )
  \frac{f_{\pi N \Delta} ( D + F )}{2 m_{\pi} f_{\pi}}
  k^{\mu} S_{N} ( p - k^{\prime} )
  \Slash{k}^{\prime} \gamma _{5}
  \mathcal{F} ( \Lambda _{\pi N N} , q^{\prime} )
  \mathcal{F} ( \Lambda _{\pi N \Delta} , q )
  .
\end{equation}

\subsection{\boldmath $\eta N \to \pi \Delta$}

The $\eta N \to \pi ^{j} \Delta$ interaction is given as
\begin{equation}
  V_{\eta N \to \pi \Delta}
  = \bar{u}_{\Delta \mu} ( - \bm{q}^{\prime} , \lambda _{\Delta ^{\prime}} )
  \bar{V}_{\text{12}}^{\mu}
  u_{N} ( - \bm{q} , \lambda _{N} )
  , 
\end{equation}
with
\begin{equation}
  \bar{V}_{\text{12}}^{\mu} = ( T^{\dagger j} )
  \left [ - \frac{f_{\pi N \Delta} ( D - 3 F )}{2 \sqrt{3} m_{\pi} f_{\eta}}
    \right ]
  k^{\prime \mu}
  \frac{S_{N} ( p + k ) + S_{N} ( p^{\prime} + k^{\prime} )}{2} 
  \Slash{k} \gamma _{5}
  \mathcal{F} ( \Lambda _{\pi N \Delta} , q^{\prime} )
  \mathcal{F} ( \Lambda _{\eta N N} , q ) .
\end{equation}

\subsection{\boldmath $\sigma N \to \pi \Delta$}

The $\sigma N \to \pi ^{j} \Delta$ interaction is given as
\begin{equation}
  V_{\sigma N \to \pi \Delta}
  = \bar{u}_{\Delta \mu} ( - \bm{q}^{\prime} , \lambda _{\Delta ^{\prime}} )
  \bar{V}_{\text{13}}^{\mu}
  u_{N} ( - \bm{q} , \lambda _{N} )
  , 
\end{equation}
with
\begin{equation}
  \bar{V}_{\text{13}}^{\mu} = ( T^{\dagger j} )
  \frac{g_{\sigma N N} f_{\pi N \Delta}}{m_{\pi}}
  k^{\prime \mu}
  \frac{S_{N} ( p + k ) + S_{N} ( p^{\prime} + k^{\prime} )}{2} 
  \mathcal{F} ( \Lambda _{\pi N \Delta} , q^{\prime} )
  \mathcal{F} ( \Lambda _{\sigma N N} , q ) .
\end{equation}

\subsection{\boldmath $\rho N \to \pi \Delta$}

The $\rho ^{i} N \to \pi ^{j} \Delta$ interaction is given as
\begin{equation}
  V_{\rho N \to \pi \Delta}
  = \bar{u}_{\Delta \mu} ( - \bm{q}^{\prime} , \lambda _{\Delta ^{\prime}} )
  \bar{V}_{\text{14}}^{\mu \nu}
  e_{\rho \nu} ( \bm{q} , \lambda _{M} )
  u_{N} ( - \bm{q} , \lambda _{N} )
  , 
\end{equation}
with
\begin{equation}
  \bar{V}_{\text{14}}^{\mu \nu} = ( T^{\dagger j} \tau ^{i} )
  \frac{g_{\rho N N} f_{\pi N \Delta}}{2 m_{\pi}}
  k^{\prime \mu}
  \frac{S_{N} ( p + k ) + S_{N} ( p^{\prime} + k^{\prime} )}{2}
  \left [ \gamma ^{\nu}
    - \frac{\kappa _{\rho}}{4 m_{N}}
    ( \gamma ^{\nu} \Slash{k} - \Slash{k} \gamma ^{\nu} ) \right ]
  \mathcal{F} ( \Lambda _{\pi N \Delta} , q^{\prime} )
  \mathcal{F} ( \Lambda _{\rho N N} , q ) .
\end{equation}

\subsection{\boldmath $\pi \Delta \to \pi \Delta$}

The $\pi ^{i} \Delta \to \pi ^{j} \Delta$ interaction is given as
\begin{equation}
  V_{\pi \Delta \to \pi \Delta} =
  \bar{u}_{\Delta \mu} ( - \bm{q}^{\prime} , \lambda _{\Delta ^{\prime}} )
  \bar{V}_{\text{15}}^{\mu \nu}
  u_{\Delta \nu} ( - \bm{q} , \lambda _{\Delta} )
  , 
\end{equation}
with
\begin{equation}
  \bar{V}_{\text{15}}^{\mu \nu} = ( T^{\dagger j} T^{i} )
  \left ( \frac{f_{\pi N \Delta}}{m_{\pi}} \right ) ^{2}
  k^{\prime \mu}
  \frac{S_{N} ( p + k ) + S_{N} ( p^{\prime} + k^{\prime} )}{2} 
  k^{\nu}
  \mathcal{F} ( \Lambda _{\pi N \Delta} , q^{\prime} )
  \mathcal{F} ( \Lambda _{\pi N \Delta} , q ) .
\end{equation}

\subsection{\boldmath $s$-channel exchange of
  bare $N^{\ast}$ and $\Delta ^{\ast}$ states}
\label{app:bare}

To take into account the bare $N^{\ast}$ and $\Delta ^{\ast}$ states
for the meson--baryon scattering in $s$ channel, I add
\begin{equation}
  V_{j k}^{\rm bare} ( E ; q^{\prime} , q )
  = 
  \frac{g_{j} g_{k}}
       {2 m_{\pi} ( E - M_{0} )}
  \left ( \frac{q^{\prime}}{m_{\pi}} \right )^{L^{\prime}}
  \left ( \frac{q}{m_{\pi}} \right )^{L}
  \mathcal{F} ( \Lambda , q^{\prime} )
  \mathcal{F} ( \Lambda , q ) ,
\end{equation}
to the corresponding partial-wave components $V_{\alpha , j k}$ in
Eq.~\eqref{eq:LSeq}.  Here, $M_{0}$ is the bare mass of the $N^{\ast}$
and $\Delta ^{\ast}$ states, $g_{j}$ is the coupling constant for the
bare state to the $j$th meson--baryon channel, and $\Lambda$ is the
cutoff.  I note that only this bare-state contribution depends on the
center-of-mass energy $E$ among the meson--baryon interaction terms.

\end{widetext}

\end{document}